\documentclass[onecolumn]{aastex63}

\received{June 15, 2020}
\revised{September 4, 2020}
\accepted{October 12, 2020}
\published{January 22, 2021}
\submitjournal{PSJ}

\usepackage{amssymb,amsmath}
\usepackage{gensymb} 
\usepackage{booktabs}

\usepackage{graphicx}
\graphicspath{{figures/}}
\makeatletter
\def\ScaleWidthIfNeeded{%
 \ifdim\Gin@nat@width>\linewidth
    \linewidth
  \else
    \Gin@nat@width
  \fi
}
\def\ScaleHeightIfNeeded{%
  \ifdim\Gin@nat@height>0.9\textheight
    0.9\textheight
  \else
    \Gin@nat@width
  \fi
}
\makeatother
\setkeys{Gin}{width=\ScaleWidthIfNeeded,height=\ScaleHeightIfNeeded,keepaspectratio}%

\newcommand{\ibid}{\textit{ibid}}
\newcommand{\pericenter}{\varpi}
\newcommand{\meanAnom}{\mathcal{M}}
\newcommand{\node}{\Omega}
\newcommand{\newtonG}{\mathcal{G}}

\newcommand{\loveJ}{\text{K}_{lmpq,\;j}}
\newcommand{\loveSgnJ}{\widetilde{\text{K}}_{lmpq,\;j}}
\newcommand{\loveI}{\text{K}_{lmpq,\;i}}
\newcommand{\loveSgnI}{\widetilde{\text{K}}_{lmpq,\;i}}

\newcommand{\spin}{\dot{\theta}}
\newcommand{\avgd}[1]{\left<#1\right>}
\newcommand{\partialDerv}[2]{\frac{\partial #1}{\partial #2}}
\newcommand{\partialDervLine}[2]{\partial #1/\partial #2}

\shorttitle{Dual Dissipation for Eccentric Non-synchronously Rotating Systems}
\shortauthors{Renaud et al.}
\begin{document}

\title{Tidal Dissipation in Dual-Body, Highly Eccentric, and Non-synchronously Rotating Systems: \\ 
Applications to Pluto-Charon and the Exoplanet TRAPPIST-1e}







\correspondingauthor{Joe Renaud}
\email{joseph.p.renaud@nasa.gov}

\author[0000-0002-8619-8542]{Joe P. Renaud}
\affiliation{Universities Space Research Association \\
7178 Columbia Gateway Dr. \\
Columbia, MD 21046, USA}
\affiliation{NASA Goddard Space Flight Center \\
8800 Greenbelt Rd. \\
Greenbelt, MD 20771, USA}
\affiliation{Sellers Exoplanet Environments Collaboration, \\
NASA Goddard Space Flight Center}

\author[0000-0001-9788-2799]{Wade G. Henning}
\affiliation{University of Maryland \\
College Park, MD 20742, USA}
\affiliation{NASA Goddard Space Flight Center \\
8800 Greenbelt Rd. \\
Greenbelt, MD 20771, USA}
\affiliation{Sellers Exoplanet Environments Collaboration, \\
NASA Goddard Space Flight Center}

\author{Prabal Saxena}
\affiliation{University of Maryland \\
College Park, MD 20742, USA}
\affiliation{NASA Goddard Space Flight Center \\
8800 Greenbelt Rd. \\
Greenbelt, MD 20771, USA}
\affiliation{Sellers Exoplanet Environments Collaboration, \\
NASA Goddard Space Flight Center}

\author[0000-0002-6220-2869]{Marc Neveu}
\affiliation{University of Maryland \\
College Park, MD 20742, USA}
\affiliation{NASA Goddard Space Flight Center \\
8800 Greenbelt Rd. \\
Greenbelt, MD 20771, USA}

\author{Amirhossein Bagheri}
\affiliation{Institute of Geophysics, ETH Z\"{u}rich \\
Z\"{u}rich, Switzerland}

\author[0000-0002-8119-3355]{Avi Mandell}
\affiliation{NASA Goddard Space Flight Center \\
8800 Greenbelt Rd. \\
Greenbelt, MD 20771, USA}
\affiliation{Sellers Exoplanet Environments Collaboration, \\
NASA Goddard Space Flight Center}

\author{Terry Hurford}
\affiliation{NASA Goddard Space Flight Center \\
8800 Greenbelt Rd. \\
Greenbelt, MD 20771, USA}
\affiliation{Sellers Exoplanet Environments Collaboration, \\
NASA Goddard Space Flight Center}

\begin{abstract}
Using the Andrade-derived Sundberg-Cooper rheology, we apply several improvements to the secular tidal evolution of TRAPPIST-1e and the early history of Pluto-Charon under the simplifying assumption of homogeneous bodies. By including higher-order eccentricity terms (up to and including $e^{20}$), we find divergences from the traditionally used $e^{2}$ truncation starting around $e=0.1$. Order-of-magnitude differences begin to occur for $e>0.6$. Critically, higher-order eccentricity terms activate additional spin-orbit resonances. Worlds experiencing non-synchronous rotation can fall into and out of these resonances, altering their long-term evolution. Non-zero obliquity generally does not generate significantly higher heating; however, it can considerably alter orbital and rotational evolution. Much like eccentricity, obliquity can activate new tidal modes and resonances. Tracking the dual-body dissipation within Pluto and Charon leads to faster evolution and dramatically different orbital outcomes. \\
Based on our findings, we recommend future tidal studies on worlds with $e\geq0.3$ to take into account additional eccentricity terms beyond $e^{2}$. This threshold should be lowered to $e>0.1$ if non-synchronous rotation or non-zero obliquity is under consideration. Due to the poor convergence of the eccentricity functions, studies on worlds that may experience very high eccentricity ($e \geq 0.6$) should include terms with high powers of eccentricity. We provide these equations up to $e^{10}$ for arbitrary obliquity and non-synchronous rotation. Finally, the assumption that short-period, solid-body exoplanets with $e\gtrsim0.1$ are tidally locked in their 1:1 spin-orbit resonance should be reconsidered. Higher-order spin-orbit resonances can exist even at these relatively modest eccentricities, while previous studies have found such resonances can significantly alter stellar-driven climate.

\end{abstract}
\keywords{Tidal friction (1698), Orbital theory (1182), Exoplanet tides (497), Exoplanet evolution (491), Trans-Neptunian objects (1705)}

\section{Introduction}\label{sec:intro}

New observations of extrasolar planets and Solar System objects are motivating a resurgence in improved modeling of tidal dissipation. Fundamental questions remain for both local and extrasolar settings. For instance, how does a system of two worlds (be it a star and an exoplanet, or a Solar System planet and its moon), each with distinct internal structures, tidally evolve on long timescales? New advancements in tidal theory \citep{BoueEfroimsky2019} as well as improved material modeling \citep{SundbergCooper2010, JacksonFaul2010, McCarthyCastillo-Rogez2013} set the stage to reexamine this and other questions with new fidelity. In this study, we couple the latest tidal evolution framework to advanced rheological modeling that describes a world's ability to dissipate tidal energy. We then apply this model to two systems that may experience strong repercussions from these changes, assuming either a homogeneous interior or (for Section \ref{sec:results:tno} \textit{only}) a simple multilayer model (both described in Section \ref{sec:method:interior_model}).\par

Tides provide a conduit to extract the energy stored in a celestial body's orbit or rotation, then transform it into internal heat via friction. This couples the thermal evolution of a world undergoing tidal friction to changes in its orbit and rotation \citep{MD}. The efficiency at which energy is converted is dependent upon the object's physical bulk properties, such as viscosity and rigidity \citep[e.g.,][]{Kaula1964}. These properties are strong functions of temperature, which further intensifies the link between the orbital, rotational, and thermal evolution. In systems of two or more worlds, the general practice in tidal theory is to first quantify which body, if any, dominates the overall rate of dissipation, and thus will govern the orbital evolution. The assumption that one body is dominating the dissipation can greatly simplify analysis, such as by reducing the number of terms in governing equations by half. However, for many real systems, both worlds may dissipate strongly enough to affect the system's evolution, as is the case for Io and Jupiter \citep{HussmannSpohn2004}. In such cases, a dual-body dissipation model is required. Binary systems, where two co-orbiting bodies have very similar mass, naturally lack one clear dominant source of dissipation. In cases such as Pluto and Charon \citep{farinella1979tidal, dobrovolskis1997dynamics, Cheng2014, BarrCollins2015}, or the early Earth and Moon \citep{touma1998resonances, CanupAsphaug2001, CukStewart2012, zahnle2015tethered, rufu2020tidal}, the threshold for when a dual-dissipation model is required is not always clear. However, \textit{without} starting from a dual-dissipation model in such circumstances, it is impossible to know if any approximation is valid. Therefore, the development and testing of the best theoretical set of governing equations available is a critical starting point. \par

Prior studies have investigated the long-term tidal evolution of planets experiencing non-synchronous rotation (NSR) \citep{Ferraz-Mello2008, Barnes2017} as well as dual-body dissipation \citep{Jackson2008ApJ678, Barnes2008AstroBio, Correia2009, Heller2011}. Yet, these and similar investigations estimate the efficiency of a world's tidal dissipation by using the Constant Time Lag (CTL) or Constant Phase Lag (CPL) models. The former assumes the efficiency can be modeled by the inverse of a single scalar \textit{tidal quality factor}, $Q^{-1}$. The latter generally assumes there is a linear relationship between dissipation efficiency and forcing frequency. This is often denoted with a frequency-dependent quality factor, $Q^{-1}(f)$. Studies have shown that the CPL and CTL models can describe the dissipation within giant gaseous planets and stars with reasonable accuracy for the frequency bands of primary interest \citep{Ferraz-Mello2020}. However, experiments on solids and semi-solids (e.g., rocks, partially melted rocks, and ices) have found that their response to shear forces (such as tidal forces) are far more complicated, requiring additional dependencies on temperature and frequency \citep[][and references therein]{Henning2009}, than can be described by the CPL and CTL models. Recent studies have begun to replace the CPL and CTL methods with more realistic rheological responses \replaced{\citep{Castillo-Rogez2011, Henning2009, Makarov2013, Behounkova2014, ShojiKurita2014, harada2014strong, BiersonNimmo2016, RenaudHenning2018, Khan2018, Bagheri2019a, Samuel2019}}{\citep{Castillo-Rogez2011, Henning2009, Makarov2013, Behounkova2014, Correia2014, harada2014strong, ShojiKurita2014, BiersonNimmo2016, RenaudHenning2018, Khan2018, Bagheri2019a, Samuel2019}}. Even more recently, such realistic rheologies have been combined with the latest advances in spin-orbit evolution modeling and applied to the dual-dissipation of Mars and its moons to determine their genesis \citep{Bagheri2019b}, as well as to the Kepler-21 exoplanet system \citep{Luna2020}. The stability of higher-order SOR's has also been recently explored as a function of rheological parameters \citep{Walterova2020}. Here, we build upon and extend this and other prior work to explore dual-dissipation in the context of icy worlds and exoplanets. We also explore the impact of higher-order eccentricity terms at arbitrary obliquity.

The model and formulae provided in this work are designed for general use. We therefore showcase the impacts of dual-body dissipation and higher-order eccentricity corrections in a general sense without focusing on a particular system's expected evolution. However, to place these results in context, we examine two scenarios that these improvements impact considerably: highly eccentric, short-period exoplanets (Section \ref{sec:result:trunc}), and the early evolution of binary Trans-Neptunian and Kuiper Belt objects (Section \ref{sec:results:tno}). The latter scenario is motivated by the concept of collisionally formed planet-moon systems \citep{CanupAsphaug2001, Canup2005, Canup2011, PahlevanStevenson2007, CukStewart2012}. Collisional formation naturally generates systems that tend to have post-collision spin rates that are highly non-synchronous to their orbital motion, as well as high initial eccentricities ($e \geq 0.1$).\par

For exoplanets, there is a strong interest in characterizing their environment from the perspectives of energy balance and chemical composition, particularly in the context of habitability \citep{WP_ExoplanetScience, WP_ExoGeoscience}. It is commonly assumed that worlds with short orbital periods ($P \lesssim 50$ days) rotate synchronously with their orbital motion \citep[e.g.][]{Jackson2008, Barnes2017, pierrehumbert2019atmospheric}. However, many phenomena may cause individual short-period exoplanets to not reside in the 1:1 spin-orbit resonance (SOR). First, orbital scattering \citep[e.g.,][]{Thommes2008, Matsumura2008}, and capture events \citep{agnor2006neptune, dos2012dynamical, woolfson2013capture} may result in quickly changing orbital frequencies which are unlikely to coincide with the prior rotation rate \citep{Vinson2017, Leconte2018}. More importantly, high eccentricity can also accelerate the spin rate of the exoplanet (or the host star, see \citet{CaroneThesis2012}) out of synchronicity into higher-order SOR's. Mercury presently resides in such a higher-order SOR, as it rotates 3 times for every 2 orbits. Even very short-period exoplanets may possess non-negligible eccentricity \citep{Bourrier2018}. Because population demographics for unseen outer perturber planets are still poorly known for many systems with short-period worlds \citep{BeckerAdams2017, PayneFordVeras2010}, the magnitude of perturbed equilibrium eccentricities remains difficult to predict, and many systems with reported $e = 0$, arrive at these values simply from assumption. Significant non-zero eccentricities can lead to similar higher-order SOR trapping as occurred for Mercury. Non-zero obliquity may also be stimulated by several phenomena, including collisions, satellites, Cassini state evolution, and secular spin-orbit resonance theory \citep{winn2005obliquity, brunini2006origin, atobe2007obliquity, miguel2010planet, rogoszinski2016tilting}. Tidal dissipation plays an important role in determining whether or not an object may become trapped in such higher, non-1:1 SOR's. \citet{Makarov2012ApJ} found that both the inclusion of higher-order eccentricity terms in the governing equations, and the utilization of advanced rheological models, are critical to determining if a given world is captured in a higher-order SOR or dissipates to its synchronous state. The influence of torques acting on a world's permanent triaxiality (generated through non-tidal effects) can also be critical in determining if a higher-order SOR is reached. We chose to focus only on the impact of tidal torques in this study, but point the interested reader to the works of \citet{rodriguez2012spin} and \citet{Margot2018} for a review of the influence of triaxiality in SOR capture. We also note throughout this work instances where our results may be altered by such further considerations. One reason it is important to constrain an exoplanet's likely spin state, especially what circumstances lead to non-1:1 SORs, is that an exoplanet's climate is dramatically altered if its rotation rate falls on a higher-order resonance \citep[e.g.][]{dobrovolskis2007spin, Wordsworth2015, Turbet2016AandA, DelGenio2019AstroBio}. \par

\begin{figure}[hbt!]
\centering
\includegraphics[width=0.75\textwidth]{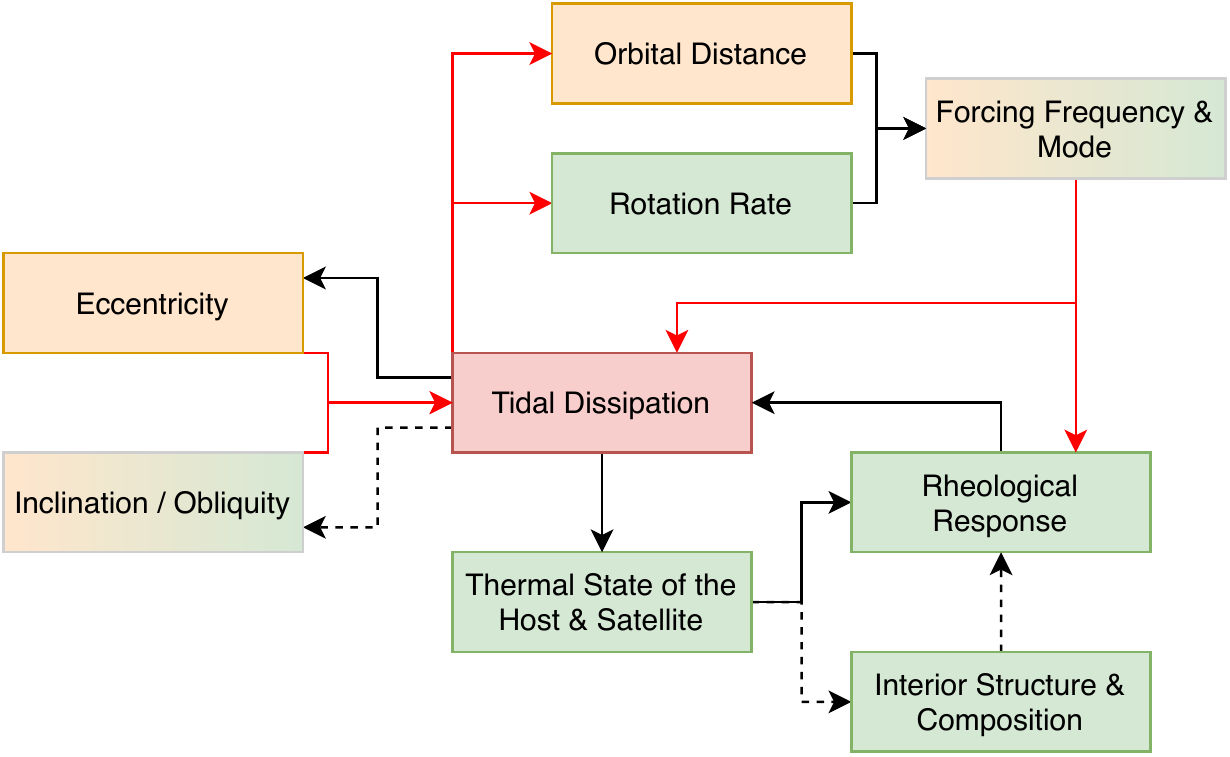}
\caption{Schematic representation of the modeling scheme used in this study. Yellow and green boxes represent, respectively, orbital and planetary properties. Arrows indicate dependencies between different concepts. Red lines are areas that this study improves upon. Dashed lines are not considered or are greatly simplified in this work and are left for future study.}
\label{fig:model}
\end{figure}

Several Trans-Neptunian and Kuiper belt objects (which we will collectively refer to as TNOs) have recently been found with relatively massive satellite(s). Besides Pluto and Charon, which we discuss in detail, some examples include Eris \& Dysnomia \citep{Brown2005Eris, Brown2006}, Haumea, Hi`iaka, \& Namaka \citep{Bouchez2005}, Orcus \& Vanth \citep{BrownSuer2007}, Makemake \& MK2 \citep{Parker2016ApJ}, Gonggong \& Xiangliu \citep{kiss2017discovery}, and potentially a newly discovered satellite of Varuna \citep{Fernandez-Valenzuela2019ApJ}. The compactness of these binary systems substantially increases their tidal susceptibility, and in some cases has distinctly slowed their rotation rates \citep[e.g.,][]{kiss2017discovery}. While in this work we focus on the Pluto-Charon system, the concepts explored are equally applicable to other TNO binaries. However, since formation hypotheses for these binaries vary widely, as do their compositions and masses, we leave their discussion to a future dedicated study. As for Pluto-Charon, the leading origin theory is that the binary formed via an impact between two bodies of roughly similar size. Details of this scenario (such as whether Charon accreted from a post-impact disk surrounding Pluto or instead remained mostly intact) remain uncertain \citep{Canup2005, Canup2011, Kenyon2019}. The other commonly considered binary formation hypotheses are co-accretion \citep{nesvorny2019trans}, capture \citep{goldreich2002formation}, and possibly fission of a fast-spinning object \citep{ortiz2012rotational}. Each of these scenarios variously affects compositions and interior structures \citep{DeschNeveu2017, BiersonNimmo2019}. For this study, we are primarily concerned with orbital and spin states. For Pluto-Charon, the post-impact mutual orbit tends to be highly eccentric ($e>0.1$), with both Pluto and Charon in non-synchronous rotation (NSR). Modern observations of Pluto-Charon show them to be in a dual-synchronous state with a very low (effectively zero) eccentricity \citep{stern2018pluto}. However, evolution from NSR and high eccentricity to their modern state relies at least in part on both bodies experiencing an epoch of significant tidal dissipation \citep{RobuchonNimmo2011, Cheng2014, BarrCollins2015, Hammond2016, DeschNeveu2017}. It is generally thought that this evolution is quick, for example \citet{Cheng2014} found the full evolution took $\sim 1$ Myr using the CTL model and close to 10 Myr for CPL. However, \citet{Saxena2018} showed that by considering NSR, the evolution can be slowed if the objects enter higher-order SOR's, but this work did not track the dissipation within both bodies simultaneously. We reexamine this problem by considering dual-body dissipation, as well as including higher-order eccentricity corrections which are necessary for the initial eccentricity values suggested by formation scenarios. \added{Concepts that this study considers and improves upon, as well as areas that are left for future work, are visualized in Figure \ref{fig:model}.}

\section{Methods}\label{sec:method}

\subsection{Tidal Efficiency}\label{sec:method:interior_model}
The efficiency of tidal dissipation is dependent upon the ability of a planet or moon's bulk to deform under tidal forcing and convert that deformation into frictional heat. Different materials, such as rock and ice, will respond to tidal forces in distinct ways. Even in a planet modeled as having homogeneous chemical composition, contrasting temperatures, pressures, and phase states will affect which frictional processes are dominating at the microscopic scale, and thus change what forcing frequencies lead to maximum local dissipation. The CPL method treats total tidal efficiency as a constant, regardless of any temperature or frequency dependence. This is generally approximated by the value $k_2/Q$ where $k_2$ is the 2$^{\text{nd}}$ order static tidal potential Love number. Love numbers quantify the response of a planet (e.g., maximum surface height change per cycle) to an external gravitational perturbation, while including both the deforming body's internal material strength, and its own self-gravity \citep{Love1909}. They may be either real values (elastic planetary response), or in the Fourier approach to viscoelastic tides, represented by a complex-valued number \citep[utilizing the correspondence principle, see:][]{CaputoMainardi1971b}. $Q$ is a constant, real-valued scalar known as the Quality Factor that is small for highly dissipative worlds and large otherwise. The CTL method imparts a dependence on a frequency, $f$, by defining $Q^{-1}(f) = f\bar{t}$ (depending upon the circumstances, this forcing frequency may be the orbital motion, rotational frequency, or a linear combination of the two) and a constant\footnote{The CTL method as presented by \citet{Wisdom2008} assumes that dissipation is \textit{linear} with frequency. This assumption is built into the equations before the relationship between $Q^{-1}$ and frequency $f$ is defined. Therefore, it is incorrect to use the CTL equations of \citet{Wisdom2008} with a $Q^{-1}(f)$ that is defined to be anything other than linear with frequency.}, $\bar{t}$. However, both of these methods, CPL and CTL, do not match laboratory experiments on solid materials whose response is tied to temperature and forcing frequency in more complex ways \citep{RajAshby1971, Karato1990, SundbergCooper2010, JacksonFaul2010, McCarthyCooper2016}. A more accurate technique is to model the tidal efficiency as the imaginary portion of the complex Love number, $-\text{Im}[k_{l}(\ldots)]$ \citep{Segatz1988}. The functional form of this multiplier is dependent upon the choice of rheology \citep{Efroimsky2012apj}. \par

The rheological response captured by the Love number is dependent upon the strength of the material, generally expressed through its rigidity and viscosity, as well as the forcing frequency. \citet{EfroimskyWilliams2009} showed\footnote{This theory was expanded in subsequent studies; see the work by \citet{EfroimskyMakarov2014} and \citet{BoueEfroimsky2019}.} that the Darwin-Kaula tidal expansion leads to a different frequency dependence on the complex Love number for each \textit{tidal mode}, $\omega_{lmpq}$. In the equations below, the subscripts $l$ and $m$ emerge from the tidal potential being a spherical harmonic across the surface of a world \citep{Darwin1880}. The two other integers $p$ and $q$ arise from transforming the viscoelastic tidal dissipation from spherical coordinates into more traditional Keplerian elements \citep{Kaula1961, Kaula1964}. This leads to series expansions in both eccentricity and relative obliquity. If we assume that there is no precession of the pericenter or orbital node and that the change in the mean anomaly can be approximated by $n$, the mean orbital motion, then we can write down the tidal mode for the $j$-th body $\omega_{lmpq,\;j}$ as,

\begin{equation}\label{eq:tidal_modes}
    \omega_{lmpq,\;j} \approx (l - 2p + q)n - m\spin_{j},
\end{equation}

\noindent where $\spin_{j}$ is the spin rate of the body. The response of a body's bulk is dependent upon the forcing frequency which is defined as the absolute value of the tidal mode, $\chi_{lmpq,\;j} \equiv \left|\omega_{lmpq,\;j}\right|$ \citep{Efroimsky2012apj}. For simplicity of notation we define,

\begin{equation}\label{eq:Love_NoSign}
    \loveI \equiv -\text{Im}\left[k_{l}\left(\chi_{lmpq}\right)\right],
\end{equation}

\noindent which is strictly positive for all forcing frequencies. However, the sign of the mode does determine the direction of tidal torques imparted on both the host and satellite. Therefore, we also define a version that carries each mode's sign,

\begin{equation}\label{eq:Love_Sign}
    \loveSgnI \equiv -\text{Sgn}\left(\omega_{lmpq}\right)\text{Im}\left[k_{l}\left(\chi_{lmpq}\right)\right].
\end{equation}

The impact of different rheological models has been studied in both rocky \citep{Henning2009, Bolmont2014, ShojiKurita2014, Padovan2014, BiersonNimmo2016, Dumoulin2017, Khan2018, Makarov2018, Margot2018, Bagheri2019a, LauFaul2019, Luna2020}, and icy \citep{Castillo-Rogez2011, McCarthyCastillo-Rogez2013, Saxena2018} worlds. The differences can be extreme depending upon the specific circumstances. Using a realistic rheological model rather than the CPL and CTL assumptions can have a major impact (e.g., tidal dissipation rates vary by over six orders of magnitude, based on uncertainties in internal material properties alone) as we will discuss in Section \ref{sec:results:tno}. However, the specific choice of \textit{which} rheological model to use will depend upon the composition and physical state of the material in question. To simplify the comparison, we choose to only use what we have found to be the most versatile model presently available, the Sundberg-Cooper rheology. This model has been found to fit the experimental rheological response of both rocks and ices \citep{SundbergCooper2010, Caswell2015, CaswellCooper2016}, uniquely well, as it describes two key phenomena in one composite model. First, it includes the `response broadening' behavior of models such as the Andrade rheology \citep{Jackson1993}. Second, it includes an experimentally observed secondary peak in dissipation as is modeled by the Burgers rheology \citep{sabadini1987consequences, Cooper2002}. Many other rheological models exist, but are often either specific to unique materials or conditions, or cumbersome to mathematically generalize. The Sundberg-Cooper model is optimal for all of these considerations. The implementation details of this model can be found in \citet{RenaudHenning2018}. \par

Using methods to estimate the dissipation for a specific body (such as solutions to equations describing the elastic deformation of \textit{layered} spherical bodies; see \citet{HenningHurford2014} and \citet{Tobie2019}) is beyond the scope of this work wherein we simply want to perform a direct comparison of the impact of orbital corrections and dual-body dissipation when using robust rheological modeling. For the same reason, we generally do not model the coupling between thermal and orbital models, however we do discuss the thermal-orbital evolution of Pluto-Charon briefly in Section \ref{sec:results:tno}. For that analysis, we follow the interior and thermal modeling that \citet{HussmannSpohn2004} applied to Europa's evolution. Otherwise, outside of Section \ref{sec:results:tno}, we assume homogeneous planets with constant dissipation throughout. For rocky worlds, we set the dissipating portion of the mantle to be modestly viscous with minimal partial melt, at a static viscosity of $\eta = 10^{22}$ Pa s, which is typical of observed values for Earth's mid-mantle \citep{MitrovicaForte2004}, and a shear modulus (or its inverse, compliance, $J$) of $\mu = J^{-1} = 50$ GPa \citep[e.g.,][]{dziewonski1981preliminary}. Such observed values are appropriate for rocky materials at temperatures near $\sim 1300$ K, with activation energies in the range 300--400 kJ mol$^{-1}$ \citep{turcotte2002geodynamics}. For icy worlds, we assume the dissipation predominately occurs in a convecting, viscoelastic ice layer experiencing temperatures just below the melting point at standard pressure, near 260--270 K (assuming a low presence of anti-freeze chemicals such as ammonia). We select a baseline ice viscosity of $\eta = 10^{14}$ Pa s and shear modulus of $\mu = J^{-1} = 3.3$ GPa, both of which are typical ``medium strength'' choices in many icy world investigations \citep{nimmo2009geodynamics, Quick2015Icarus, kamata2017interior, Rhoden2017, spencer2020pluto}. Dissipation in the rocky core of icy Solar System moons has been found to be negligible in comparison to ice shell dissipation \citep[e.g.,][]{ojakangas1989thermal}. However, the slope of viscosity versus temperature for low-pressure water ice is very steep and may lead to an ice shell that is less dissipative than a rocky core. This will be particularly true for a warm core that is well insulated from the overlying ice shell (perhaps, as may be the case for Ganymede, by a layer of high-pressure ice), or else for a core that contains a significant fraction of water or is porous leading to a lower effective viscosity \citep{Roberts2015, Choblet2017}. For this reason, the thermal evolution discussed in Section \ref{sec:results:tno:time} tracks the dissipation in both the icy and rocky layers, however we do not consider possible effects due to porosity in this work. \par

Additional, equally critical but less familiar, material parameters for the Sundberg-Cooper rheology include the Andrade exponent, $\alpha$, and timescale ratio, $\zeta$. More experimental work is needed to constrain these parameters at planetary temperatures and pressures, particularly for ices. However, several studies have constrained the Andrade exponent for silicate material to the range $0.2 < \alpha < 0.5$ with the majority of the findings converging to $\alpha_{avg} \approx 0.30$ \citep{Tan2001, Jackson2002, Jackson2004, WebbJackson2003}, we use this value throughout this study. The Andrade timescale ratio is less constrained and there is some preliminary evidence that its value may depend upon temperature \citep{Bunton2001, SundbergCooper2010}, forcing frequency \citep{Karato1990}, and melt fraction \citep{Jackson2004}. For this study, we invoke a commonly used estimate that the Andrade timescale, $\tau_{\text{An}}$, is equal to the \textit{Maxwell timescale}, $\tau_{\text{Max}}$, leading to $\zeta \equiv \tau_{\text{An}}/\tau_{\text{Max}} = 1$. \citet{RenaudHenning2018} found that variations in $\zeta$ do not tend to change tidal outcomes until the value becomes several orders of magnitude away from unity. Such large variations do not appear to be supported by laboratory experiments. The inverse of these timescales represents the frequency that a material experiences its peak dissipation (analogous to the resonant frequency in a harmonic oscillator). The Sundberg-Cooper rheology exhibits two dissipation peaks, the first occurs at the inverse Maxwell time, $\tau_{\text{Max}}^{-1} = (J\eta)^{-1}$. The location of its second, smaller peak is set by the inverse \textit{Voigt-Kelvin time}, $\tau_{\text{VK}}^{-1} = (\delta J\eta_{\text{p}})^{-1}$. We set $\delta J$ = 0.2 $J$ and $\eta_{\text{p}}$ = 0.02 $\eta$ which mimics the medium strength material used by \citet{Henning2009} for comparison purposes. We note that \citet{SundbergCooper2010} found the ratio $\delta J / J$ as large as 1.91. Like the Andrade parameters, these properties are understudied at planetary conditions and may themselves be dependent upon temperature and frequency. However, we have found that variations in the Voigt-Kelvin properties result in relatively minor changes in tidal outcomes compared with variations in other parameters such as the baseline viscosity $\eta$. We choose to keep both the Andrade and Voigt-Kelvin properties the same for ices and silicates so that more direct comparisons can be made between the objects examined. \par

The homogeneous assumption will, in general, overestimate the amount of dissipation within a planet since its entire volume, in reality, will not be uniform in its dissipation \citep[e.g.,][]{Tobie2019}. This will be particularly true for thin viscoelastic ice shells which, under a homogeneous ice model, will appear to dissipate a large amount of energy compared to a layered model. To account for this in Pluto-Charon, we multiply the global Love number by a tidal volume fraction, $f_{\text{TVF}} = V_{\text{Tidal}} / V_{\text{Planet}}$. Here, $V_{\text{tidal}}$ is equal to the volume of material that is participating in dissipation within the planet (equivalent to the $V_{\text{conv}}$ used by \citet{HussmannSpohn2004}). For the time evolution of Pluto-Charon (Section \ref{sec:results:tno:time}) there are two volume fractions: one equal to the rocky core's volume (which remains static) and one for the dynamic, viscoelastic icy shell that can grow or shrink depending on the energy budget \citep{HussmannSpohn2004}. These tidal volume fractions scale their respective Love numbers calculated for the high-viscosity core and the low-viscosity ice. For Sections \ref{sec:results:tno:dual} and \ref{sec:results:tno:obliquity} we present results that are snapshots in time and thus do not rely on a dynamic volume fraction for the icy shell. Instead, we use a constant value of $f_{\text{TVF}} = 10\%$ which is the maximum viscoelastic volume we find in our time studies. Therefore, results in those sections should be considered upper bounds on the possible solid-body dissipation. \par

An important caveat to the above discussion is the possibility of tides in fluid layers (be it gaseous envelopes, or oceans/pockets of liquid water or magma). Indeed, for the modern Earth, tidal motion in our ocean, including baroclinic waves, as well as flow through straits, bays, and across sea-floor topography, leads to much more dissipation than the tidal deformation of the Earth's solid interior \citep[e.g.,][]{EgbertRay2000}. The tidal response of fluid layers has a complex and non-linear dependence upon ocean depth, density, composition, and the nature of interfaces with solid features. All of these influence mechanical wave velocities, with frequencies that can be resonant with tidal forcing \citep{Tyler2014, Tyler2015, HayMatsuyama2017, Auclair-Desrotour2019, GreenWayBarnes2019, HayMatsuyama2019, Tyler2020}. Incorporating such models with the orbital improvements we present here is beyond the scope of this work. Instead, we will assume negligible liquid ocean tidal dissipation on the objects discussed. For exoplanets, this is akin to assuming no ocean exists. Pluto and Charon, on the other hand, certainly had substantial liquid oceans in their past, and at least for Pluto, may still today \citep{Nimmo2016}. For these worlds, we do not model the dissipation of any sub-surface ocean, so our work should be considered a lower bound on total tidal heating and an upper bound on evolution timescales. \par

\subsection{Orbital and Rotational Evolution due to Tidal Forces}\label{sec:method:orbital}

We follow the methods of \citet{BoueEfroimsky2019} to calculate changes in semi-major axis, eccentricity, and spin rate. Below we provide a summary of that work. We do not presume the ratio of relative masses of the bodies to be in a certain range, but for ease of explanation we will assume that the \textit{host} object will always be the object with the most mass of the pair --- we prescribe its properties with the subscript $h$. We give the properties of the orbiting \textit{satellite} the subscript $s$. When discussing an arbitrary object, we use the subscript $j$, and for its opposite tidal partner, $k$. \par

The orbital evolution equations utilize a frame of reference that is coprecessing with the primary body\footnote{As discussed in \citet{BoueEfroimsky2019}, this is a modification to the frame of reference used by \citet{Kaula1964} which was fixed to the primary's equator at a specific time. The difference between these two frames of reference will be important when considering the \textit{change} in obliquity (or inclination).}. By this nomenclature, ``primary'' refers to whichever object is dissipating tidal energy. In the dual-dissipation model, we assume \textit{both} objects are dissipating, thus there will be two frames of reference, one coprecessing with each object. Two separate computations of the system's tidal potential, $U$, are then performed, once in each object's frame of reference. As we neglect relativistic corrections, frame choices are based on what is most useful for mathematically describing the geometry of tidal deformation for each body. The tidal potential contains both rapidly and slowly oscillating terms in addition to secular terms. As we are only concerned with the long-term evolution of these systems, we use a tidal potential that has been averaged over both an orbital cycle (eliminating short-period oscillations) and apsidal precession (eliminating long-period oscillations) \citep{EfroimskyMakarov2014}. To avoid confusion, we prescribe all variables which are orbit-averaged in this way with angle brackets, $\avgd{X}$. \par

\explain{Throughout this revised manuscript we have made sure that all formulae that are derivatives with respect to time are described as \textit{orbit-averaged} and are denoted with angle brackets. This is to emphasize that these formulae are based on the tidal potential which has been averaged over both an orbital and apsidal cycle \citep[this is further discussed in][]{EfroimskyMakarov2014}.}

The semi-major axis, $a$, (and through it, the orbital mean motion, $n$) and eccentricity, $e$, will change depending upon the dissipation exhibited by both the host and satellite. Utilizing the conservation of energy and of angular momentum, while assuming a closed system, we can write down the orbital derivatives as,

\begin{equation}\label{eq:da_dt}
    \avgd{\dot{a}} = -\frac{2}{na}\frac{M_{h} + M_{s}}{M_{h}M_{s}} \partialDerv{}{\meanAnom}\left(M_{s}\avgd{U_{h}} + M_{h}\avgd{U_{s}}\right),
\end{equation}

\begin{equation}\label{eq:de_dt}
    \avgd{\dot{e}} = \frac{\sqrt{1 - e^2}}{nea^2}\frac{M_{h} + M_{s}}{M_{h}M_{s}}\left(M_{s}\partialDerv{\avgd{U_{h}}}{\pericenter_{h}} + M_{h}\partialDerv{\avgd{U_{s}}}{\pericenter_{s}} - \sqrt{1 - e^2}\partialDerv{}{\meanAnom}\left(M_{s}\avgd{U_{h}} + M_{h}\avgd{U_{s}}\right)\right),
\end{equation}

\noindent where $\meanAnom$ is the mean anomaly of the satellite's orbit and $\pericenter_{j}$ is the $j$-th body's argument of pericenter. $M$ represents the mass of either the host or the satellite. \par

Each body's change in spin rate can be found by utilizing its polar moment of inertia $C_j$ and the partial derivative of the tidal potential with respect to the orbital line of nodes as reckoned from the object's equator, $\Omega_{j}$,

\begin{equation}\label{eq:change_spin}
    \avgd{\ddot{\theta}_{j}} = \frac{M_{k}}{C_{j}}\partialDerv{\avgd{U_{j}}}{\Omega_{j}}.
\end{equation}

True spin synchronization, such as into 1:1 spin-orbit resonance, generally occurs for planetary bodies with nonzero triaxiality (also referred to as possessing a permanent quadrupole moment, such as is severely established by the lunar dichotomy on Earth's Moon). Otherwise, as found by \citet{hut1981tidal}, objects with triaxiality below a certain threshold, to the limit of a perfect sphere, will evolve to equilibrium rotation states known variously as pseudo or quasi synchronous rotation \citep[see also,][]{MD, Heller2011, Makarov2013}. The degree to which pseudo-synchronous rotation rates are offset from perfect resonance is a function of eccentricity \citep{hut1981tidal} \added{and viscoelastic response \citep{Correia2014}}. Permanent quadrupole terms may also lead to the induction and maintenance of physical librations, which further complicate SOR capture \citep[e.g.,][]{rodriguez2012spin, Margot2018}. Although we here invoke a constant polar moment of inertia for both host and satellite, this does \textit{not} also imply assuming each object is a perfect sphere. What we do assume is that non-tidal triaxiality is in a generally low range, so that equilibrium rotation rates can still be near exact SOR states, yet physical librations are not of high magnitude. Pluto and Charon are both good candidates to possess such non-zero, yet low-to-moderate triaxiality \citep{mckinnon2014internal}, to help achieve near-exact SOR's (including non-1:1 states). Although no polar flattening was detected for either body to the detection limit of $\approx0.5$\% in \emph{New Horizons} images \citep{Nimmo2017}, the Sputnik Planitia region of the anti-Charon hemisphere of Pluto has been interpreted as having higher density than the rest of Pluto's crust, implying a non-spherically symmetric mass distribution \citep{Keane2016, Nimmo2016}. TRAPPIST-1e, with a putative solid surface (able to not relax rapidly to hydrostatic equilibrium), is similarly a reasonable candidate for the same assumptions. \par

While we do consider the effects of tidal dissipation due to non-zero obliquity, we do not calculate or track the \textit{change} of obliquity over time. An interested reader can reference Eq. 118 and Appendix F of \citet{BoueEfroimsky2019} and the work by \citet{Luna2020}. \par

For each body, we calculate the tidal potential derivatives and the tidal heating as\footnote{These formulae make the following assumptions: There is no precession of the node ($\dot{\Omega} \approx 0$), only the \textit{secular} evolution is considered (equations are averaged over the orbital and apsidal precession cycles), and we are not including the role of either body's triaxiality nor relativistic effects.}, 

\begin{equation}\label{eq:dissipation}
\begin{aligned}
    \left[ \begin{array}{c}
        \partialDervLine{\avgd{U_{j}}}{\mathcal{M}} \\
        \partialDervLine{\avgd{U_{j}}}{\pericenter_{j}} \\
        \partialDervLine{\avgd{U_{j}}}{\Omega_{j}} \\
        \avgd{\dot{E}_{j}}
    \end{array} \right]
= &\frac{\newtonG{}M_{k}}{a}\sum_{l=2}^{\infty}\left(\frac{R_{j}}{a}\right)^{2l + 1}\sum_{m=0}^{l}\frac{\left(l-m\right)!}{\left(l+m\right)!}\left(2-\delta_{0m}\right)\sum_{p=0}^{l}F_{lmp}^{2}\left(I_{j}\right) \\
& \times \sum_{q=-\infty}^{\infty}G_{lpq}^{2}\left(e\right) 
    \left[ \begin{array}{c}
        (l-2p+q)\loveSgnJ \\
        (l-2p)\loveSgnJ \\
        m\loveSgnJ \\
        \chi_{lmpq,\;j}M_{k}\loveJ
    \end{array} \right],
\end{aligned}
\end{equation}

\noindent where $\newtonG$ is Newton's gravitational constant, $R$ and $I$ represent, respectively, the radius and obliquity of either world, and $\delta_{0m}$ is the Kronecker delta function which is equal to 1 for $m = 0$ and 0 otherwise. Here, $F$ and $G$ are the inclination and eccentricity functions \citep[see, e.g.,][]{Kaula1964} whose definitions can be found in Appendix \ref{sec:app:eccen_inclin}. \par


It is important to note that the eccentricity functions $G(e)$ cannot, in general, be written down exactly. They require a truncation at the desired power of $e$ which in turn leaves the tidal heating and tidal potential derivatives reliant on the same approximation. Historically, the eccentricity functions have been found to converge very poorly\footnote{In part because coefficients that precede higher-order terms grow in magnitude and can balance the additional powers of eccentricity.}, requiring high-order truncations to adequately account for tidal effects at high eccentricities \citep{Bagheri2019b}. This motivates us to quantify the effects of loosening the truncation restrictions on the eccentricity functions. We provide the tidal heating and potential derivative equations up to and including $e^{10}$ terms in Appendix \ref{sec:app:truncation} for NSR tides, and Appendix \ref{sec:app:sync} for tidally-locked worlds. This truncation level matches that presented by \citet{Wisdom2008} who utilized a CTL dissipation model (as we discuss in the following section). Utilizing the $e^{10}$ truncation results, under NSR, leads to 44 unique tidal modes and 37 unique forcing frequencies (ignoring both frequencies which are zero and modes corresponding to $F$ and $G$ functions that vanish). As we will show in Section \ref{sec:result:trunc}, for worlds experiencing NSR tides at very high eccentricity ($e>0.6$), we find that the $e^{10}$ truncation level is insufficient to fully capture the orbital and rotational evolution. We therefore also explore the impacts of eccentricity terms up to and including $e^{20}$. This quite high truncation is important for the early evolution of Pluto and Charon which can spend a significant amount of time in NSR while simultaneously experiencing eccentricities greater than around 0.6 (see Section \ref{sec:results:tno:time}. For the $e^{20}$ truncation level, there are 74 unique frequencies. Presenting this many terms in a tabular format, as we do for $e^{10}$ (see Table \ref{tab:dissipation_order10}), becomes quite cumbersome. However, the above formula, and $F(I)$ \& $G(e)$ presented in Appendix \ref{sec:app:eccen_inclin}, allows one to calculate the dissipation equations to a desired truncation level, even beyond $e^{20}$. Including terms beyond $e^{20}$ may be important for worlds experiencing eccentricities greater than 0.8, especially if they are also in NSR. For example, this may have been the case for the early evolution of Triton \citep[e.g.,][]{RufuCanup2017} as well as for asteroids and comets. However, increasing the truncation level also increases the number of tidal modes leading to greater computational time. For this reason, we recommend future studies to determine the minimum truncation level required for a particular problem. We have found that terms beyond $e^{20}$ (terms up to and including $e^{22}$ were tested) do not produce a significant difference for the systems examined in this work. \par

The complex Love number must be calculated at each frequency for both worlds before Eqs. \ref{eq:dissipation} can be calculated. Even after a world has reached synchronous rotation, it will still be subjected to multiple tidal modes of the form $\pm 1n$, $\pm 2n$, $\pm 3n$, and so on (see Appendix \ref{sec:app:sync}). As long as sign conventions are carefully followed, these formulae are equally valid for negative (retrograde) spin rates and orbital motions. \par

\subsubsection{Comparison to Other Formulations}\label{sec:method:orbital:others}
The orbital and rotational evolution of a dual-body system has been studied by others using different assumptions than the ones we use here. In this section we highlight differences between our formulation and the often-used equations of \citet{GoldreichSoter1966} and \citet{Wisdom2008}. \par

\vspace{2mm}\noindent\textbf{Comparison to Goldreich \& Soter (1966)} \\
\citet{GoldreichSoter1966} wrote down the change in eccentricity as (see Eq. 22--25 in \ibid. We have modified these equations to match our notation),

\begin{subequations}\label{eq:goldreich:edot}
    \begin{align}
        \dot{e} =& \; \dot{e}_{h} + \dot{e}_{s}, \label{eq:goldreich:edot:sum} \\ 
        \frac{2}{3}\frac{\dot{e}_{h}}{e} =& \; \frac{19}{4}\frac{k_{2,\,h}}{Q_{h}}\frac{M_{s}}{M_{h}}\frac{nR_{h}^{5}}{a^{5}}\text{Sgn}\left(2\spin - 3n\right)
        \label{eq:goldreich:edot:host} \\ 
        \frac{2}{3}\frac{\dot{e}_{s}}{e} =& -7\frac{k_{2,\,s}}{Q_{s}}\frac{M_{s}}{M_{h}}\frac{nR_{s}^{5}}{a^{5}}.
        \label{eq:goldreich:edot:satellite}
    \end{align}
\end{subequations}

Combining Eqs. \ref{eq:goldreich:edot:host} and \ref{eq:goldreich:edot:satellite} with Eq. \ref{eq:goldreich:edot:sum} results in a popular formulation for dual-body eccentricity changes \citep[e.g.,][]{Barnes2008AstroBio, ShojiKurita2014}\footnote{Note that these and other authors have replaced the host body's contribution coefficient of 19/4 with -19/4. Using Eqs. 23 \& 24 in \citet{GoldreichSoter1966}, this replacement is only valid when the host's spin rate is $\spin_{h} < 3/2 n$. We drop the sign dependence in Eq. \ref{eq:goldreich:edot:full} and set the satellite and host's contribution to $\dot{e}$ to be opposite, which is valid for $\spin_{h} > 3/2 n$ as is the case for the Earth-Moon system. However, for studies where long-term spin rate changes are considered, the sign dependence should be left in place.},

\begin{equation}\label{eq:goldreich:edot:full}
    \dot{e} = \frac{3}{2}\frac{ne}{a^{5}}\left[-7\frac{M_{h}}{M_{s}}\frac{R_{s}^{5}k_{2,\,s}}{Q_{s}} + \frac{19}{4}\frac{M_{s}}{M_{h}}\frac{R_{h}^{5}k_{2,\,h}}{Q_{h}}\right]
\end{equation}

The derivation of this equation is described by \citet{Goldreich1963}. It is based on a theory of tides that considers four different tidal modes which correspond to the following forcing frequencies,

\begin{subequations}\label{eq:goldreich:tidal_lags}
    \begin{align}
        2\epsilon_{0} \rightarrow& \; 2\spin - 2n, \\
        2\epsilon_{1} \rightarrow& \; 2\spin - 3n, \\
        2\epsilon_{2} \rightarrow& \; 2\spin - n, \\
        2\epsilon_{3} \rightarrow& \; \frac{3}{2}n.
    \end{align}
\end{subequations}

To arrive at Eq. \ref{eq:goldreich:edot:full}, the author assumes that the satellite is in synchronous rotation which sets $\epsilon_{0} = 0$ and $\epsilon_{1} = -\epsilon_{2}$. After those simplifications are made, they then use a CPL model to equate $\epsilon_{2} = \epsilon_{3}$. This is in contrast to the tidal host, which the authors assume is \textit{not} in synchronous rotation (this was done to mimic the Earth-Moon system). Instead, the CPL model is applied right away, equating all the tidal lags to one another: $\epsilon_{0} = \epsilon_{1} = \epsilon_{2} = \epsilon_{3}$. This results in the host and satellite contributions to $\dot{e}$ having different coefficients. Furthermore, each of these tidal modes will, in general, carry a unique sign. These signs are lost once the CPL model is applied. \par

The orbital evolution model developed by \citet{GoldreichSoter1966} requires the satellite's rotation rate to be synchronized to the orbital motion. It also uses a CPL method to estimate the material response of the world. Lastly, it truncates the dissipation equations to only include $e^{2}$ terms (and only considers the quadrupole terms). For these reasons, we do not find it suitable for the scenarios we explore in this work. \par

For completeness, we note that the above equations of Goldreich and Soter can be retrieved from Eq. \ref{eq:de_dt} and Appendix \ref{sec:app:truncation} (or Appendix \ref{sec:app:sync} for the spin-synchronous satellite term), by setting $l = 2$ and neglecting powers of eccentricity above $e^2$. In Appendix \ref{sec:app:truncation}, the above four modes considered by \citet[][their equation 7]{Goldreich1963} correspond to ($l$, $m$, $p$, $q$) = (2,2,0,0) for $\epsilon_0$, (2,2,0,1) for $\epsilon_1$, (2,2,0,-1) for $\epsilon_2$, and (2,0,1,-1 \& 1) for $\epsilon_3$, respectively, in the limits of $e^2$ truncation and $M_h \gg M_s$:

\begin{equation}\label{eq:goldreich:eq7}
    \frac{\dot{e}}{e} \propto \left[\epsilon_0 - \frac{49}{4}\epsilon_1 + \frac{1}{4}\epsilon_{2} + \frac{3}{2}\epsilon_{3}\right]
\end{equation}

The first coefficient of 7 in Eq. \ref{eq:goldreich:edot:full} results from the satellite being synchronous to the orbital motion, for which \citet{Goldreich1963} set $\epsilon_0$ = 0 and $\epsilon_1 = -\epsilon_2$ in Eqs. \ref{eq:goldreich:tidal_lags} to obtain [$25/2 ~\epsilon_2 + 3/2 ~\epsilon_3$] for the term in brackets, and \emph{only then} making the CPL assumption that $\epsilon_2 = \epsilon_3$ to obtain a coefficient $14 ~\epsilon$ which corresponds to the coefficient of 7 when accounting for the factor of 2 difference between \citet{Goldreich1963}'s tidal lags and the definition of $\loveI$ in Eq. \ref{eq:Love_NoSign}. \par

The second coefficient of 19/4 in Eq. \ref{eq:goldreich:edot:full} results from immediately applying the CPL assumption to Eq. \ref{eq:goldreich:eq7}, setting $\epsilon_0$ = $\epsilon_1$ = $\epsilon_2$ = $\epsilon_3$ to obtain a coefficient $19/2 ~\epsilon$, which corresponds to the 19/4 term again accounting for the above factor-of-2 difference. \par

\vspace{2mm}\noindent\textbf{Comparison to Wisdom (2008)} \\
\citet{Wisdom2008} derived the tidal heating of a satellite subjected to an arbitrary eccentricity and obliquity. The resulting dissipation equations rely on Hansen coefficients, as does the model we use in this work. However, there are two limitations with the model of \citet{Wisdom2008}. First, it was derived assuming a CTL method which assumes the dissipation efficiency is \textit{linear} with frequency (via $-\operatorname{Im}[\bar{k}_{l}(\omega)]$). While this is an improvement over the CPL method, it has still been found to not match the real response of planetary materials across the frequency domain. Second, it assumes the satellite is either in synchronous rotation or \textit{equilibrium rotation} (sometimes referred to as `pseudo-synchronous' rotation). This equilibrium rotation rate varies with eccentricity and obliquity, and may depend upon the tidal potential when dissipation is strong \replaced{\citep{Makarov2015}}{\citep{Correia2014, Makarov2015}}. This is in contrast to our method which calculates the change in spin rate with time, see Eq. \ref{eq:change_spin}. While the tidal potential will vary with eccentricity and obliquity, their values do not immediately change the ``instantaneous'' (yet still orbit-averaged) spin rate as they would under an equilibrium model. Further discussion surrounding the CTL method and the applicability of pseudo-synchronous rotation can be found in \citet{Makarov2013}. \par

The equations we use also reproduce the model used by \citet{Wisdom2008} if we apply the same key assumption that tidal dissipation varies linearly with frequency. The spin-synchronous tidal heating rate (truncated to $e^{10}$ and computed at zero obliquity) found by \citet[][see Eqs. 21 and 26]{Wisdom2008} can be compared to Eq. \ref{eq:app:sync:heating} if one sets $\operatorname{K}_{j}(a|n|) = ak_{2}/Q$ for $a \in \{1,2,3,4,5\}$. For example, the coefficient of the $e^{4}$ term in Eq. \ref{eq:app:sync:heating} matches the one presented in Eqs. 21 and 26 of \citet{Wisdom2008} by setting $\operatorname{K}_{j}(|n|) = k_{2}/Q$ and $\operatorname{K}_{j}(2|n|) = 2k_{2}/Q$ and noting that \citet{Wisdom2008} has factored out an overall multiple of 7 that we do not in Eq. \ref{eq:app:sync:heating}.

\subsection{Implementation Details}\label{sec:method:implementation}

The tidal heating, orbital, and rotational changes are calculated by summing up the contribution of each tidal mode in the tidal potential and heating equations (see Appendix \ref{sec:app:truncation}). The coefficients of the eccentricity and inclination functions are pre-calculated using the equations presented in Appendix \ref{sec:app:eccen_inclin} (see also the Appendix of \citet{Veras2019}). Calculations are performed using the NumPy software package \citep{NumPy}. The time integration discussed in Section \ref{sec:results:tno:time} was performed using a 3$^{\text{rd}}$-order Bogacki-Shampine method provided by the Julia language's differential equation package \citep{JuliaDiffeq}. \par

Much of the work discussed in this study requires only the mass, radius, and orbital separation of the planets under consideration. We provide these key properties in Table \ref{tab:methods:planet_params}. The thermal evolution model used for Pluto-Charon necessitates knowledge of these worlds' ice layer thickness. For this, we use an ice shell thickness of 337 km and 197 km for Pluto and Charon, respectively \citep{Nimmo2017}.

\begin{table}[hbt!]
\begin{tabular}{lcccc}
\hline
\textbf{Parameter}                                & \textbf{TRAPPIST-1}          & \textbf{TRAPPIST-1e}            & \textbf{Pluto}       & \textbf{Charon}       \\ \hline
\multicolumn{1}{l|}{Radius [$10^{6}$ m]}          & 81.4$^{[d]}$                & 5.995$^{[a,\;b]}$               & 1.1883$^{[e]}$      & 0.606$^{[e]}$         \\
\multicolumn{1}{l|}{Mass [$10^{24}$ kg]}          & $1.6 \times 10^{5}$$^{[d]}$ & 4.6$^{[c]}$                     & 0.01328$^{[e,\;f]}$ & 0.001603$^{[e,\;f]}$ \\
\multicolumn{1}{l|}{Semi-Major Axis [$10^{6}$ m]} & ---                          & $4,380.7$$^{[c]}$ & ---                  & 19.596$^{[f]}$       \\ \hline \hline
\end{tabular}
\caption{Throughout this study we apply our tidal model to two different systems: TRAPPIST-1e orbiting its star, and Charon orbiting Pluto. Here we present the properties of these worlds used in our calculations. References: $a$) \citet{Delrez2018}, $b$) \citet{Kane2018}, $c$) \citet{Grimm2018}, $d$) \citet{Gillon2017}, $e$) \citet{Nimmo2017}, and $f$) \citet{Brozovic2015}.}
\label{tab:methods:planet_params}
\end{table}

\section{Results and Discussion}\label{sec:results}

\subsection{Tidal Dynamics of an Eccentric TRAPPIST-1e}\label{sec:result:trunc}

There are a growing number of detections of super-Earth or smaller, short-period ($P \lesssim 50$ days) exoplanets that have large eccentricities ($e \geq 0.1$). Many of these worlds are part of multi-planet systems, such as L98-59, Kepler-444, K2-136, and TOI-700 \citep{Kostov2019, Campante2015, Mann2018, Gilbert2020, Rodriguez2020}. In these cases, in a manner analogous to the Galilean satellites, near-mean motion resonances (MMRs), secular resonances, and secular perturbations may all be driving non-zero eccentricities which will further drive tidal dissipation \citep{VanEylen2019}. 

\subsubsection{Dissipation at Zero Obliquity and Synchronous Rotation}\label{sec:result:exoplanet:sync}

\begin{figure}[hbt!]
\centering
\includegraphics[width=1\textwidth]{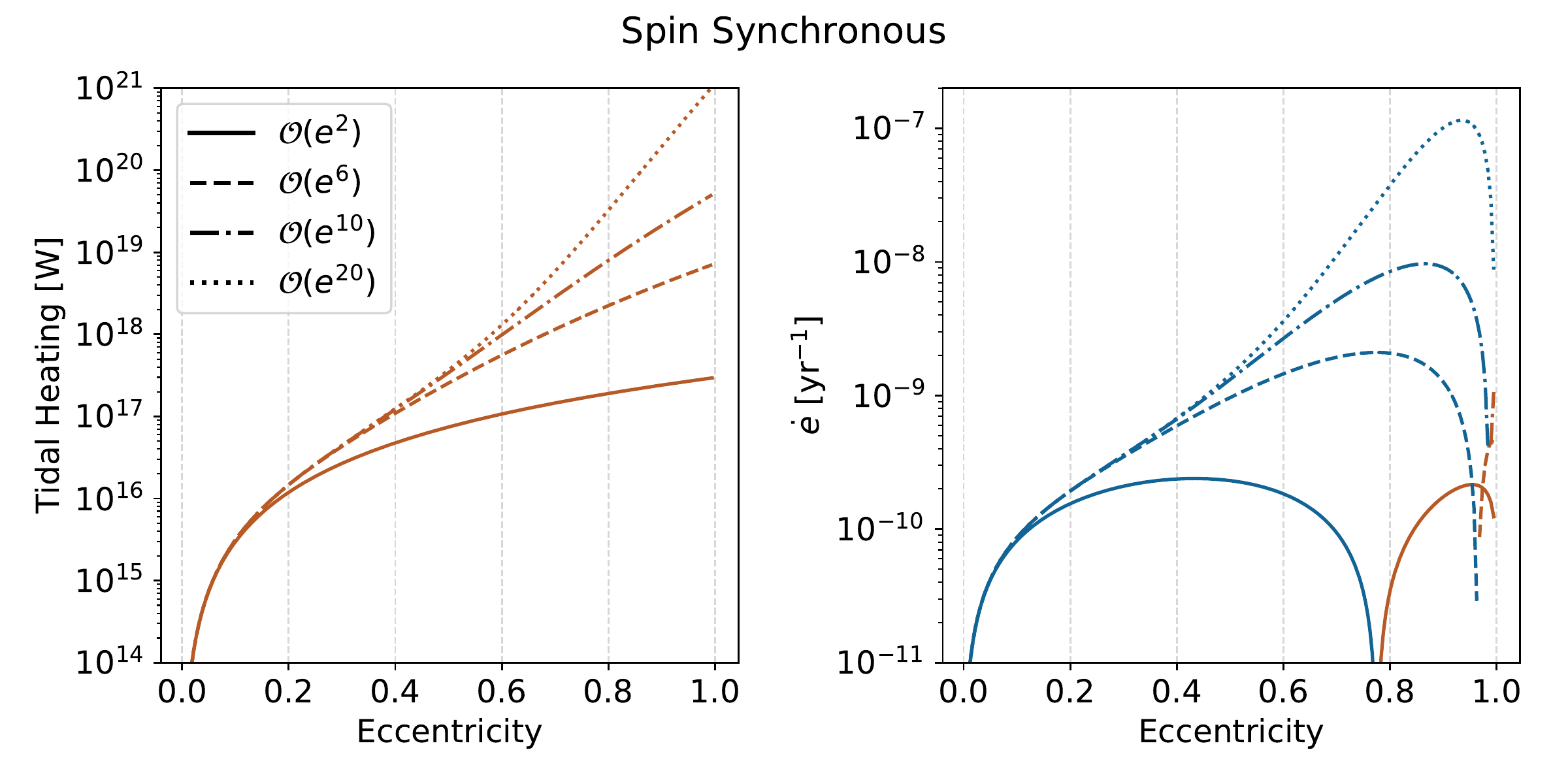}
\caption{Tidal heating (left) and the time derivative of eccentricity (right) are plotted as functions of eccentricity for four different eccentricity truncation levels. The sign of the eccentricity derivative is denoted by the change in color: orange indicates positive (growing) eccentricity while blue is negative (circularizing). Differences between the commonly used $e^{2}$ truncation and higher orders start near an eccentricity of 0.1 with significant (nearly an order-of-magnitude) divergences at $e > 0.3$. Interestingly, restricting $\dot{e}$ to $e^{2}$ results in a flip in sign at an eccentricity just below 0.8; this feature is rectified once higher-order terms are included. This suggests that the use of lower truncation levels at high eccentricities can change both magnitude and \textit{direction} of long-term orbital evolution. Calculations were made using rock-like material parameters, as well as the planetary and orbital parameters for TRAPPIST-1e.}
\label{fig:exoplanet:sync}
\end{figure}

Here we investigate the impact that using higher-order eccentricity terms have on tidal dissipation. To provide context to the results, we choose to look at the exoplanet TRAPPIST-1e \citep{Gillon2017} as it mimics a scaled-up version of the tidally active moon Io \citep{luger2017seven, Turbet2018, barr2018interior} including MMR perturbations (planetary properties can be found in Table \ref{tab:methods:planet_params})\footnote{TRAPPIST-1e is much further away from its star than Io is from Jupiter, but since the star is nearly one hundred times more massive than Jupiter, the planet has an orbital period of the same order as Io. TRAPPIST-1e's larger radius, and therefore larger volume contributes to dissipation, which also partially makes up for its slower orbital period.}. In Figure \ref{fig:exoplanet:sync}, we calculate tidal heating and $\dot{e}$ using different truncation levels in the dissipation equations. In this figure the planet is assumed to be spin-synchronous to the observed orbital period of $6.099$ days \citep{Delrez2018}. Differences between truncation levels appear around $e = 0.1$ and can lead to order-of-magnitude changes in both heating and orbital evolution once $e > 0.3$. A striking finding is that the commonly used $e^{2}$ truncation predicts the \textit{sign} of the eccentricity derivative to flip (noted by the change in color from blue to orange in Figure \ref{fig:exoplanet:sync}) just below $e = 0.8$. This feature is completely rectified at truncation level $e^{10}$ and higher. Figure \ref{fig:exoplanet:sync} shows the tidal heating and change in eccentricity as snapshots in time. Since tidal heating is only a function of even powers of $e$ and not $\dot{e}$, and since its value is always positive, it does not experience the same dramatic changes seen at very high eccentricity ($e>0.6$) that $\dot{e}$ does. \par

\subsubsection{Heating Rates from Obliquity Tides at Synchronous Rotation}\label{sec:result:trunc:obliquity}

The obliquity of exoplanets is currently unknown. \citet{Heller2011} found that any non-zero obliquity in a short-period exoplanet, with no moons, would likely align perpendicular to its orbital plane quickly (as a point of comparison, Venus is presently misaligned from retrograde-perpendicular by 2.64$^{\degree}$). We leave a detailed discussion regarding obliquity alignment timescales, as well as stable Cassini states with dissipation \citep{peale2006proximity, fabrycky2007cassini}, for future study. However, before alignment, or following collisional perturbation, obliquity will affect both the orbital and rotational evolution, as well as provide additional interior heating. In \citet{Saxena2018} it was demonstrated that, for Trans-Neptunian Objects, tides due to obliquity or an inclined orbit generate significantly less dissipation than those due to NSR or eccentricity. \citet{Heller2011} also found that obliquity tides require low eccentricity ($e < 0.3$) before they have significant impact. Both of these works estimated tidal heating due to obliquity to be\footnote{\citet{Heller2011} looked at both the CPL and CTL models, the latter of which defines heating due to obliquity tides to be proportionate to $\cos^{2}{(I)} / (1 + \cos^{2}{(I)})$ rather than $\sin^{2}{(I)}$.} $\propto\sin^{2}{(I)}$. This estimate is valid for low-obliquity, spin-synchronous worlds in a highly circular orbit. However, for large obliquity, or a world in NSR, then higher-order obliquity terms are required. Adding these corrections may be required even for low obliquity due to cross-terms between the inclination and eccentricity functions. For example, the tidal mode corresponding to $l$, $m$, $p$, $q$ = 2, 0, 0, -1 in Table \ref{tab:dissipation_order10} has its lowest-order term $\propto e^{2}\sin^{4}(I)$ which grows quickly with eccentricity as long as $I\notin\{0, \pi\}$. Conversely, while eccentricity can enhance obliquity tides, it is not a requirement: the mode corresponding to $l$, $m$, $p$, $q$ = 2, 0, 0, 0 has a term $\propto \sin^{4}(I)$, independent of eccentricity (Table \ref{tab:dissipation_order10}). Unlike the eccentricity functions $G$, the inclination functions $F$ do not contain any infinite summations (assuming a fixed maximum tidal harmonic order, $l$). Therefore, it is possible to write down an exact $F$ formula. Since an exoplanet's obliquity is unknown, we do not make any assumptions on its magnitude and therefore leave the inclination functions general (see Table \ref{tab:dissipation_order10}). \par

\begin{figure}[hbt!]
\centering
\includegraphics[width=.6\textwidth]{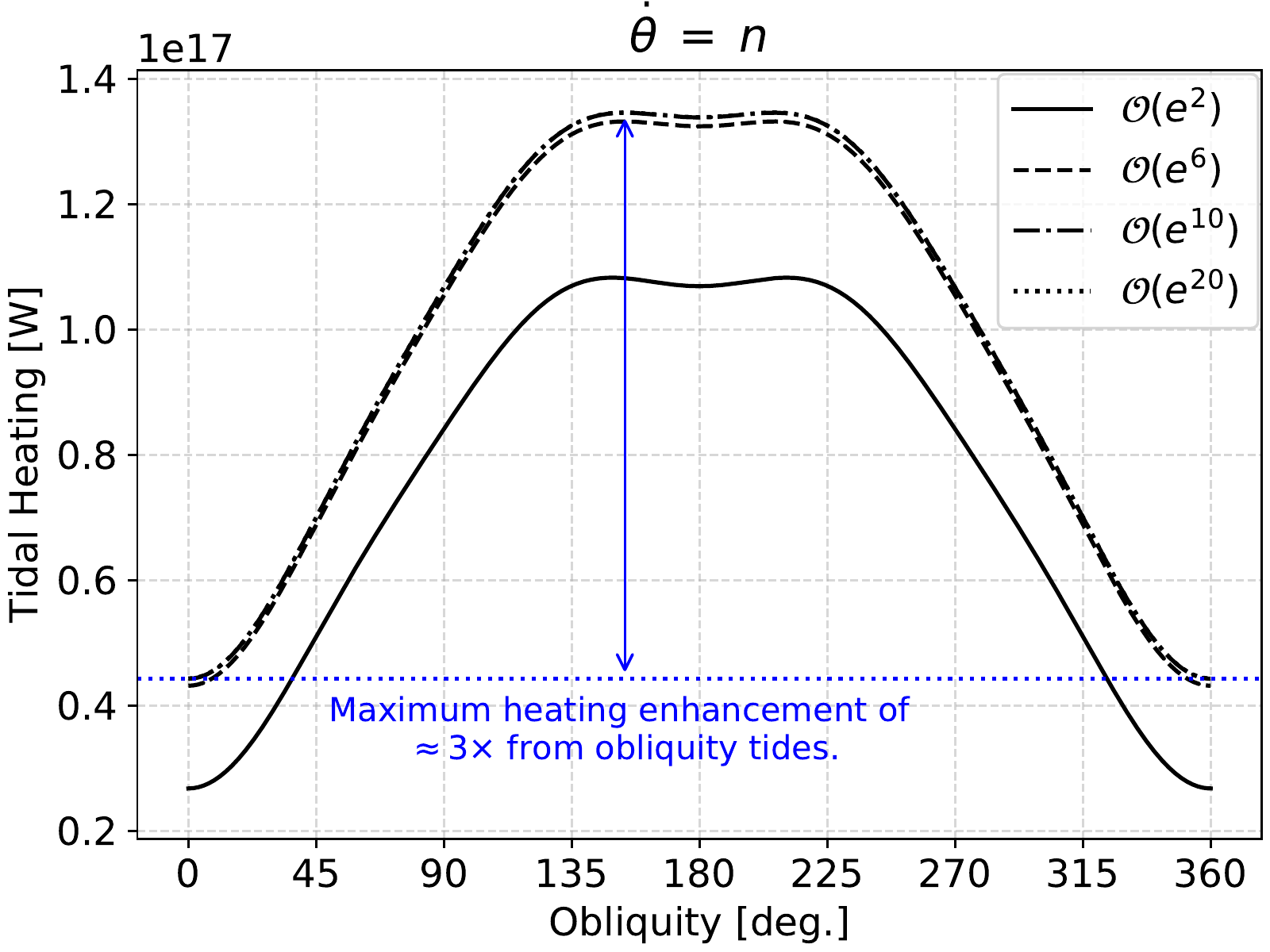}
\caption{Tidal heating is calculated for TRAPPIST-1e as a function of obliquity. As in Figure \ref{fig:exoplanet:sync}, four different eccentricity truncation levels are shown, and we assume the exoplanet is in its 1:1 spin-orbit resonance. The value of eccentricity is fixed to 0.3. At zero obliquity, $I = 0^{\degree}$, this constant eccentricity imparts $\approx 27$ PW of heating when only using the $e^{2}$ terms and $\approx 44$ PW when using terms up to and including $e^{20}$. Obliquity tides reach a peak enhancement of heating on either side of $180^{\degree}$. For the $e^{20}$ truncation, obliquity tides provide approximately 3 times the heating as the baseline eccentricity tides (indicated by the blue arrow and annotation). At $I=180^{\degree}$, the planet is flipped relative to its orbital plane. Mathematically, this is equivalent to a world that has zero obliquity, but is in a negative (retrograde) rotation rate. This retrograde motion will quickly synchronize to the orbital motion, leading to large rotational torques and, therefore, NSR heating.}
\label{fig:exoplanet:obliquity:heating}
\end{figure}

In Figure \ref{fig:exoplanet:obliquity:heating} we calculate tidal heating for TRAPPIST-1e across the obliquity domain. A constant eccentricity of $e = 0.3$ provides the planet with a considerable amount of internal heating even when obliquity is zero. At this large eccentricity, the importance of higher-order eccentricity terms remains. Without these higher-order corrections, the amount of heat the planet experiences is underestimated by between a factor of 1.25 and 1.65 depending upon the obliquity. As obliquity increases (up to $180^{\degree}$), its impact on tidal heating tends to lower this enhancement factor that higher eccentricity truncations provide. At their peak, obliquity tides can increase the planet's heating by about a factor of $3$. This peak in heating occurs on either side of $180^{\degree}$, which indicates a near-total flip in the planet's rotation axis. By definition, this is equivalent to a world with a near-zero obliquity and a retrograde spin rate ($\spin=-n$). This equivalence suggests that obliquity tides, after a certain angle, effectively become NSR tides. This may be important if the rotation rate is found to evolve faster than obliquity damping. In this case, if a short-period planet were to reach an obliquity beyond 90$^{\degree}$ our results show the following would happen: obliquity would evolve to 180$^{\degree}$, representing effective retrograde rotation. This is still highly dissipative and not stable, however further axis-angle change would not resolve the condition. Instead, the rate of spin would evolve via NSR terms, declining through zero, and returning to prograde rotation without axis reorientation. This may be one mechanism where slow rotator planets, temporarily below 1:1 SOR, could exist. Torques leading to such outcomes will always compete with other torques, such as from atmospheric flow.


We find that tides at low obliquities ($I < 45^{\degree}$) can provide exoplanets with a modest enhancement of heating ($< 2 \times$). This is in contrast to the orders of magnitude higher heating that can result from increases in eccentricity or for a mismatch of spin and orbital frequency. It therefore may be feasible to neglect heating due to obliquity tides when the planet has even a large obliquity ($I \geq 45^{\degree}$). However, even though the impact on heating may be modest, the impact of obliquity tides on the \textit{derivatives of the tidal potential} can be quite dramatic as we will discuss in Section \ref{sec:result:trunc:NSR_Obliquity}. \par

\subsubsection{Non-synchronous Rotation at High Eccentricity}\label{sec:result:trunc:NSR}

In the previous section, we examined the dissipation for TRAPPIST-1e with its spin rate locked to its orbital motion. Loosening this restriction enables NSR dissipation, which can both generate a large amount of heat and create a further coupling between the material response and orbital evolution via the rheological dependence on many tidal modes (see Eq. \ref{eq:tidal_modes}). An additional coupling occurs between eccentricity (and obliquity) and spin rate \citep[e.g.][]{Makarov2012ApJ}: A high eccentricity can lead to higher-order spin-orbit resonance trapping and can even accelerate a planet's spin rate out of the 1:1 resonance\footnote{Even notwithstanding the further issue of possible pseudo-synchronous rotation for worlds with triaxial moments of inertia.}. For example, Mercury's high eccentricity is key to maintaining the planet's 3:2 SOR \citep[e.g.,][]{CorreiaLaskar2009, Noyelles2014, MakarovEfroimsky2014}. \par

\begin{figure}[hbt!]
\centering
\includegraphics[width=1\textwidth]{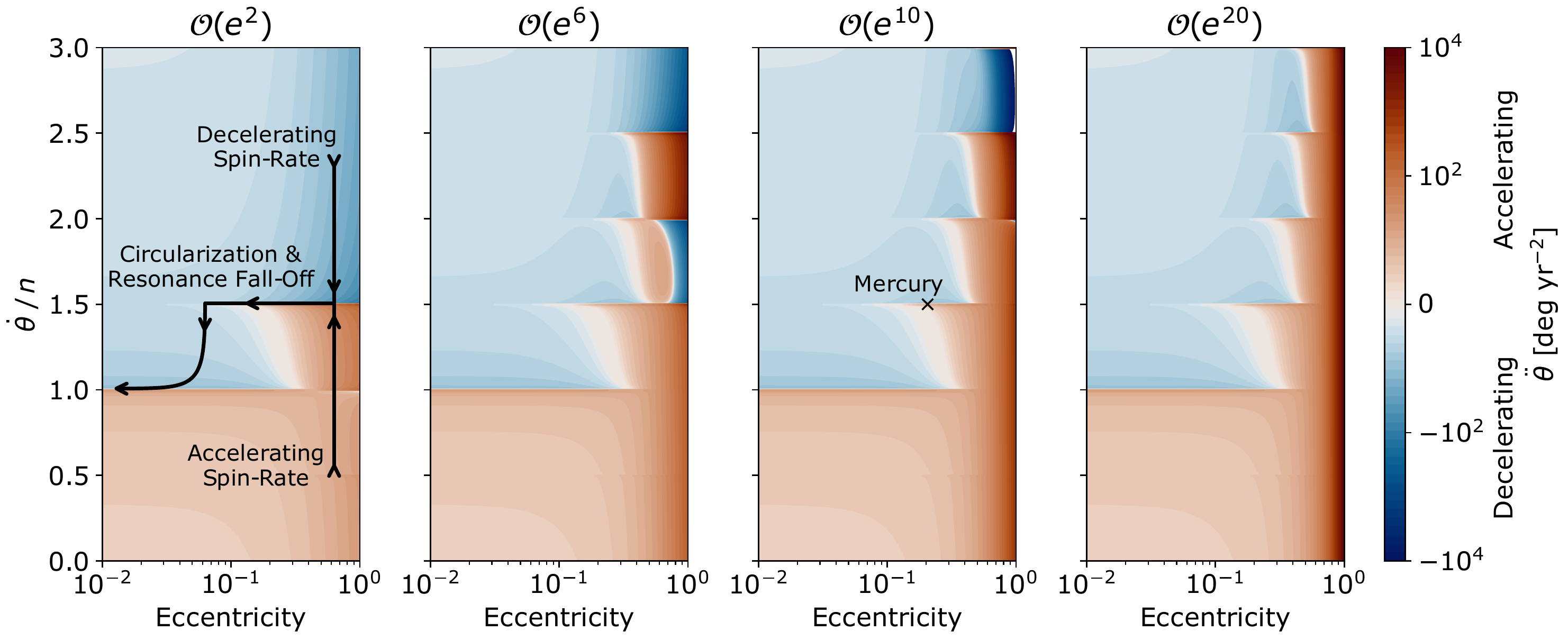}
\caption{TRAPPIST-1e's spin rate derivative (in log scale) is shown via contours over a phase space of the ratio of spin rate, $\spin$, to orbital motion, $n$, and eccentricity. Spin rate acceleration and deceleration are denoted by red and blue regions, respectively. Darker regions indicate faster changes. Areas where the two colors meet are possible spin-orbit resonances. For example, the 1:1 resonance exists at very low eccentricity (including $e=0$), whereas the 3:2 resonance requires $e \geq 0.1$. Moving from left to right, in each subplot we loosen the truncation on eccentricity as noted by the subplot title. Neglecting higher-order eccentricity terms results in spuriously missing higher-order spin-orbit resonances, and in determining the \textit{wrong sign} for the spin rate derivative at $e > 0.5$. In real systems, eccentricity will evolve simultaneously with changes in spin rate (though at different rates and, possibly, signs). Excluding any external perturbations, tidal dissipation tends to drive a planet to the left of each subplot (low eccentricity) and towards the 1:1 spin-orbit resonance line. The black arrows in the first subplot show two of the many trajectories time evolution may follow, for a world that starts with a super-synchronous (top-down arrow) or sub-synchronous (bottom-up arrow) spin rate. If however, $e$ is externally perturbed, such as by mean motion resonances, then overall leftward migration may be replaced by left-right oscillatory motion. This may cause an object to periodically both fall off and rise back onto any given SOR ledge (see text for details). 
}
\label{fig:exoplanet:nsr:contour}
\end{figure}

In Figure \ref{fig:exoplanet:nsr:contour}, we explore the role of higher-order eccentricity truncations, again using parameters that match TRAPPIST-1e for context. The contours show the acceleration (red) and deceleration (blue) of the planet's spin rate. Where the two colors meet are regions of potentially constant spin rate. A clear convergence can be seen at the 1:1 SOR at low eccentricity. This tidally-locked spin rate is the ultimate end state (assuming low triaxiality and no external perturbations) of tidal evolution and is often assumed for short-period exoplanets. Higher-order SOR's can be seen as \textit{ledges}, where a planet can be trapped if the eccentricity is high enough. The present-day position of Mercury (in its 3:2 SOR) is shown in the third subpanel (but applies to all subpanels equally). \par

In the absence of MMR's or other eccentricity pumping mechanisms, dissipation inside a planet trapped in a higher-order SOR may continue to circularize its orbit. Eventually, the eccentricity will be low enough that the planet's spin rate will fall off any ledge and quickly reach the ledge below. Arrows in the first subplot of Figure \ref{fig:exoplanet:nsr:contour} show notional trajectories in the time evolution of a planetary body through this phase space, as described above, if eccentricity is free to circularize under the influence of tidal dissipation in the satellite alone, without significant external perturbations. Initially vertical trajectories imply spin rate evolution is more rapid in these regions than the eccentricity evolution. Once an SOR ledge is reached, the trajectory becomes temporarily horizontal under the influence of free eccentricity circularizing, yet spin rate being in a stable steady state (again notwithstanding complications not included here due to possible high or low triaxiality \citep{hut1981tidal, rodriguez2012spin}). Once eccentricity falls below a critical value for each ledge, rapid spin rate evolution again occurs, until the next ledge is reached. The critical eccentricity value is a function of the planet's tidal efficiency (viscoelastic properties and internal structure) and obliquity as seen in Figures \ref{fig:high_order:obliquity:polar_torque} and \ref{fig:high_order:obliquity_NSR}. \par

However, the complete absence of external perturbers, may be a somewhat rare condition, as evidenced by satellite multiplicity in our Solar System (including the Pluto system), and in exoplanet systems \citep{limbach2015exoplanet, sandford2019multiplicity}. The eccentricity of Mercury itself evolves in a \textit{chaotic} manner influenced by both the Sun and the multitude of other Solar System bodies external to its orbit \citep[e.g.,][]{ward1976secular, burns1979rotation, CorreiaLaskar2009, lithwick2011theory, boue2012simple}. This helps explain why Mercury remains in 3:2 SOR today, as its eccentricity is not free to evolve to zero under the influence of tides alone. Therefore, objects in similar settings, with eccentricity forced by external perturbations (such as MMR's, secular perturbations, or secular resonances), may not follow the horizontal component of the trajectory in Figure \ref{fig:exoplanet:nsr:contour}, but may instead experience a left-right oscillatory motion. Consider a world trapped on an SOR ledge subject to such oscillatory eccentricity evolution. If $e$ falls below the critical lower threshold value for that ledge, the world's spin rate will begin to evolve rapidly down to the next lower ledge. If, however $e$ oscillates to a high value, it may move underneath the ledge of a higher-order SOR. Note that (for these rheological parameter values) each shelf has some degree of overhang, as seen in Figure \ref{fig:exoplanet:nsr:contour}. If $e$ oscillation moves sufficiently far under any overhang to cross from a zone bound by both blue above and red below (convergent evolution to an SOR ledge), then into a zone with all red (SOR ledge instability, and spin rate acceleration), then the system will begin to spin-up the object again, possibly long enough to return it to a higher-order ledge. Such behavior may cycle numerous times, if supported by the range of forced eccentricity values of a given multi-body planetary system. Ledge progression is therefore not uniformly downward, in the same way that $e$ evolution in complex systems is not uniformly decreasing. This evolution will be further complicated by including the effects of triaxiality \citep{Margot2018}. \par

Accounting for additional eccentricity terms (for $e \geq 0.1$) and associated tidal modes allows for trapping in higher-order SOR's (shown in subplots 2--4 of Figure \ref{fig:exoplanet:nsr:contour}). Additionally, an inadequately low truncation level for a given eccentricity can spuriously predict the wrong sign in rotation-rate acceleration, as can be seen by comparing the region between the 2:1 and 3:2 SOR's across subplots 2 and 4 of Figure \ref{fig:exoplanet:nsr:contour}. \par

The importance of these differences is debatable since spin rate changes are generally much faster than the evolution of other orbital elements \citep{Correia2009}. A planet may only experience some of these regions, which can lead to stark climate differences, for relatively short time periods. However, whether or not a planet becomes \textit{trapped} at a higher-order SOR can be critical for its long-term thermal-orbital evolution. For high eccentricities, these regimes described by higher-order truncations suggest that an initially slow-rotator planet's spin rate will tend to always increase until it reaches the lowest-order resonance associated with its (changing) eccentricity. By neglecting higher-order terms, this evolution would stop at a much lower resonance, greatly changing the outcome of its long-term evolution. Even temporary high tidal heating on such a ledge may dramatically alter later events, such as by mediating the onset of plate tectonics, ocean condensation, mantle outgassing, and secondary atmosphere formation \citep{barnes2013tidal, airapetian2020impact}. Temporary high-order SOR capture can be thermally akin to a prolonged, or late-stage, epoch of short-lived radioisotope (e.g., $^{26}\text{Al}$) decay. \par

\subsubsection{Impacts of Non-zero Obliquity on NSR}\label{sec:result:trunc:NSR_Obliquity}

\begin{figure}[hbt!]
\centering
\includegraphics[width=0.6\textwidth]{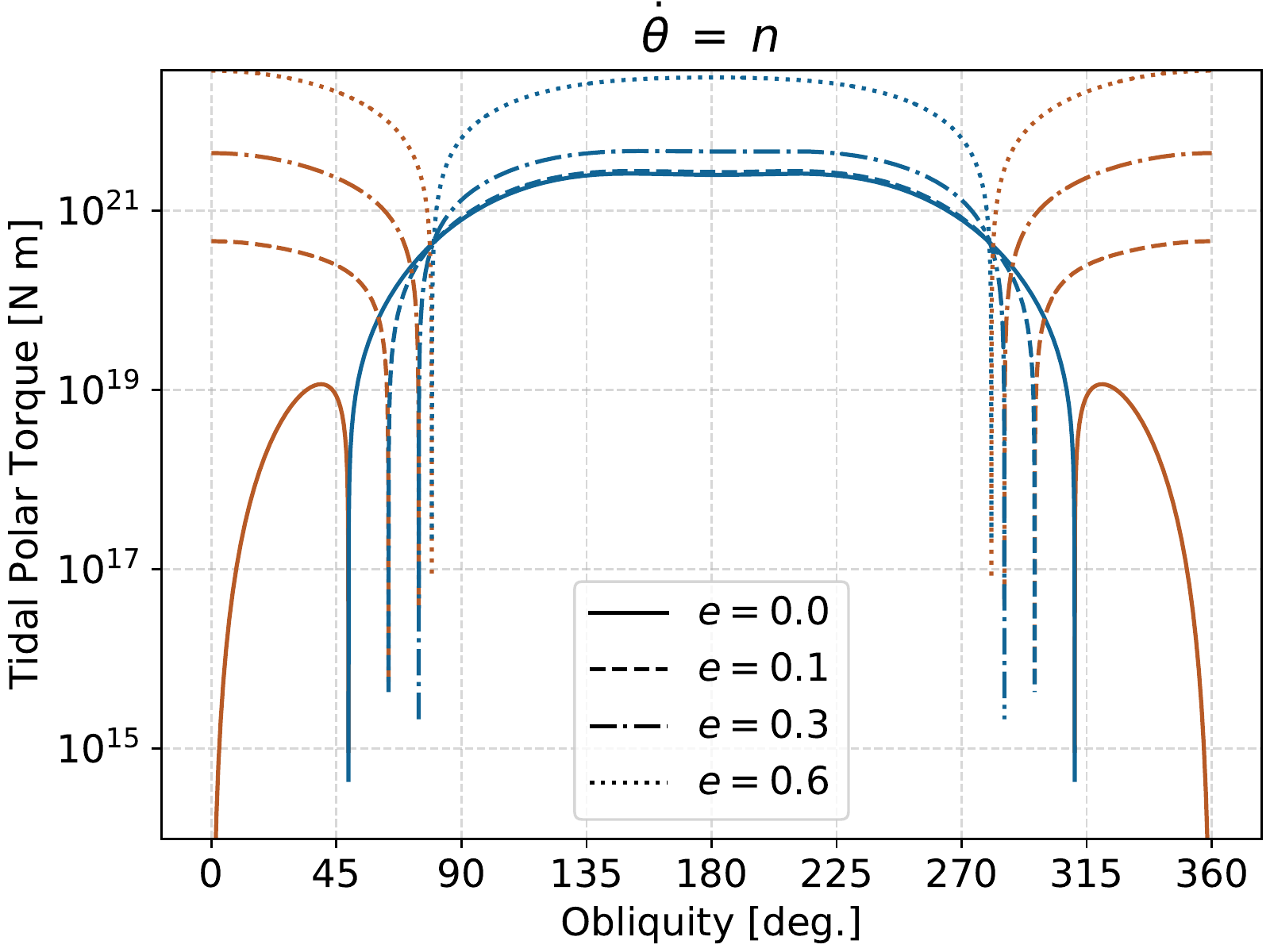}
\caption{The tidal polar torque ($\avgd{\tau_{z}} = C\avgd{\ddot{\theta}} = M_{\star} \, \partial{}\avgd{U}/\partial{}\node$) is calculated for TRAPPIST-1e across the obliquity domain. Four different eccentricities are shown in various line styles (all calculations use the $e^{20}$ truncation level). Red and blue colored lines indicate, respectively, positive and negative torques (relative to the orbital motion). A flip in the torque's direction (indicated by the spikes of $\left|\tau_{z}\right| \rightarrow 0$ and the change in line color) marks a transition in spin rate evolution. At low obliquity the positive rotation rate is defined as prograde to the orbital motion. As obliquity increases a flip occurs where the still positive rotation rate is now considered retrograde to the orbital motion. This flip of signs occurs between $0^{\degree}\pm(45^{\degree}$--$90^{\degree})$, depending on eccentricity. It marks the critical obliquity angle where the planet is considered flipped relative to the orbital plane (from a spin-dynamics perspective).}
\label{fig:high_order:obliquity:polar_torque}
\end{figure}

In Section \ref{sec:result:exoplanet:sync}, we found that obliquity tides typically cause only modest (rather than order-of-magnitude) increases in tidal heating, even when considering equations that do not approximate the obliquity dependence. The discussion around Figure \ref{fig:exoplanet:obliquity:heating} hinted that the addition of obliquity-activated tidal modes may impact the orbital and rotational evolution of a body. To explore this further, we calculate the tidal polar torque (which governs the change in rotation rate) across an arbitrary obliquity domain in Figure \ref{fig:high_order:obliquity:polar_torque}. We find that obliquity can alter the change in spin rate by orders of magnitude. At high obliquities ($I > 45^{\degree}$) the rotation axis reaches a critical point where the spin rate is no longer considered prograde to the orbital motion (denoted by the sharp spikes and change in line color in Figure \ref{fig:high_order:obliquity:polar_torque}). A high eccentricity produces a large baseline torque which necessitates larger obliquities to cause this flip in definition. There is an asymptotic relationship between eccentricity and the critical obliquity at which this flip occurs, as $e \rightarrow 1, I_{\text{crit}} \rightarrow 90^{\degree}$. \par 

\begin{figure}[hbt!]
\centering
\includegraphics[width=1\textwidth]{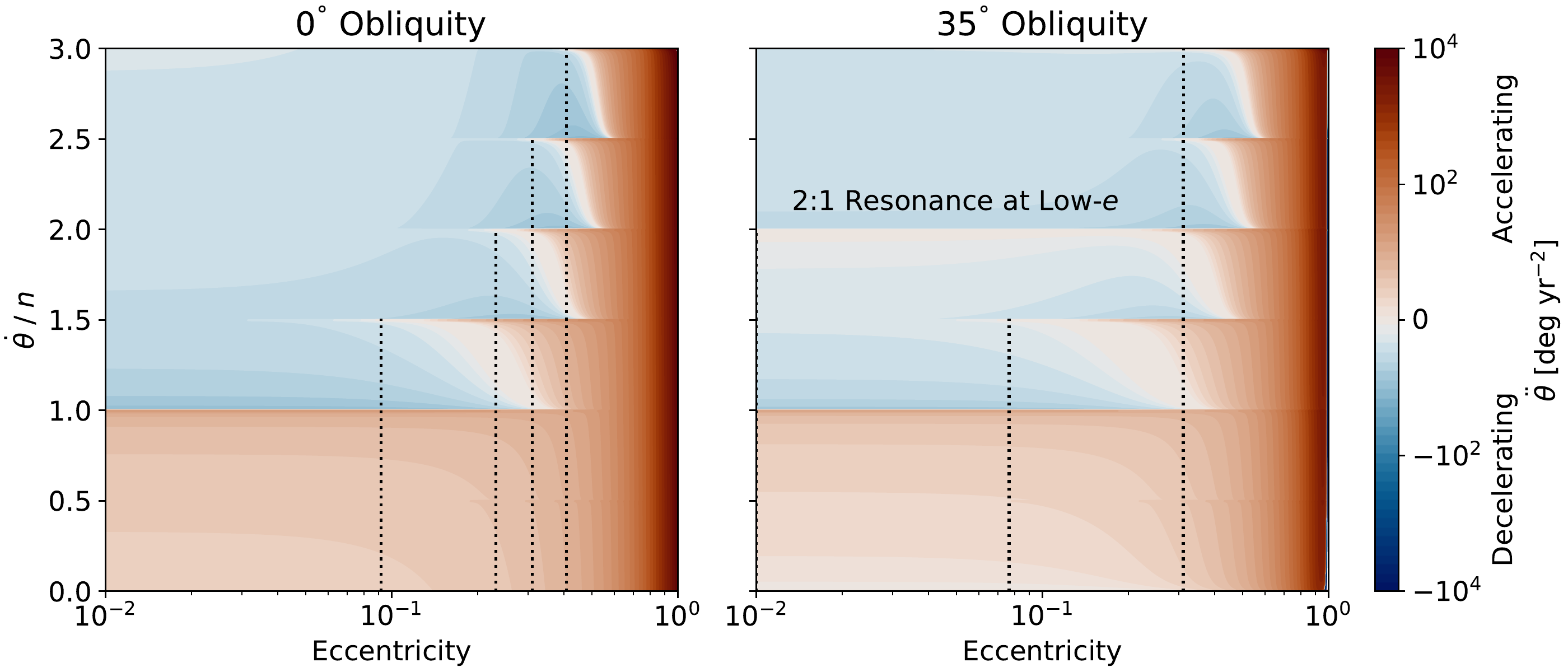}
\caption{As in Figure \ref{fig:exoplanet:nsr:contour}, we calculate TRAPPIST-1e's change in rotation over a phase space of eccentricity and ratio of spin rate to orbital motion. Eccentricity terms are truncated at $e^{20}$. The left subplot assumes no obliquity and matches the right-most plot of Figure \ref{fig:exoplanet:nsr:contour}. In the right plot, an obliquity of $35^{\degree}$ is imparted on the exoplanet. The vertical dotted lines mark, approximately, the minimum eccentricity required for a higher-order spin-orbit resonance. A non-zero obliquity acts to slightly lower the minimum eccentricity required for the higher-order resonances. More importantly, an obliquity of $35^{\degree}$ enables the 2:1 resonance to occur at very low eccentricity.}
\label{fig:high_order:obliquity_NSR}
\end{figure}

To show the impact that obliquity has on higher-order spin-orbit resonances, in Figure \ref{fig:high_order:obliquity_NSR} we reproduce the contours of Figure \ref{fig:exoplanet:nsr:contour} but with TRAPPIST-1e's obliquity constant at $0^{\degree}$ (left subplot) and $35^{\degree}$ (right subplot). First, we find obliquity tides have a slight damping effect on the underlying spin rate derivative contours, leading to an overall decrease in dissipation across the phase space. For most of the SOR's, this results in a very minor change in the minimum eccentricity required for planet spin-trapping (indicated by vertical dotted lines). However, a dramatic transformation occurs for the 2:1 SOR. Obliquity tides counteract the regular NSR tides in this region and create a resonance ledge that \textit{extends} all the way to $e=0$. A planet that either initially has a very high spin rate, or has a very large eccentricity that induces a high spin rate, will become trapped on this ledge until its obliquity is dissipated. Depending on the relative rates of eccentricity and obliquity damping, a planet may transition from the 2:1 to 1:1 SOR, completely bypassing the 3:2 resonance. \par

\begin{figure}[hbt!]
\centering
\includegraphics[width=1\textwidth]{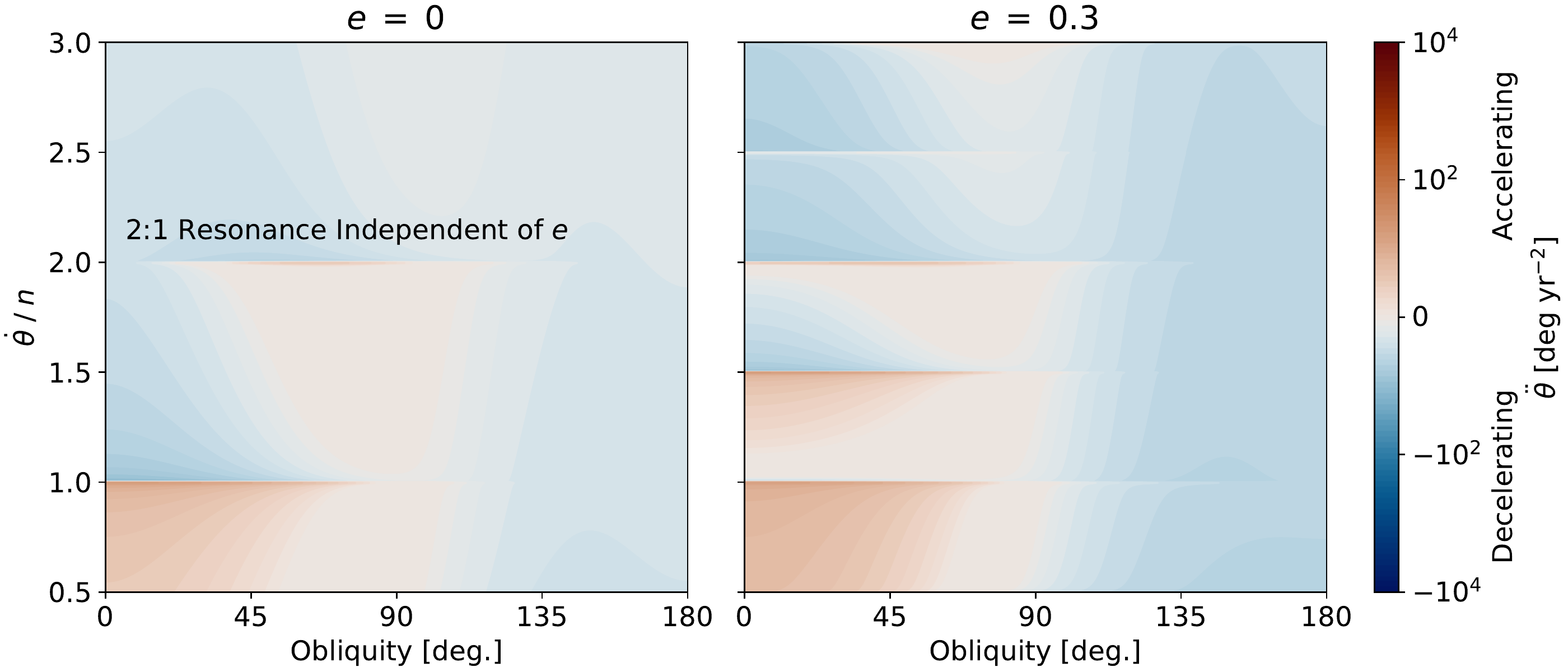}
\caption{Rotation rate derivatives are calculated in a phase space of spin rate divided by orbital motion versus obliquity. Left: The eccentricity is set to zero. The 1:1 resonance falls off at very high obliquity as the spin rate is no longer considered prograde to the orbit, thereby breaking the 1:1. The 2:1 spin-orbit resonance appears between $23^{\degree}$--$113^{\degree}$ indicating that this resonance is possible for zero or near-zero eccentricity as was found in Figure \ref{fig:high_order:obliquity_NSR}. Right: The eccentricity is set to 0.3. The 2:1 and 3:2 spin-orbit resonances are now present for all obliquities less than around 90$^{\degree}$.}
\label{fig:high_order:obliquity:contours}
\end{figure}

The presence of the 2:1 SOR at low eccentricity raises the question of what is the minimum obliquity to induce such a feature. In Figure \ref{fig:high_order:obliquity:contours} we again calculate the change in spin rate as contours except now across the obliquity domain, rather than the eccentricity domain (one may imagine such plots as differing slices through a 3-dimensional cube of ledge structures). As discussed in the previous section, ledges of SOR trapping can be found in both subplots of Figure \ref{fig:high_order:obliquity:contours}. These ledges drop off at the same critical obliquities near $I=90^{\degree}$ as found in Figure \ref{fig:high_order:obliquity:polar_torque}. In the left subplot, where eccentricity is set to zero, the 2:1 SOR has a ledge between $I = 23^{\degree}$--$113^{\degree}$. This is the same feature that leads to the 2:1 SOR ledge found at low eccentricity for $I = 35^{\degree}$ in Figure \ref{fig:high_order:obliquity_NSR}. This indicates that the minimum obliquity to induce the low-$e$ 2:1 SOR is around $I = 23^{\degree}$. By increasing the eccentricity (right subplot), this ledge extends leftward to low obliquity, including $I=0^{\degree}$, implying that a large eccentricity can allow for the 2:1 SOR trapping regardless of obliquity, as was found in the previous section. \par

\subsubsection{Viscosity Variations}\label{sec:result:exoplanet:viscosity}

\begin{figure}[hbt!]
\centering
\includegraphics[width=.6\textwidth]{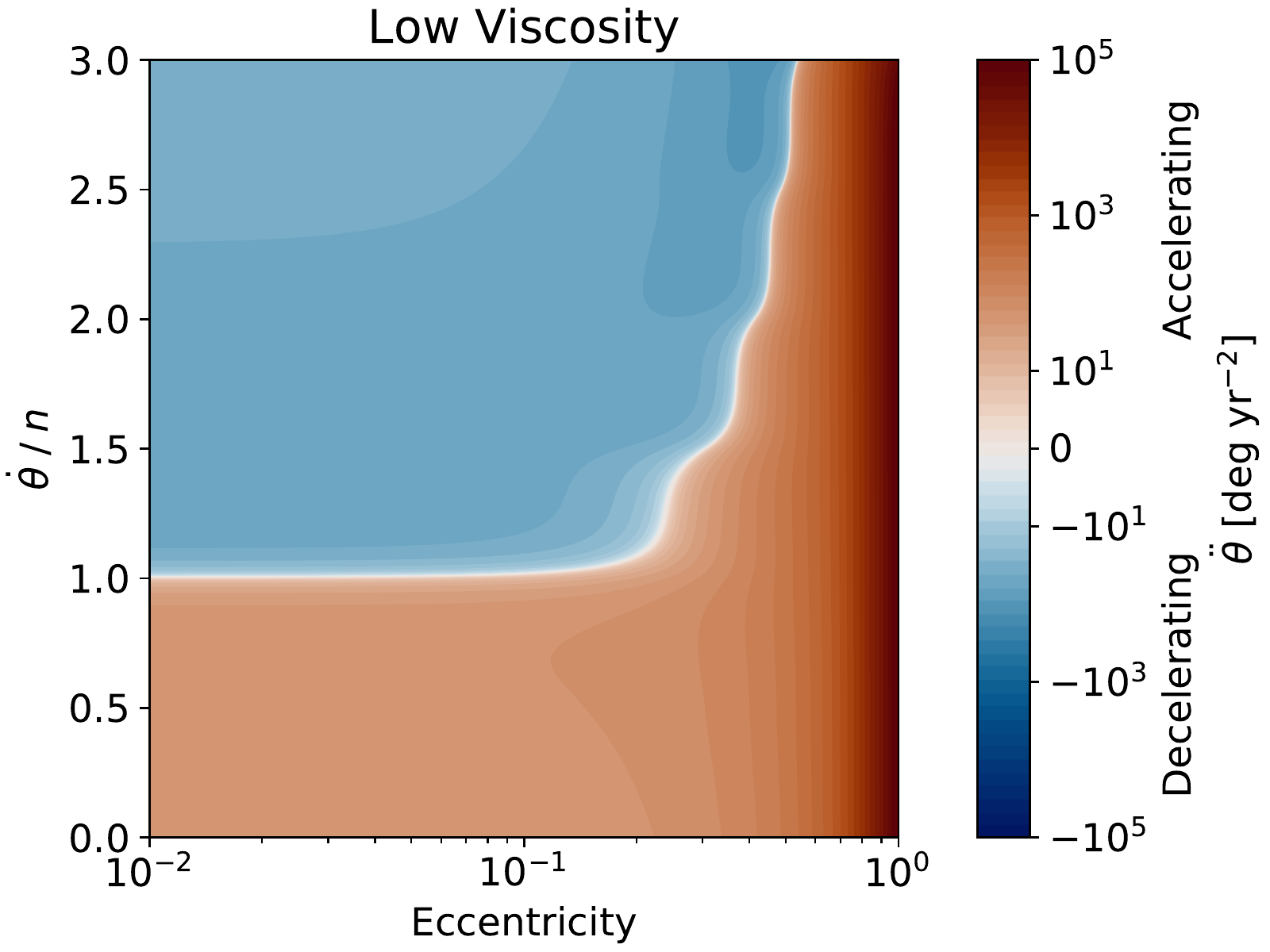}
\caption{Spin rate derivative (relative to orbital motion) as a function of eccentricity for the same system as in Figure \ref{fig:exoplanet:nsr:contour}, but with a viscosity of $10^{14}$ Pa s (previously $10^{22}$ Pa s) and a shear modulus of 1 GPa (previously 50 GPa). This mimics either a planet with severe partial melting within silicate layers, or an H$_2$O-rich world with a significant amount of dissipative ice. The sharp ledges seen in the previous figure are smoothed out. Rather than experiencing long-term capture into 3:2 or another higher-order spin-orbit resonance, a planet is more likely to continuously transition to lower spin rates unless the eccentricity is held constant or semi-constant (due to perhaps a mean motion resonance with other bodies).}
\label{fig:high_order:low_visc_NSR}
\end{figure}

Up until now, we have assumed TRAPPIST-1e's bulk was rocky and responded to tidal forces with a modest viscosity of $\eta = 10^{22}$ Pa s and shear modulus of 50 GPa. Tidal dissipation, and therefore the shape of the spin rate acceleration contours, is highly sensitive to viscosity. For example, \citet{Walterova2020} showed that the stability of higher-order SOR's is a complicated function of rheological properties (such as viscosity) and eccentricity. We also find this in Figure \ref{fig:high_order:low_visc_NSR} where we present the same NSR calculations except for a much lower viscosity of $\eta = 10^{14}$ Pa s and a shear rigidity of 1 GPa. These low values may be appropriate for a tidally dominating upper mantle that is partially melted (e.g., at a $\sim 3\%$ volume fraction, \citet{shankland1981geophysical, berckhemer1982shear, sato1991viscosity}) induced by tidal or other endogenic heat sources, or else, a super-Earth with a high-pressure ice mantle of thickness $\gtrsim 1000$ km \citep{fu2009interior, noack2016water}. This lower viscosity smooths out the spin-orbit ledges seen in the previous figure, making spin-orbit resonance trapping much less likely. Instead, if a planet is imparted with a large eccentricity, then the initially high rotation rate will continuously decrease, slowing but not stopping at higher-order SORs. If circularization is halted due to, for instance, MMR with another planet, then the spin rate could still become held at a value outside of the 1:1 SOR. However, unlike the high-viscosity case, any change in $e$ will always result in a direct change to $\spin$. This outcome is somewhat analogous to the phenomenon discussed in \citet{Makarov2013}, whereby a sufficiently partially-molten state for a planet may also interrupt the conditions for pseudosynchronous rotation. \par

The homogeneous model used in this study does not capture the effect of multiple layers of material each with a different (by orders of magnitude) viscosity and rigidity \citep[e.g.,][]{Tobie2019}. These layers will each have a unique resonant forcing frequency (or multiple ones depending upon the rheology) which will result in a peak for that layer's tidal response. It is expected that this will alter the figures and analysis presented in this section. However, any additional frequency peaks will result in a \textit{more} complex picture rather than a simpler one. \citet{HenningHurford2014} found that analyzing the frequency response of the most-dissipative layer (e.g., any asthenosphere), somewhat regardless of its volume fraction, is generally key to obtaining the true full orbital behavior, however this might be overturned by the profound differences between Figures \ref{fig:exoplanet:nsr:contour}, \ref{fig:high_order:obliquity_NSR}, and \ref{fig:high_order:low_visc_NSR}. The improvements discussed here still serve as a foundation for future studies. \par

\subsection{Dual-Body Dissipation Applied to Pluto-Charon}\label{sec:results:tno}

Several Trans-Neptunian Objects and Kuiper Belt Objects (collectively ``TNOs'') have been found with large satellites. Some theories suggest that these systems formed through a collisional process. Such an origin can initialize these worlds with high eccentricity and spin rates. Any high-energy initial state for Pluto-Charon has tidally dissipated to the low-eccentricity, dual-synchronous state observed today. This damping process has largely erased initial orbital and rotation conditions of such systems. However, we can deduce some information from observations of other, non-binary TNOs, as well as the formation process itself \citep[][and references therein]{Kenyon2019}:

\begin{itemize}
    \item The initial spin rates of both objects in a binary system are unlikely to have been equal to one another or to their initial orbital motion.
    \item The initial eccentricity could be large and the initial obliquity of the bodies would be random. 
\end{itemize}

For these reasons, the tidal evolution of TNO binaries must be reexamined using insights found in the previous section regarding NSR and the inclusion of higher-order eccentricity terms. Furthermore, while a simple tidal response assumption such as CPL may be acceptable for estimating dissipation within a star or a gas giant, it does not accurately describe the better-known rheological response to tidal forcing of solid materials (e.g., ice and rock) inside \textit{both} objects in a TNO system like Pluto-Charon. Since dissipation of both binary members affects long-term dynamical evolution (e.g., changes in $a$ and $e$) of the system (Section \ref{sec:method:orbital}), we must consider dissipation inside both worlds simultaneously (\textit{dual-dissipation}). \par

\subsubsection{Effect of Dual Dissipation on the Time Derivative of Eccentricity}\label{sec:results:tno:dual}

\begin{figure}[hbt!]
\centering
\includegraphics[width=1\textwidth]{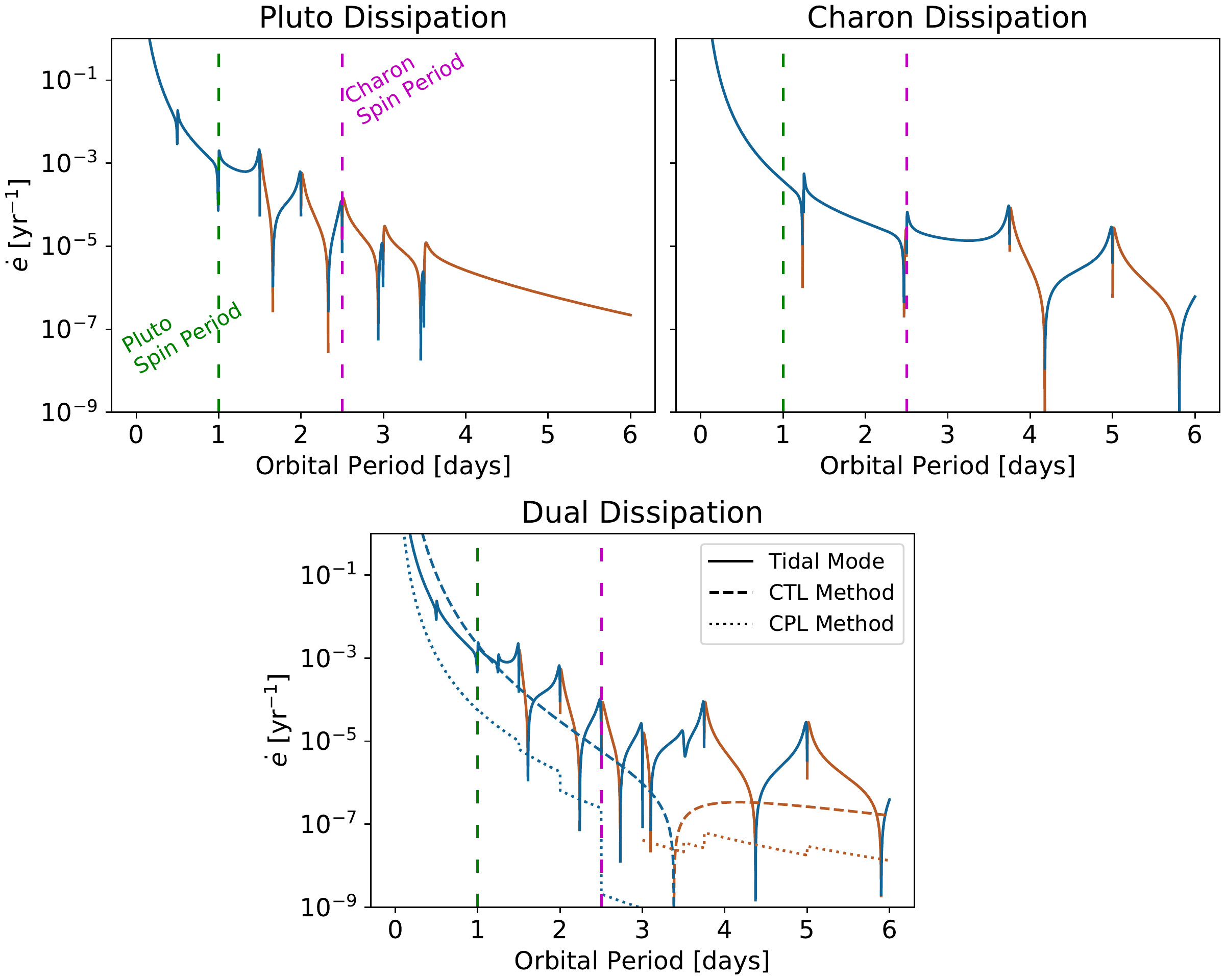}
\caption{The time derivative of Pluto-Charon's mutual eccentricity is shown as a function of the orbit's period. Blue indicates a shrinking $e$ (negative derivative), whereas orange shows a growing $e$ (positive derivative). Pluto and Charon's spin periods are arbitrarily set to 1 day and 2.5 days, respectively (indicated by the vertical green and magenta dashed lines), as a possible state in their early evolution before they reached their current spin-synchronous value of 6.39 days. Either world is in NSR for orbital periods outside of these vertical lines. In the top two subplots, tidal dissipation is restricted to Pluto (left subplot) and Charon (right subplot). The dual-dissipation model (where tides are calculated for both worlds) is shown in the bottom subplot where, in addition to the canonical Sundberg-Cooper model, CTL and CPL models with dual dissipation are shown for comparison. For these models, a fixed-$Q$ of 100 was used for both Pluto and Charon. A value of $e = 0.3$ is assumed along with no obliquity in either body. The derivatives calculated with a multi-mode rheological model capture a far more robust picture of spin-orbit resonances compared to CTL or CPL assumptions.}
\label{fig:tno:dual}
\end{figure}

On the surface, the dual-dissipation model is simply an addition of the two individual planets' dissipation terms into the disturbing potential \citep{Heller2011, BoueEfroimsky2019}. However, since the change in the orbital motion is now dependent upon the dissipation of \textit{both} solid worlds, a further level of coupling occurs when using a frequency-dependent rheology due to the complex Love number's dependence on the orbital motion via the tidal modes \citep{Efroimsky2012apj}. Such coupling can be seen in Figure \ref{fig:tno:dual}, which shows the dependence of Pluto-Charon's mutual $\dot{e}$ on the orbit's period. When dissipation inside both Pluto and Charon is accounted for (bottom subplot), the resulting $\dot{e}$ is a superposition of the effects found when dissipation is restricted to Pluto and Charon (top subplots). \par

In Figure \ref{fig:tno:dual}, the spin periods of Pluto and Charon were fixed arbitrarily to create a phase space cross-section for illustration. In practice (Section \ref{sec:results:tno:time}) they evolve at different rates, potentially stalling temporarily as one or both worlds encounter higher-order spin-orbit resonances depending on their $e$, $I$, and interior state (Section \ref{sec:result:trunc}). It is therefore important to capture all peaks and troughs in $\dot{e}$ as seen in Figure \ref{fig:tno:dual}. In general, lower-fidelity methods (CPL or CTL models, dissipation in a single body) tend to underpredict dissipation (depending on the choice of $Q$) and decrease the range of eccentricity values which lead to spin-orbit trapping. \par

\subsubsection{Additional Effects due to Non-zero Obliquity}\label{sec:results:tno:obliquity}

\begin{figure}[hbt!]
\centering
\includegraphics[width=1\textwidth]{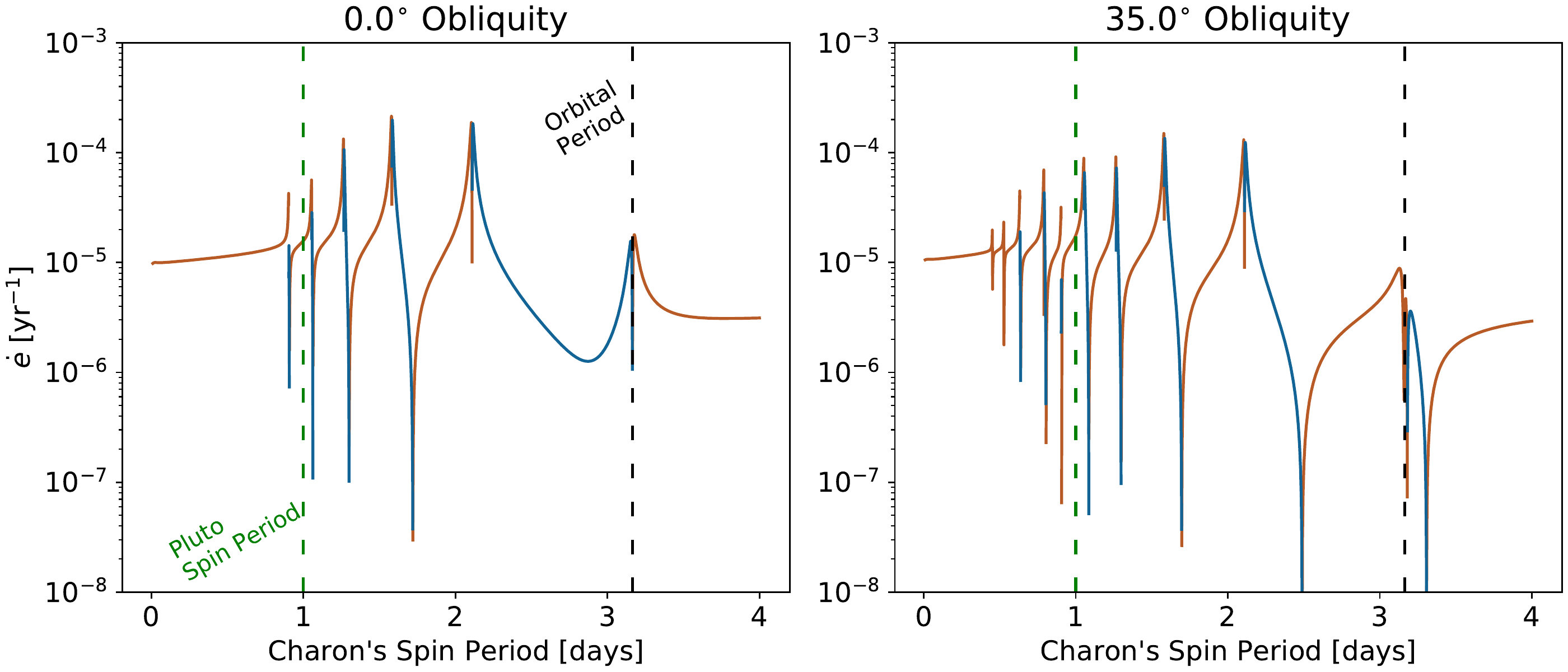}
\caption{To show how obliquity can induce additional spin-orbit resonances, Pluto and Charon's mutual $\dot{e}$ is shown as a function of Charon's spin period. Pluto's spin period is set to 1 day (vertical green dashed line) and their orbital period to half its modern value ($\approx 3.2$ days; vertical black dashed line). These are indicative of a snapshot early in the system's evolution. As in Figure \ref{fig:tno:dual}, blue indicates negative $\dot{e}$ and orange marks positive. In the left subplot, both Pluto and Charon have zero obliquity and no orbital inclination. In the right subplot, Pluto remains at zero obliquity while Charon's is increased to 35$^{\degree}$. Charon's obliquity tides activate new tidal modes which impart additional peaks and troughs in the eccentricity derivative. Thus, obliquity tides can cause order-of-magnitude differences in the orbital and rotational evolution, even though they may not have a significant direct impact on internal heating (Figure \ref{fig:exoplanet:obliquity:heating}).}
\label{fig:tno:obliquity}
\end{figure}

Just as Pluto's other moons are observed to have high obliquity relative to the Pluto-Charon orbital plane \citep{Weaver2016}, it is also likely that Pluto and/or Charon's obliquity was initially non-zero\footnote{Pluto's smaller satellites most likely have large modern-day obliquities due to their much weaker tidal dissipation owing to their small size and therefore cold internal temperatures. They also orbit at least twice as far from Pluto as Charon does, thereby decreasing their tidal susceptibility by a factor of $> 64$. An interested reader may review the work of \citet{Correia2015} and \citet{Quillen2017} for more information on these moons' evolution.}. As discussed in Sections \ref{sec:result:trunc:obliquity} \& \ref{sec:result:trunc:NSR_Obliquity}, non-zero obliquity can lead to modest increases in tidal heating and potentially dramatic changes in rotational and orbital evolution. To explore the impact of obliquity tides on the Pluto-Charon system, we calculate $\dot{e}$ at a possible snapshot in time of Pluto and Charon's early evolution (Figure \ref{fig:tno:obliquity}). Since Charon's $\spin$ likely evolved more quickly than Pluto's, or their mutual $n$, we choose it as a free parameter across the x-axis. The eccentricity is fixed to 0.3, therefore numerous tidal modes exist even for the $I=0^{\degree}$ case (left subplot). By setting Charon's obliquity to 35$^{\degree}$ (right subplot) we find several new modes that increase the number of potential spin-orbit couplings as well as altering the ones present for no obliquity. \par


\subsubsection{Time Evolution of Pluto-Charon using a Dual-Dissipation Model}\label{sec:results:tno:time}

The discussion so far has focused on static snapshots to show complexity changes when additional tidal modes become active. In reality, all system parameters (orbital motion, eccentricity, spin rates, etc.) evolve in time. They also strongly depend on the viscosity and rigidity of both worlds which in turn depend upon interior structure and thermal state. Several recent studies have looked at this coupled thermal-orbital evolution for Pluto-Charon \replaced{\citep{RobuchonNimmo2011, Cheng2014, BarrCollins2015, Hammond2016, Quillen2017, Saxena2018}}{\citep{RobuchonNimmo2011, Cheng2014, BarrCollins2015, Hammond2016, Quillen2017, Saxena2018, Arakawa2019}}, but these generally do not consider the dual-body dissipation for a highly eccentric and non-synchronously rotating system\footnote{\citet{Cheng2014} did consider the dissipation within both bodies and tracked the non-synchronous spin rate, including the effect of Pluto's rotational flattening. However, this study utilized the CTL and CPL model which do not model the real response of these worlds' bulk to tidal forces. In NSR situations especially, frequency response is critical, as the forcing frequency spans many values within a time simulation.}. The range of likely initial conditions and possible interior configurations and compositions creates a large parameter space that deserves a dedicated study. However, to show how some of the concepts discussed here can change the long-term evolution, we show one example evolution scenario for Pluto-Charon. The interior and thermal evolution of Pluto and Charon follows the methods discussed by \citet{HussmannSpohn2004} in the context of Europa. This thermal model is coupled with the orbital evolution described in Section \ref{sec:method:orbital}. For this particular example, we assume Pluto and Charon both start at a spin rate higher than their initial orbital mean motion, which in turn is faster than the modern-day value, indicative of a closer-in Charon ($a_{\text{Initial}} = 6R_{\text{Pluto}}$). To approximate a possible post-collision state, Charon's initial orbital eccentricity is set to $e = 0.5$. We do not, however, model the impact of obliquity tides in this example ($I_{\text{Pluto}} = I_{\text{Charon}} = 0^{\degree}$). Evidence is beginning to show that Pluto may have initially been warm \citep{Bierson2020}, so we start both Pluto and Charon with relatively warm interiors. Primordial concentrations of radioactive isotopes in their rocky cores help to sustain this warm state regardless of tidal dissipation. \par

\begin{figure}[hbt!]
\centering
\includegraphics[width=1\textwidth]{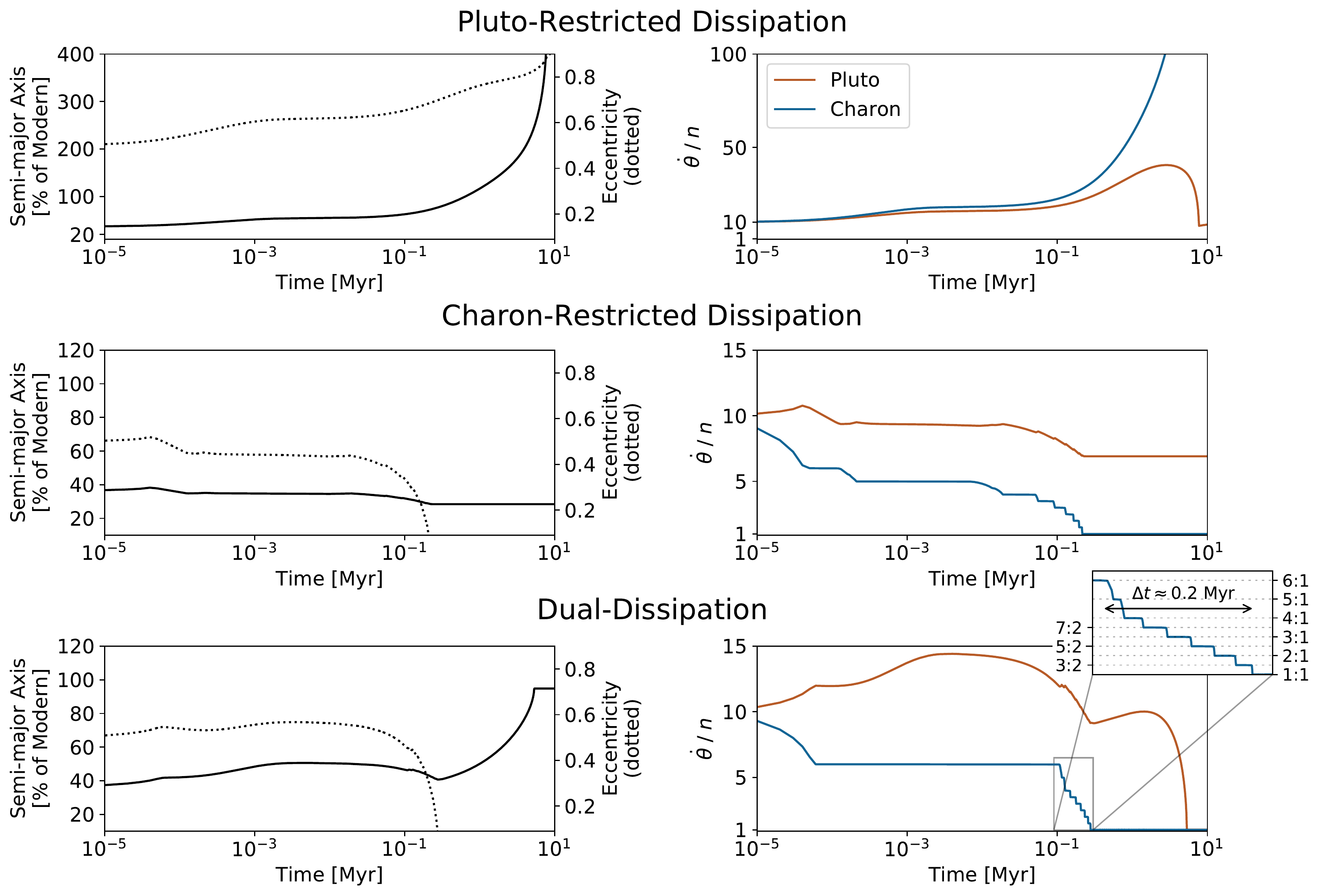}
\caption{The orbital (left) and rotational (right) evolution is shown for an example set of initial conditions for the Pluto-Charon system. The initial orbital parameters are $a_{\text{Initial}} = 6R_{\text{Pluto}}$, $e_{\text{Initial}} = 0.5$, and no mutual inclinations or obliquities. The spin rates of Pluto and Charon are both set to $10 \times$ the initial orbital frequency. The orbits and spin rates quickly evolve within the first 10 years (not shown) to the values shown at the left of each plot. Top and middle rows: Dissipation is turned off within, respectively, Charon and Pluto. Last Row: Both Pluto and Charon are allowed to simultaneously dissipate tidal energy. The inset plot in the bottom right highlights Charon's spin rate falling from the 6:1 spin-orbit resonance onto lower-order spin-orbit resonances, a process which lasts about 200 kyr. Scales are set generally to help comparisons between models. Exact arrival at present-day parameters is not sought, as the goal is to visualize, from among countless possible time histories, the essential role of model fidelity.}
\label{fig:tno:timedomain}
\end{figure}

In Figure \ref{fig:tno:timedomain}, we compare the dual-body dissipation model (bottom row) to models where only one body is dissipating tidal energy. Not accounting for the simultaneous dissipation within both bodies can lead to significantly different orbital and rotational outcomes. As seen in the dual-dissipation model, Pluto and Charon reach their dual-synchronous end state in just over 5 million years. \par

The system experiences two phases during dynamical evolution. First, Charon's spin rate evolves towards the 1:1 SOR. Its evolution is slowed by encountering higher-order SOR's. At the same time, the orbit circularizes ($e \rightarrow 0$). As the eccentricity decreases, Charon is no longer able to remain in the higher-order resonances (equivalent to falling off the ledges described in the last subplot of Figure \ref{fig:exoplanet:nsr:contour}). Charon's tidal dissipation also acts to contract the mutual orbit. \par

Second, after Charon has reached 1:1 SOR, the spin rate of Pluto is next driven toward synchronization with the orbital motion, $n$. At this point, $e = 0$. Therefore, Pluto does not encounter any higher-order SOR's. Angular momentum is transferred from Pluto's fast spin rate into the mutual orbit, expanding the semi-major axis. This increase in orbital separation is \textit{not} seen when dissipation is restricted to Charon. For the case where only Pluto is dissipating (top row), orbital expansion begins immediately and is not counteracted by Charon's dissipation\footnote{In this scenario, eccentricity increases throughout the time domain and, after a million years, becomes very large. This level of eccentricity likely requires even higher-order truncations than the $e^{20}$ terms we use here. Therefore, the Pluto-restricted dissipation (top row) results should be taken with some skepticism after $\approx 1$ Myr.}. After a million years, the orbital separation is so large that tidal dissipation has dropped significantly (recall that tidal dissipation $\propto a^{-6}$; See Eq. \ref{eq:dissipation}). For the dual-dissipation case, on the other hand, after 250,000 years, $a$ remains at about 40\% of its modern value ($a \approx 6.6R_{\text{Pluto}}$) and begins to expand due to Pluto's super-synchronous spin rate. Unlike the Pluto-restricted case, the orbital separation never becomes so large that dissipation ceases. Pluto's spin rate continues to decrease toward synchronization until, after around 5 Myr, the system has reached the circular, dual synchronous state that we find it in today. \par

Thus, accounting for dissipation in both Pluto and Charon results in a significantly different (and more complete) picture of the binary's orbital and rotational evolution. This is only one example to illustrate how much dynamical evolution can vary dramatically depending on a full portrait of tidal modes and sources, as well as both initial conditions and interior states. In particular, the effect of a heterogeneous interior (here comprising a silicate-metal-organic-rich core, liquid ocean, and ice-rich shell) warrants further study, including both fluid tide dissipation and effects on dissipation in the ice shell arising from how an ocean mechanically decouples the shell from the core. Although we have assumed here that dissipation in such a heterogenous interior would be decreased by the viscoelastic volume fraction $f_{\text{TVF}}$ relative to that computed for a homogeneous interior, in reality material temperatures, mechanical boundary conditions, and compositions may affect tidal dissipation in different ways (although use of the volume fraction assumes we are focusing on whatever material layer is \textit{most} dissipative from a temperature/composition perspective). Accurately capturing these effects requires utilizing the tracked interior structure, in a fully multilayer tidal computation (bottom right box of Fig. \ref{fig:model}), along with the high-degree multi-modal aspects of this study. \par

\section{Conclusions}\label{sec:conclusion}

We have found that using traditional tidal evolution formulae, which truncate eccentricity functions to $e^{2}$, on planets and moons in highly eccentric orbits ($e \geq 0.1$) can lead to significant changes to spin rate evolution and modest errors in heating rates. These errors can increase by orders of magnitude for very high eccentricity ($e \geq 0.6$). Specifically, the time derivative of eccentricity using the $e^{2}$ truncation predicts a flip in direction (from a circularizing orbit to one with a growing eccentricity) around $e = 0.8$ that is not seen when higher-order terms are included. These errors are compounded when the world is allowed to rotate non-synchronously. NSR can lead to spin-orbit trappings which are sensitive to eccentricities as low as $e=0.1$ for the cases tested here. Higher-order eccentricity terms not only activate new spin-orbit resonances but also alter the path a planet may take as it falls onto them. Eccentricities of around $e=0.4$ can accelerate a planet's spin rate out of the 1:1 SOR to frequencies several times the orbital motion. \par 

A world that experiences a new-onset secular perturbation, secular resonance, or mean-motion resonance, imparting significant eccentricity, might be knocked out of its tidally locked state from eccentricity-induced NSR alone. This highly eccentric, NSR state can generate a large amount of tidal heating in the planet's interior. Any NSR state will quickly evolve to a lower dissipative, SOR state (1:1 or higher depending upon the specifics). There is a possible testable bias for higher-order SOR's to be found in younger star systems, which experience greater levels of eccentricity-enhancing mechanisms, including planet-planet mergers, migrations, secular resonance crossings, and changing orbital resonance states. For exoplanets that become trapped in a higher-order SOR for long periods of time, the climate could be dramatically altered from the new solar incidence \citep{Turbet2016AandA, DelGenio2019AstroBio}. For these reasons, care should be taken when assuming 1:1 tidal locking for short-period exoplanets that exhibit an eccentricity greater than about 0.1. This assumption should be seriously reexamined for worlds with $e \geq 0.3$. For $e < 0.1$, we expect results derived with the common truncation to $e^{2}$ to remain reasonably valid for most systems, although this depends upon a world's viscoelastic state. For instance, \citet{Walterova2020} found that the 3:2 SOR is possible for eccentricities as low as 0.08 depending on the viscoelastic state of the exoplanet. This also does not preclude the potential need to consider higher-eccentricity terms when studying the world's past evolution when eccentricities may have been higher. Observing an exoplanet's spin rate is currently a difficult proposition. Observational evidence of spin rates has been reported for exoplanet's larger than Jupiter \citep[e.g.,][]{Snellen2014, Zhou2016}, but probing rocky super-Earth or smaller worlds we discuss here will require new advancements in observing technology and techniques. However, a short-period exoplanet observed to have both a low eccentricity and a non-1:1 spin rate could be evidence for a dynamically young system experiencing orbital perturbations or resonances. Alternatively, such an exoplanet may exhibit a higher-order SOR due to a significant triaxiality, obliquity, or due to it being a poor dissipator of tidal energy.\par

The \textit{rotational and orbital model} described in this work is \textit{equally} suited to the long-term evolution of stars and gaseous planets (such as close-in hot Jupiters). Thus, our recommendations regarding eccentricity truncation levels to use are generally extensible to such worlds. However, these worlds' tidal dissipation mechanisms are poorly understood at present. In general, using a CTL or CPL model for gas giants, rather than the Sundberg-Cooper rheology we employ for rocky and icy worlds in this study, is a reasonable approach, yet studies such as \citet{storch2014viscoelastic} suggest viscoelastic gas giant layers could alternatively play dominant roles in dissipation. Similar questions arise from \citet{Lainey2020} for Saturn. Considering giant planets of decreasing size, there must be a transition mass range at which viscoelastic layers become important, and then dominant. This mass range may even be near that of Neptune itself \citep{remus2012anelastic, zeng2014effect}, and vary as a planet ages. Thus, the possible applicability of this works' findings to ice giant planets may be of high relevance considering the number of mini-Neptune worlds now observed \citep{dressing2013occurrence}.  Overall, since the likelihood of capture into higher-order SOR's is highly dependent upon the tidal efficiency, the findings presented throughout this work should be reexamined when discussing gaseous planets or stars. \par

For very high eccentricities, like those expected in some TNO binary formation scenarios, the use of high powers of eccentricity in tidal dissipation equations must be considered. This is due to the poor convergence of the eccentricity functions, $G(e)$, used in the Darwin-Kaula tidal theory \citep{Bagheri2019b}. This poor convergence is magnified when studying worlds experiencing NSR. In Appendix \ref{sec:app:truncation} we provide the tidal dissipation equations for NSR with eccentricity terms up to and including $e^{10}$ at an arbitrary obliquity. The effects of terms up to and including $e^{20}$ are discussed throughout this study. We also provide the synchronous-rotation equations assuming no obliquity in Appendix \ref{sec:app:sync}. \par

Obliquity tides can increase tidal heating (a maximum of $3 \times$ higher heating for TRAPPIST-1e assuming synchronous rotation and rock-like material properties) but are generally still a weaker source of heat than a high eccentricity or out-of-resonance spin rate. However, even modest obliquities do alter rotational geometry in a manner that can have a significant impact on tidal torques. Therefore, orbital and rotation evolution are highly sensitive to non-zero obliquity. Obliquity tides also activate new tidal modes, some of which can be scaled up by any non-zero eccentricity. If the eccentricity is large, then even a small obliquity can lead to new and powerful tidal modes. We also find that obliquity can greatly enhance the chance of trapping into the 2:1 spin-orbit resonance. The likelihood of this trapping will depend on the relative damping rates between obliquity and eccentricity, which will be the topic of a future study. Also left for future work is an assessment of the impact of the common simplifying assumption we invoke, of averaged orbital motions (i.e., equating true and mean anomalies), on tidal dissipation at high eccentricity. \par

The rotational evolution of any planet will be further complicated by effects that we do not consider in this work. Prime among them is the influence of triaxiality which can induce additional torques that may allow for easier capture into higher-order SOR's at lower eccentricities \citep{rodriguez2012spin, MakarovEfroimsky2014, FrouardEfroimsky2017}. However, the eccentricity, obliquity, and NSR effects discussed in this work will still be present during any triaxiality-induced evolution. Therefore, understanding the impact of these latter effects in isolation is an important step towards building a complete picture of the rotational and orbital evolution of planets and moons. \par

All results from this study point to more complexity and nuance in the rotational, orbital, and thermal history of worlds, as can manifest in dynamical cascades through many high-order spin orbit resonances (as in the lower right-hand inset of Figure \ref{fig:tno:timedomain}; see also the work by \citet{Correia2014}). At each successive SOR, tidal heating drops, suggesting (laterally inhomogeneous) pulsations in both heat and stress. Such rapid changes ($\sim$ 0.01--0.05 Myrs each in Figure \ref{fig:tno:timedomain}) might act akin to `freeze thaw cycles', which exacerbate fracturing on Earth, or akin to cyclic working of fractures on other icy moons. \citet{rhoden2020charon} found that fractures on Charon do not match diurnal tidal stress patterns. That result, combined with dynamical trends in this work, suggest fracturing still visible on Charon could, in part, have been influenced by a complex SOR cascade. \par

We have implemented a new dual-dissipation model that simultaneously tracks tides within both the host and satellite of a binary system using semi-analytical equations and the Sundberg-Cooper rheology. This model was applied to Pluto-Charon based on its presumed energetic origin, but is equally applicable to any satellite orbiting a central host (such as Neptune \& Triton, the Earth \& Moon, or an exoplanet around a highly dissipative star). This dual-dissipation model creates a second-order thermal-orbital feedback in which one planet's spin rate can cause orbital changes which will in turn cause thermal variations inside the opposite planet. For Pluto-Charon, we find that the dual-dissipation model generates faster orbital changes than found by \citet{Saxena2018}. However, the potential for interim higher-order SOR trapping may still extend the time it takes for the system to reach its dual-synchronous state as is observed today. Not considering dual-body dissipation leads to dramatically different orbital outcomes. Lastly, we confirm others' work that the CTL and CPL methods have a very different response to NSR tides than a rheology-based viscoelastic model. The CTL and CPL methods, while perhaps remaining applicable to the poorly understood dissipation within gaseous and stellar bodies, are not accurate models of the long-term evolution of rocky and icy worlds that experience varying forcing frequencies. \par

\acknowledgments \label{sec:acknowledgments}
We would like to thank Michael Efroimsky for his guidance on implementing the dual-dissipation model for NSR tides. Valeri Makarov and Dimitri Veras assisted with the eccentricity and inclination functions. We would also like to thank Julien Frouard, Alyssa Rhoden, Robert Tyler, and Eric Wolf for thought-provoking discussions. Lastly, we are grateful for the helpful comments and suggestions provided by this manuscript's anonymous reviewers. This work was supported by the NASA Habitable Worlds Program (NH16ZDA001N-HW) as well as from the Sellers Exoplanet Environments Collaboration at NASA Goddard Space Flight Center. M.N., P.S., and W.H. acknowledge support from the CRESST-II cooperative agreement between NASA Goddard Space Flight Center and the University of Maryland, College Park. A.B. acknowledges support by the Swiss National Science Foundation (SNSF project 172508 ``Mapping the internal structure of Mars").

\software{
    NumPy \citep{NumPy},
    SymPy \citep{SymPy},
    Julia DiffEq \citep{JuliaDiffeq}, Matplotlib \citep{Matplotlib}, and
    reduced-color-perception conscious color maps \citep{ScientificColorMaps}.
}

\newpage

\appendix \label{sec:appendix}

\section{Dissipation Formulae for Arbitrary Obliquity and Eccentricity Terms Truncated to the 10th Power}\label{sec:app:truncation}

In Table \ref{tab:dissipation_order10}, we provide the tidal potential derivatives and heating equation for arbitrary obliquity. The dissipation equations were calculated using Eqs. \ref{eq:dissipation}. We have dropped all terms containing powers of $e^{12}$ and higher, as well as any terms that would result in the tidal heating and potential derivative equations equaling zero. These derivations assume no pericenter or nodal precession and only consider the secular evolution. Lastly, only quadrupole terms ($l = 2$) are provided (see Appendix \ref{sec:app:high_l} for a discussion about $l > 2$). \par

Using Table \ref{tab:dissipation_order10}, one can find the derivative of the tidal potential with respect to the variable $Y$ by (where $Y \in \{\meanAnom, \pericenter, \node\}$),

\begin{equation}\label{eq:app:potential_derivative_eq}
    \partialDerv{\avgd{U_{j}}}{Y} = \frac{3}{2}\frac{GM_{k}R_{j}^{5}}{a^{6}} \sum_{i} C_{Y,\;i} \;F^{2}_{i} G^{2}_{i}\, \widetilde{\text{K}}_{j}\left(\chi_{i}\right),
\end{equation}

\noindent where the summation index $i$ is over each non-zero tidal mode, presented as individual rows in Table \ref{tab:dissipation_order10}. For each mode $\omega_i$, $C_{Y,\;i}$ is the respective derivative's coefficient (3rd column in Table \ref{tab:dissipation_order10}).  \par

Tidal heating $\avgd{\dot{E}_{j}}$ can be found in a similar fashion (note the additional factor of $M_{k}$),

\begin{equation}\label{eq:app:heating_eq}
    \avgd{\dot{E}_{j}} = \frac{3}{2}\frac{GM^{2}_{k}R_{j}^{5}}{a^{6}} \sum_{i} \chi_{i} \;C_{\avgd{\dot{E}},\;i} \;F^{2}_{i} G^{2}_{i} \text{K}_{j}\left(\chi_{i}\right).
\end{equation}

Tidal heating also carries an additional factor of the tidal forcing frequency, $\chi_{i}$, which is defined as the absolute value of each mode, $\left|\omega_{i}\right|$. The Love number in Eq. \ref{eq:app:potential_derivative_eq} contains the sign of the tidal mode (denoted by the tilde above the $\text{K}$), $\text{Sgn}(\omega_{i})$, whereas the Love number in Eq. \ref{eq:app:heating_eq} has no sign dependence and is strictly greater than or equal to zero for all modes. \par

The functional forms of both the potential derivatives and tidal heating are the same if one is considering the host or satellite, however the subscripts $j$ always refer to the object for which tidal heating is being computed, and $k$ the opposite object. These must be swapped when moving from one world to the other.

\small
\begin{longrotatetable}
\begin{longtable}[c]{@{}c|c|cccc|c|l@{}}
    \caption{Tidal heating and potential derivative terms were calculated using Eq. \ref{eq:dissipation} assuming arbitrary obliquity and truncating eccentricity terms up to and including $e^{10}$. These terms can be collapsed into a single value for the potential derivatives and tidal heating using Eqs. \ref{eq:app:potential_derivative_eq} and \ref{eq:app:heating_eq} respectively. The square of the eccentricity and inclination functions were calculated using the formulae discussed in Appendix \ref{sec:app:eccen_inclin}.  \label{tab:dissipation_order10}} \\ \toprule
    	Mode & Signature & \multicolumn{4}{c|}{Coefficients, $C_{Y}$} & Inclination Function & Eccentricity Function \\
    $\omega_{j}$ & $l$, $m$, $p$, $q$ & $\partialDerv{\avgd{U_{j}}}{\meanAnom}$ & $\partialDerv{\avgd{U_{j}}}{\pericenter_{j}}$ & $\partialDerv{\avgd{U_{j}}}{\node_{j}}$ & $\avgd{\dot{E}_{j}}$ & $F^{2}(I_{j})$ & $G^{2}(e)$ \\ \midrule
    \midrule           
    \endfirsthead      
    \toprule
    \multicolumn{8}{c}{continue table}\\
    Mode & Signature & \multicolumn{4}{c|}{Coefficients, $C_{Y}$} & Inclination Function & Eccentricity Function \\
    $\omega_{j}$ & $l$, $m$, $p$, $q$ & $\partialDerv{\avgd{U_{j}}}{\meanAnom}$ & $\partialDerv{\avgd{U_{j}}}{\pericenter_{j}}$ & $\partialDerv{\avgd{U_{j}}}{\node_{j}}$ & $\avgd{\dot{E}_{j}}$ & $F^{2}(I_{j})$ & $G^{2}(e)$ \\ \midrule
    \midrule           
    \endhead      
    \midrule
    \multicolumn{8}{c}{table continues} \\ 
    \midrule
    \endfoot      
    \bottomrule
    \endlastfoot  

    $-3n$ & 2, 0, 0, -5 & $-2$ & $\frac{4}{3}$ & $0$ & $\frac{2}{3}$ & 
    \begin{math}
    \begin{aligned}
        & \frac{9 \sin^{4}{\left(I_{j}/2 \right)} \cos^{4}{\left(I_{j}/2 \right)}}{4} 
    \end{aligned}
    \end{math} & 
    \begin{minipage}{80mm}
    \vspace{2mm}    \begin{math}
    \begin{aligned}
        & \frac{6561 e^{10}}{1638400} 
    \end{aligned}
    \end{math}
    \vspace{2mm}
    \end{minipage} \\ \hline

    $-2n$ & 2, 0, 0, -4 & $- \frac{4}{3}$ & $\frac{4}{3}$ & $0$ & $\frac{2}{3}$ & 
    Same as above & 
    \begin{minipage}{80mm}
    \vspace{2mm}    \begin{math}
    \begin{aligned}
        & \frac{7 e^{10}}{2880} + \frac{e^{8}}{576} 
    \end{aligned}
    \end{math}
    \vspace{2mm}
    \end{minipage} \\ \hline

    $-n$ & 2, 0, 0, -3 & $- \frac{2}{3}$ & $\frac{4}{3}$ & $0$ & $\frac{2}{3}$ & 
    Same as above & 
    \begin{minipage}{80mm}
    \vspace{2mm}    \begin{math}
    \begin{aligned}
        & \frac{619 e^{10}}{983040} + \frac{11 e^{8}}{18432} + \frac{e^{6}}{2304} 
    \end{aligned}
    \end{math}
    \vspace{2mm}
    \end{minipage} \\ \hline

    $n$ & 2, 0, 0, -1 & $\frac{2}{3}$ & $\frac{4}{3}$ & $0$ & $\frac{2}{3}$ & 
    Same as above & 
    \begin{minipage}{80mm}
    \vspace{2mm}    \begin{math}
    \begin{aligned}
        & \frac{2639 e^{10}}{491520} + \frac{113 e^{8}}{18432} + \frac{13 e^{6}}{768} - \frac{e^{4}}{16} \\ 
        & + \frac{e^{2}}{4} 
    \end{aligned}
    \end{math}
    \vspace{2mm}
    \end{minipage} \\ \hline

    $2n$ & 2, 0, 0, \;0 & $\frac{4}{3}$ & $\frac{4}{3}$ & $0$ & $\frac{2}{3}$ & 
    Same as above & 
    \begin{minipage}{80mm}
    \vspace{2mm}    \begin{math}
    \begin{aligned}
        & - \frac{3481 e^{10}}{19200} + \frac{2881 e^{8}}{2304} - \frac{155 e^{6}}{36} + \frac{63 e^{4}}{8} \\ 
        & - 5 e^{2} + 1 
    \end{aligned}
    \end{math}
    \vspace{2mm}
    \end{minipage} \\ \hline

    $3n$ & 2, 0, 0, \;1 & $2$ & $\frac{4}{3}$ & $0$ & $\frac{2}{3}$ & 
    Same as above & 
    \begin{minipage}{80mm}
    \vspace{2mm}    \begin{math}
    \begin{aligned}
        & \frac{4654389 e^{10}}{163840} - \frac{132635 e^{8}}{2048} + \frac{21975 e^{6}}{256} \\ 
        & - \frac{861 e^{4}}{16} + \frac{49 e^{2}}{4} 
    \end{aligned}
    \end{math}
    \vspace{2mm}
    \end{minipage} \\ \hline

    $4n$ & 2, 0, 0, \;2 & $\frac{8}{3}$ & $\frac{4}{3}$ & $0$ & $\frac{2}{3}$ & 
    Same as above & 
    \begin{minipage}{80mm}
    \vspace{2mm}    \begin{math}
    \begin{aligned}
        & - \frac{43773 e^{10}}{80} + \frac{83551 e^{8}}{144} - \frac{1955 e^{6}}{6} + \frac{289 e^{4}}{4} 
    \end{aligned}
    \end{math}
    \vspace{2mm}
    \end{minipage} \\ \hline

    $5n$ & 2, 0, 0, \;3 & $\frac{10}{3}$ & $\frac{4}{3}$ & $0$ & $\frac{2}{3}$ & 
    Same as above & 
    \begin{minipage}{80mm}
    \vspace{2mm}    \begin{math}
    \begin{aligned}
        & \frac{587225375 e^{10}}{196608} - \frac{27483625 e^{8}}{18432} + \frac{714025 e^{6}}{2304} 
    \end{aligned}
    \end{math}
    \vspace{2mm}
    \end{minipage} \\ \hline

    $6n$ & 2, 0, 0, \;4 & $4$ & $\frac{4}{3}$ & $0$ & $\frac{2}{3}$ & 
    Same as above & 
    \begin{minipage}{80mm}
    \vspace{2mm}    \begin{math}
    \begin{aligned}
        & - \frac{7369791 e^{10}}{1280} + \frac{284089 e^{8}}{256} 
    \end{aligned}
    \end{math}
    \vspace{2mm}
    \end{minipage} \\ \hline

    $7n$ & 2, 0, 0, \;5 & $\frac{14}{3}$ & $\frac{4}{3}$ & $0$ & $\frac{2}{3}$ & 
    Same as above & 
    \begin{minipage}{80mm}
    \vspace{2mm}    \begin{math}
    \begin{aligned}
        & \frac{52142352409 e^{10}}{14745600} 
    \end{aligned}
    \end{math}
    \vspace{2mm}
    \end{minipage} \\ \hline

    $-5n$ & 2, 0, 1, -5 & $- \frac{10}{3}$ & $0$ & $0$ & $\frac{2}{3}$ & 
    \begin{math}
    \begin{aligned}
        & \frac{\left(3 \sin^{2}{\left(I_{j} \right)} - 2\right)^{2}}{16} 
    \end{aligned}
    \end{math} & 
    \begin{minipage}{80mm}
    \vspace{2mm}    \begin{math}
    \begin{aligned}
        & \frac{3143529 e^{10}}{65536} 
    \end{aligned}
    \end{math}
    \vspace{2mm}
    \end{minipage} \\ \hline

    $-4n$ & 2, 0, 1, -4 & $- \frac{8}{3}$ & $0$ & $0$ & $\frac{2}{3}$ & 
    Same as above & 
    \begin{minipage}{80mm}
    \vspace{2mm}    \begin{math}
    \begin{aligned}
        & \frac{9933 e^{10}}{1280} + \frac{5929 e^{8}}{256} 
    \end{aligned}
    \end{math}
    \vspace{2mm}
    \end{minipage} \\ \hline

    $-3n$ & 2, 0, 1, -3 & $-2$ & $0$ & $0$ & $\frac{2}{3}$ & 
    Same as above & 
    \begin{minipage}{80mm}
    \vspace{2mm}    \begin{math}
    \begin{aligned}
        & \frac{6019881 e^{10}}{327680} + \frac{20829 e^{8}}{2048} + \frac{2809 e^{6}}{256} 
    \end{aligned}
    \end{math}
    \vspace{2mm}
    \end{minipage} \\ \hline

    $-2n$ & 2, 0, 1, -2 & $- \frac{4}{3}$ & $0$ & $0$ & $\frac{2}{3}$ & 
    Same as above & 
    \begin{minipage}{80mm}
    \vspace{2mm}    \begin{math}
    \begin{aligned}
        & \frac{12027 e^{10}}{640} + \frac{1661 e^{8}}{128} + \frac{63 e^{6}}{8} + \frac{81 e^{4}}{16} 
    \end{aligned}
    \end{math}
    \vspace{2mm}
    \end{minipage} \\ \hline

    $-n$ & 2, 0, 1, -1 & $- \frac{2}{3}$ & $0$ & $0$ & $\frac{2}{3}$ & 
    Same as above & 
    \begin{minipage}{80mm}
    \vspace{2mm}    \begin{math}
    \begin{aligned}
        & \frac{3240741 e^{10}}{163840} + \frac{28403 e^{8}}{2048} + \frac{2295 e^{6}}{256} \\ 
        & + \frac{81 e^{4}}{16} + \frac{9 e^{2}}{4} 
    \end{aligned}
    \end{math}
    \vspace{2mm}
    \end{minipage} \\ \hline

    
    $n$ & 2, 0, 1, \;1 & $\frac{2}{3}$ & $0$ & $0$ & $\frac{2}{3}$ & 
    Same as above & 
    \begin{minipage}{80mm}
    \vspace{2mm}    \begin{math}
    \begin{aligned}
        & \frac{3240741 e^{10}}{163840} + \frac{28403 e^{8}}{2048} + \frac{2295 e^{6}}{256} \\ 
        & + \frac{81 e^{4}}{16} + \frac{9 e^{2}}{4} 
    \end{aligned}
    \end{math}
    \vspace{2mm}
    \end{minipage} \\ \hline

    $2n$ & 2, 0, 1, \;2 & $\frac{4}{3}$ & $0$ & $0$ & $\frac{2}{3}$ & 
    Same as above & 
    \begin{minipage}{80mm}
    \vspace{2mm}    \begin{math}
    \begin{aligned}
        & \frac{12027 e^{10}}{640} + \frac{1661 e^{8}}{128} + \frac{63 e^{6}}{8} + \frac{81 e^{4}}{16} 
    \end{aligned}
    \end{math}
    \vspace{2mm}
    \end{minipage} \\ \hline

    $3n$ & 2, 0, 1, \;3 & $2$ & $0$ & $0$ & $\frac{2}{3}$ & 
    Same as above & 
    \begin{minipage}{80mm}
    \vspace{2mm}    \begin{math}
    \begin{aligned}
        & \frac{6019881 e^{10}}{327680} + \frac{20829 e^{8}}{2048} + \frac{2809 e^{6}}{256} 
    \end{aligned}
    \end{math}
    \vspace{2mm}
    \end{minipage} \\ \hline

    $4n$ & 2, 0, 1, \;4 & $\frac{8}{3}$ & $0$ & $0$ & $\frac{2}{3}$ & 
    Same as above & 
    \begin{minipage}{80mm}
    \vspace{2mm}    \begin{math}
    \begin{aligned}
        & \frac{9933 e^{10}}{1280} + \frac{5929 e^{8}}{256} 
    \end{aligned}
    \end{math}
    \vspace{2mm}
    \end{minipage} \\ \hline

    $5n$ & 2, 0, 1, \;5 & $\frac{10}{3}$ & $0$ & $0$ & $\frac{2}{3}$ & 
    Same as above & 
    \begin{minipage}{80mm}
    \vspace{2mm}    \begin{math}
    \begin{aligned}
        & \frac{3143529 e^{10}}{65536} 
    \end{aligned}
    \end{math}
    \vspace{2mm}
    \end{minipage} \\ \hline

    $-7n$ & 2, 0, 2, -5 & $- \frac{14}{3}$ & $- \frac{4}{3}$ & $0$ & $\frac{2}{3}$ & 
    \begin{math}
    \begin{aligned}
        & \frac{9 \sin^{4}{\left(I_{j}/2 \right)} \cos^{4}{\left(I_{j}/2 \right)}}{4} 
    \end{aligned}
    \end{math} & 
    \begin{minipage}{80mm}
    \vspace{2mm}    \begin{math}
    \begin{aligned}
        & \frac{52142352409 e^{10}}{14745600} 
    \end{aligned}
    \end{math}
    \vspace{2mm}
    \end{minipage} \\ \hline

    $-6n$ & 2, 0, 2, -4 & $-4$ & $- \frac{4}{3}$ & $0$ & $\frac{2}{3}$ & 
    Same as above & 
    \begin{minipage}{80mm}
    \vspace{2mm}    \begin{math}
    \begin{aligned}
        & - \frac{7369791 e^{10}}{1280} + \frac{284089 e^{8}}{256} 
    \end{aligned}
    \end{math}
    \vspace{2mm}
    \end{minipage} \\ \hline

    $-5n$ & 2, 0, 2, -3 & $- \frac{10}{3}$ & $- \frac{4}{3}$ & $0$ & $\frac{2}{3}$ & 
    Same as above & 
    \begin{minipage}{80mm}
    \vspace{2mm}    \begin{math}
    \begin{aligned}
        & \frac{587225375 e^{10}}{196608} - \frac{27483625 e^{8}}{18432} + \frac{714025 e^{6}}{2304} 
    \end{aligned}
    \end{math}
    \vspace{2mm}
    \end{minipage} \\ \hline

    $-4n$ & 2, 0, 2, -2 & $- \frac{8}{3}$ & $- \frac{4}{3}$ & $0$ & $\frac{2}{3}$ & 
    Same as above & 
    \begin{minipage}{80mm}
    \vspace{2mm}    \begin{math}
    \begin{aligned}
        & - \frac{43773 e^{10}}{80} + \frac{83551 e^{8}}{144} - \frac{1955 e^{6}}{6} + \frac{289 e^{4}}{4} 
    \end{aligned}
    \end{math}
    \vspace{2mm}
    \end{minipage} \\ \hline

    $-3n$ & 2, 0, 2, -1 & $-2$ & $- \frac{4}{3}$ & $0$ & $\frac{2}{3}$ & 
    Same as above & 
    \begin{minipage}{80mm}
    \vspace{2mm}    \begin{math}
    \begin{aligned}
        & \frac{4654389 e^{10}}{163840} - \frac{132635 e^{8}}{2048} + \frac{21975 e^{6}}{256} \\ 
        & - \frac{861 e^{4}}{16} + \frac{49 e^{2}}{4} 
    \end{aligned}
    \end{math}
    \vspace{2mm}
    \end{minipage} \\ \hline

    $-2n$ & 2, 0, 2, \;0 & $- \frac{4}{3}$ & $- \frac{4}{3}$ & $0$ & $\frac{2}{3}$ & 
    Same as above & 
    \begin{minipage}{80mm}
    \vspace{2mm}    \begin{math}
    \begin{aligned}
        & - \frac{3481 e^{10}}{19200} + \frac{2881 e^{8}}{2304} - \frac{155 e^{6}}{36} + \frac{63 e^{4}}{8} \\ 
        & - 5 e^{2} + 1 
    \end{aligned}
    \end{math}
    \vspace{2mm}
    \end{minipage} \\ \hline

    $-n$ & 2, 0, 2, \;1 & $- \frac{2}{3}$ & $- \frac{4}{3}$ & $0$ & $\frac{2}{3}$ & 
    Same as above & 
    \begin{minipage}{80mm}
    \vspace{2mm}    \begin{math}
    \begin{aligned}
        & \frac{2639 e^{10}}{491520} + \frac{113 e^{8}}{18432} + \frac{13 e^{6}}{768} - \frac{e^{4}}{16} \\ 
        & + \frac{e^{2}}{4} 
    \end{aligned}
    \end{math}
    \vspace{2mm}
    \end{minipage} \\ \hline

    $n$ & 2, 0, 2, \;3 & $\frac{2}{3}$ & $- \frac{4}{3}$ & $0$ & $\frac{2}{3}$ & 
    Same as above & 
    \begin{minipage}{80mm}
    \vspace{2mm}    \begin{math}
    \begin{aligned}
        & \frac{619 e^{10}}{983040} + \frac{11 e^{8}}{18432} + \frac{e^{6}}{2304} 
    \end{aligned}
    \end{math}
    \vspace{2mm}
    \end{minipage} \\ \hline

    $2n$ & 2, 0, 2, \;4 & $\frac{4}{3}$ & $- \frac{4}{3}$ & $0$ & $\frac{2}{3}$ & 
    Same as above & 
    \begin{minipage}{80mm}
    \vspace{2mm}    \begin{math}
    \begin{aligned}
        & \frac{7 e^{10}}{2880} + \frac{e^{8}}{576} 
    \end{aligned}
    \end{math}
    \vspace{2mm}
    \end{minipage} \\ \hline

    $3n$ & 2, 0, 2, \;5 & $2$ & $- \frac{4}{3}$ & $0$ & $\frac{2}{3}$ & 
    Same as above & 
    \begin{minipage}{80mm}
    \vspace{2mm}    \begin{math}
    \begin{aligned}
        & \frac{6561 e^{10}}{1638400} 
    \end{aligned}
    \end{math}
    \vspace{2mm}
    \end{minipage} \\ \hline

    $-\dot{\theta}_{j} - 3n$ & 2, 1, 0, -5 & $- \frac{2}{3}$ & $\frac{4}{9}$ & $\frac{2}{9}$ & $\frac{2}{9}$ & 
    \begin{math}
    \begin{aligned}
        & 9 \sin^{2}{\left(I_{j}/2 \right)} \cos^{6}{\left(I_{j}/2 \right)} 
    \end{aligned}
    \end{math} & 
    \begin{minipage}{80mm}
    \vspace{2mm}    \begin{math}
    \begin{aligned}
        & \frac{6561 e^{10}}{1638400} 
    \end{aligned}
    \end{math}
    \vspace{2mm}
    \end{minipage} \\ \hline

    $-\dot{\theta}_{j} - 2n$ & 2, 1, 0, -4 & $- \frac{4}{9}$ & $\frac{4}{9}$ & $\frac{2}{9}$ & $\frac{2}{9}$ & 
    Same as above & 
    \begin{minipage}{80mm}
    \vspace{2mm}    \begin{math}
    \begin{aligned}
        & \frac{7 e^{10}}{2880} + \frac{e^{8}}{576} 
    \end{aligned}
    \end{math}
    \vspace{2mm}
    \end{minipage} \\ \hline

    $-\dot{\theta}_{j} - n$ & 2, 1, 0, -3 & $- \frac{2}{9}$ & $\frac{4}{9}$ & $\frac{2}{9}$ & $\frac{2}{9}$ & 
    Same as above & 
    \begin{minipage}{80mm}
    \vspace{2mm}    \begin{math}
    \begin{aligned}
        & \frac{619 e^{10}}{983040} + \frac{11 e^{8}}{18432} + \frac{e^{6}}{2304} 
    \end{aligned}
    \end{math}
    \vspace{2mm}
    \end{minipage} \\ \hline

    $-\dot{\theta}_{j} + n$ & 2, 1, 0, -1 & $\frac{2}{9}$ & $\frac{4}{9}$ & $\frac{2}{9}$ & $\frac{2}{9}$ & 
    Same as above & 
    \begin{minipage}{80mm}
    \vspace{2mm}    \begin{math}
    \begin{aligned}
        & \frac{2639 e^{10}}{491520} + \frac{113 e^{8}}{18432} + \frac{13 e^{6}}{768} - \frac{e^{4}}{16} \\ 
        & + \frac{e^{2}}{4} 
    \end{aligned}
    \end{math}
    \vspace{2mm}
    \end{minipage} \\ \hline

    $-\dot{\theta}_{j} + 2n$ & 2, 1, 0, \;0 & $\frac{4}{9}$ & $\frac{4}{9}$ & $\frac{2}{9}$ & $\frac{2}{9}$ & 
    Same as above & 
    \begin{minipage}{80mm}
    \vspace{2mm}    \begin{math}
    \begin{aligned}
        & - \frac{3481 e^{10}}{19200} + \frac{2881 e^{8}}{2304} - \frac{155 e^{6}}{36} + \frac{63 e^{4}}{8} \\ 
        & - 5 e^{2} + 1 
    \end{aligned}
    \end{math}
    \vspace{2mm}
    \end{minipage} \\ \hline

    $-\dot{\theta}_{j} + 3n$ & 2, 1, 0, \;1 & $\frac{2}{3}$ & $\frac{4}{9}$ & $\frac{2}{9}$ & $\frac{2}{9}$ & 
    Same as above & 
    \begin{minipage}{80mm}
    \vspace{2mm}    \begin{math}
    \begin{aligned}
        & \frac{4654389 e^{10}}{163840} - \frac{132635 e^{8}}{2048} + \frac{21975 e^{6}}{256} \\ 
        & - \frac{861 e^{4}}{16} + \frac{49 e^{2}}{4} 
    \end{aligned}
    \end{math}
    \vspace{2mm}
    \end{minipage} \\ \hline

    $-\dot{\theta}_{j} + 4n$ & 2, 1, 0, \;2 & $\frac{8}{9}$ & $\frac{4}{9}$ & $\frac{2}{9}$ & $\frac{2}{9}$ & 
    Same as above & 
    \begin{minipage}{80mm}
    \vspace{2mm}    \begin{math}
    \begin{aligned}
        & - \frac{43773 e^{10}}{80} + \frac{83551 e^{8}}{144} - \frac{1955 e^{6}}{6} + \frac{289 e^{4}}{4} 
    \end{aligned}
    \end{math}
    \vspace{2mm}
    \end{minipage} \\ \hline

    $-\dot{\theta}_{j} + 5n$ & 2, 1, 0, \;3 & $\frac{10}{9}$ & $\frac{4}{9}$ & $\frac{2}{9}$ & $\frac{2}{9}$ & 
    Same as above & 
    \begin{minipage}{80mm}
    \vspace{2mm}    \begin{math}
    \begin{aligned}
        & \frac{587225375 e^{10}}{196608} - \frac{27483625 e^{8}}{18432} + \frac{714025 e^{6}}{2304} 
    \end{aligned}
    \end{math}
    \vspace{2mm}
    \end{minipage} \\ \hline

    $-\dot{\theta}_{j} + 6n$ & 2, 1, 0, \;4 & $\frac{4}{3}$ & $\frac{4}{9}$ & $\frac{2}{9}$ & $\frac{2}{9}$ & 
    Same as above & 
    \begin{minipage}{80mm}
    \vspace{2mm}    \begin{math}
    \begin{aligned}
        & - \frac{7369791 e^{10}}{1280} + \frac{284089 e^{8}}{256} 
    \end{aligned}
    \end{math}
    \vspace{2mm}
    \end{minipage} \\ \hline

    $-\dot{\theta}_{j} + 7n$ & 2, 1, 0, \;5 & $\frac{14}{9}$ & $\frac{4}{9}$ & $\frac{2}{9}$ & $\frac{2}{9}$ & 
    Same as above & 
    \begin{minipage}{80mm}
    \vspace{2mm}    \begin{math}
    \begin{aligned}
        & \frac{52142352409 e^{10}}{14745600} 
    \end{aligned}
    \end{math}
    \vspace{2mm}
    \end{minipage} \\ \hline

    $-\dot{\theta}_{j} - 5n$ & 2, 1, 1, -5 & $- \frac{10}{9}$ & $0$ & $\frac{2}{9}$ & $\frac{2}{9}$ & 
    \begin{math}
    \begin{aligned}
        & \frac{9 \sin^{2}{\left(2 I_{j} \right)}}{16} 
    \end{aligned}
    \end{math} & 
    \begin{minipage}{80mm}
    \vspace{2mm}    \begin{math}
    \begin{aligned}
        & \frac{3143529 e^{10}}{65536} 
    \end{aligned}
    \end{math}
    \vspace{2mm}
    \end{minipage} \\ \hline

    $-\dot{\theta}_{j} - 4n$ & 2, 1, 1, -4 & $- \frac{8}{9}$ & $0$ & $\frac{2}{9}$ & $\frac{2}{9}$ & 
    Same as above & 
    \begin{minipage}{80mm}
    \vspace{2mm}    \begin{math}
    \begin{aligned}
        & \frac{9933 e^{10}}{1280} + \frac{5929 e^{8}}{256} 
    \end{aligned}
    \end{math}
    \vspace{2mm}
    \end{minipage} \\ \hline

    $-\dot{\theta}_{j} - 3n$ & 2, 1, 1, -3 & $- \frac{2}{3}$ & $0$ & $\frac{2}{9}$ & $\frac{2}{9}$ & 
    Same as above & 
    \begin{minipage}{80mm}
    \vspace{2mm}    \begin{math}
    \begin{aligned}
        & \frac{6019881 e^{10}}{327680} + \frac{20829 e^{8}}{2048} + \frac{2809 e^{6}}{256} 
    \end{aligned}
    \end{math}
    \vspace{2mm}
    \end{minipage} \\ \hline

    $-\dot{\theta}_{j} - 2n$ & 2, 1, 1, -2 & $- \frac{4}{9}$ & $0$ & $\frac{2}{9}$ & $\frac{2}{9}$ & 
    Same as above & 
    \begin{minipage}{80mm}
    \vspace{2mm}    \begin{math}
    \begin{aligned}
        & \frac{12027 e^{10}}{640} + \frac{1661 e^{8}}{128} + \frac{63 e^{6}}{8} + \frac{81 e^{4}}{16} 
    \end{aligned}
    \end{math}
    \vspace{2mm}
    \end{minipage} \\ \hline

    $-\dot{\theta}_{j} - n$ & 2, 1, 1, -1 & $- \frac{2}{9}$ & $0$ & $\frac{2}{9}$ & $\frac{2}{9}$ & 
    Same as above & 
    \begin{minipage}{80mm}
    \vspace{2mm}    \begin{math}
    \begin{aligned}
        & \frac{3240741 e^{10}}{163840} + \frac{28403 e^{8}}{2048} + \frac{2295 e^{6}}{256} \\ 
        & + \frac{81 e^{4}}{16} + \frac{9 e^{2}}{4} 
    \end{aligned}
    \end{math}
    \vspace{2mm}
    \end{minipage} \\ \hline

    $-\dot{\theta}_{j}$ & 2, 1, 1, \;0 & $0$ & $0$ & $\frac{2}{9}$ & $\frac{2}{9}$ & 
    Same as above & 
    \begin{minipage}{80mm}
    \vspace{2mm}    \begin{math}
    \begin{aligned}
        & - \frac{1}{\left(e^{2} - 1\right)^{3}} 
    \end{aligned}
    \end{math}
    \vspace{2mm}
    \end{minipage} \\ \hline

    $-\dot{\theta}_{j} + n$ & 2, 1, 1, \;1 & $\frac{2}{9}$ & $0$ & $\frac{2}{9}$ & $\frac{2}{9}$ & 
    Same as above & 
    \begin{minipage}{80mm}
    \vspace{2mm}    \begin{math}
    \begin{aligned}
        & \frac{3240741 e^{10}}{163840} + \frac{28403 e^{8}}{2048} + \frac{2295 e^{6}}{256} \\ 
        & + \frac{81 e^{4}}{16} + \frac{9 e^{2}}{4} 
    \end{aligned}
    \end{math}
    \vspace{2mm}
    \end{minipage} \\ \hline

    $-\dot{\theta}_{j} + 2n$ & 2, 1, 1, \;2 & $\frac{4}{9}$ & $0$ & $\frac{2}{9}$ & $\frac{2}{9}$ & 
    Same as above & 
    \begin{minipage}{80mm}
    \vspace{2mm}    \begin{math}
    \begin{aligned}
        & \frac{12027 e^{10}}{640} + \frac{1661 e^{8}}{128} + \frac{63 e^{6}}{8} + \frac{81 e^{4}}{16} 
    \end{aligned}
    \end{math}
    \vspace{2mm}
    \end{minipage} \\ \hline

    $-\dot{\theta}_{j} + 3n$ & 2, 1, 1, \;3 & $\frac{2}{3}$ & $0$ & $\frac{2}{9}$ & $\frac{2}{9}$ & 
    Same as above & 
    \begin{minipage}{80mm}
    \vspace{2mm}    \begin{math}
    \begin{aligned}
        & \frac{6019881 e^{10}}{327680} + \frac{20829 e^{8}}{2048} + \frac{2809 e^{6}}{256} 
    \end{aligned}
    \end{math}
    \vspace{2mm}
    \end{minipage} \\ \hline

    $-\dot{\theta}_{j} + 4n$ & 2, 1, 1, \;4 & $\frac{8}{9}$ & $0$ & $\frac{2}{9}$ & $\frac{2}{9}$ & 
    Same as above & 
    \begin{minipage}{80mm}
    \vspace{2mm}    \begin{math}
    \begin{aligned}
        & \frac{9933 e^{10}}{1280} + \frac{5929 e^{8}}{256} 
    \end{aligned}
    \end{math}
    \vspace{2mm}
    \end{minipage} \\ \hline

    $-\dot{\theta}_{j} + 5n$ & 2, 1, 1, \;5 & $\frac{10}{9}$ & $0$ & $\frac{2}{9}$ & $\frac{2}{9}$ & 
    Same as above & 
    \begin{minipage}{80mm}
    \vspace{2mm}    \begin{math}
    \begin{aligned}
        & \frac{3143529 e^{10}}{65536} 
    \end{aligned}
    \end{math}
    \vspace{2mm}
    \end{minipage} \\ \hline

    $-\dot{\theta}_{j} - 7n$ & 2, 1, 2, -5 & $- \frac{14}{9}$ & $- \frac{4}{9}$ & $\frac{2}{9}$ & $\frac{2}{9}$ & 
    \begin{math}
    \begin{aligned}
        & 9 \sin^{6}{\left(I_{j}/2 \right)} \cos^{2}{\left(I_{j}/2 \right)} 
    \end{aligned}
    \end{math} & 
    \begin{minipage}{80mm}
    \vspace{2mm}    \begin{math}
    \begin{aligned}
        & \frac{52142352409 e^{10}}{14745600} 
    \end{aligned}
    \end{math}
    \vspace{2mm}
    \end{minipage} \\ \hline

    $-\dot{\theta}_{j} - 6n$ & 2, 1, 2, -4 & $- \frac{4}{3}$ & $- \frac{4}{9}$ & $\frac{2}{9}$ & $\frac{2}{9}$ & 
    Same as above & 
    \begin{minipage}{80mm}
    \vspace{2mm}    \begin{math}
    \begin{aligned}
        & - \frac{7369791 e^{10}}{1280} + \frac{284089 e^{8}}{256} 
    \end{aligned}
    \end{math}
    \vspace{2mm}
    \end{minipage} \\ \hline

    $-\dot{\theta}_{j} - 5n$ & 2, 1, 2, -3 & $- \frac{10}{9}$ & $- \frac{4}{9}$ & $\frac{2}{9}$ & $\frac{2}{9}$ & 
    Same as above & 
    \begin{minipage}{80mm}
    \vspace{2mm}    \begin{math}
    \begin{aligned}
        & \frac{587225375 e^{10}}{196608} - \frac{27483625 e^{8}}{18432} + \frac{714025 e^{6}}{2304} 
    \end{aligned}
    \end{math}
    \vspace{2mm}
    \end{minipage} \\ \hline

    $-\dot{\theta}_{j} - 4n$ & 2, 1, 2, -2 & $- \frac{8}{9}$ & $- \frac{4}{9}$ & $\frac{2}{9}$ & $\frac{2}{9}$ & 
    Same as above & 
    \begin{minipage}{80mm}
    \vspace{2mm}    \begin{math}
    \begin{aligned}
        & - \frac{43773 e^{10}}{80} + \frac{83551 e^{8}}{144} - \frac{1955 e^{6}}{6} + \frac{289 e^{4}}{4} 
    \end{aligned}
    \end{math}
    \vspace{2mm}
    \end{minipage} \\ \hline

    $-\dot{\theta}_{j} - 3n$ & 2, 1, 2, -1 & $- \frac{2}{3}$ & $- \frac{4}{9}$ & $\frac{2}{9}$ & $\frac{2}{9}$ & 
    Same as above & 
    \begin{minipage}{80mm}
    \vspace{2mm}    \begin{math}
    \begin{aligned}
        & \frac{4654389 e^{10}}{163840} - \frac{132635 e^{8}}{2048} + \frac{21975 e^{6}}{256} \\ 
        & - \frac{861 e^{4}}{16} + \frac{49 e^{2}}{4} 
    \end{aligned}
    \end{math}
    \vspace{2mm}
    \end{minipage} \\ \hline

    $-\dot{\theta}_{j} - 2n$ & 2, 1, 2, \;0 & $- \frac{4}{9}$ & $- \frac{4}{9}$ & $\frac{2}{9}$ & $\frac{2}{9}$ & 
    Same as above & 
    \begin{minipage}{80mm}
    \vspace{2mm}    \begin{math}
    \begin{aligned}
        & - \frac{3481 e^{10}}{19200} + \frac{2881 e^{8}}{2304} - \frac{155 e^{6}}{36} + \frac{63 e^{4}}{8} \\ 
        & - 5 e^{2} + 1 
    \end{aligned}
    \end{math}
    \vspace{2mm}
    \end{minipage} \\ \hline

    $-\dot{\theta}_{j} - n$ & 2, 1, 2, \;1 & $- \frac{2}{9}$ & $- \frac{4}{9}$ & $\frac{2}{9}$ & $\frac{2}{9}$ & 
    Same as above & 
    \begin{minipage}{80mm}
    \vspace{2mm}    \begin{math}
    \begin{aligned}
        & \frac{2639 e^{10}}{491520} + \frac{113 e^{8}}{18432} + \frac{13 e^{6}}{768} - \frac{e^{4}}{16} \\ 
        & + \frac{e^{2}}{4} 
    \end{aligned}
    \end{math}
    \vspace{2mm}
    \end{minipage} \\ \hline

    $-\dot{\theta}_{j} + n$ & 2, 1, 2, \;3 & $\frac{2}{9}$ & $- \frac{4}{9}$ & $\frac{2}{9}$ & $\frac{2}{9}$ & 
    Same as above & 
    \begin{minipage}{80mm}
    \vspace{2mm}    \begin{math}
    \begin{aligned}
        & \frac{619 e^{10}}{983040} + \frac{11 e^{8}}{18432} + \frac{e^{6}}{2304} 
    \end{aligned}
    \end{math}
    \vspace{2mm}
    \end{minipage} \\ \hline

    $-\dot{\theta}_{j} + 2n$ & 2, 1, 2, \;4 & $\frac{4}{9}$ & $- \frac{4}{9}$ & $\frac{2}{9}$ & $\frac{2}{9}$ & 
    Same as above & 
    \begin{minipage}{80mm}
    \vspace{2mm}    \begin{math}
    \begin{aligned}
        & \frac{7 e^{10}}{2880} + \frac{e^{8}}{576} 
    \end{aligned}
    \end{math}
    \vspace{2mm}
    \end{minipage} \\ \hline

    $-\dot{\theta}_{j} + 3n$ & 2, 1, 2, \;5 & $\frac{2}{3}$ & $- \frac{4}{9}$ & $\frac{2}{9}$ & $\frac{2}{9}$ & 
    Same as above & 
    \begin{minipage}{80mm}
    \vspace{2mm}    \begin{math}
    \begin{aligned}
        & \frac{6561 e^{10}}{1638400} 
    \end{aligned}
    \end{math}
    \vspace{2mm}
    \end{minipage} \\ \hline

    $-2\dot{\theta}_{j} - 3n$ & 2, 2, 0, -5 & $- \frac{1}{6}$ & $\frac{1}{9}$ & $\frac{1}{9}$ & $\frac{1}{18}$ & 
    \begin{math}
    \begin{aligned}
        & 9 \cos^{8}{\left(I_{j}/2 \right)} 
    \end{aligned}
    \end{math} & 
    \begin{minipage}{80mm}
    \vspace{2mm}    \begin{math}
    \begin{aligned}
        & \frac{6561 e^{10}}{1638400} 
    \end{aligned}
    \end{math}
    \vspace{2mm}
    \end{minipage} \\ \hline

    $-2\dot{\theta}_{j} - 2n$ & 2, 2, 0, -4 & $- \frac{1}{9}$ & $\frac{1}{9}$ & $\frac{1}{9}$ & $\frac{1}{18}$ & 
    Same as above & 
    \begin{minipage}{80mm}
    \vspace{2mm}    \begin{math}
    \begin{aligned}
        & \frac{7 e^{10}}{2880} + \frac{e^{8}}{576} 
    \end{aligned}
    \end{math}
    \vspace{2mm}
    \end{minipage} \\ \hline

    $-2\dot{\theta}_{j} - n$ & 2, 2, 0, -3 & $- \frac{1}{18}$ & $\frac{1}{9}$ & $\frac{1}{9}$ & $\frac{1}{18}$ & 
    Same as above & 
    \begin{minipage}{80mm}
    \vspace{2mm}    \begin{math}
    \begin{aligned}
        & \frac{619 e^{10}}{983040} + \frac{11 e^{8}}{18432} + \frac{e^{6}}{2304} 
    \end{aligned}
    \end{math}
    \vspace{2mm}
    \end{minipage} \\ \hline

    $-2\dot{\theta}_{j} + n$ & 2, 2, 0, -1 & $\frac{1}{18}$ & $\frac{1}{9}$ & $\frac{1}{9}$ & $\frac{1}{18}$ & 
    Same as above & 
    \begin{minipage}{80mm}
    \vspace{2mm}    \begin{math}
    \begin{aligned}
        & \frac{2639 e^{10}}{491520} + \frac{113 e^{8}}{18432} + \frac{13 e^{6}}{768} - \frac{e^{4}}{16} \\ 
        & + \frac{e^{2}}{4} 
    \end{aligned}
    \end{math}
    \vspace{2mm}
    \end{minipage} \\ \hline

    $-2\dot{\theta}_{j} + 2n$ & 2, 2, 0, \;0 & $\frac{1}{9}$ & $\frac{1}{9}$ & $\frac{1}{9}$ & $\frac{1}{18}$ & 
    Same as above & 
    \begin{minipage}{80mm}
    \vspace{2mm}    \begin{math}
    \begin{aligned}
        & - \frac{3481 e^{10}}{19200} + \frac{2881 e^{8}}{2304} - \frac{155 e^{6}}{36} + \frac{63 e^{4}}{8} \\ 
        & - 5 e^{2} + 1 
    \end{aligned}
    \end{math}
    \vspace{2mm}
    \end{minipage} \\ \hline

    $-2\dot{\theta}_{j} + 3n$ & 2, 2, 0, \;1 & $\frac{1}{6}$ & $\frac{1}{9}$ & $\frac{1}{9}$ & $\frac{1}{18}$ & 
    Same as above & 
    \begin{minipage}{80mm}
    \vspace{2mm}    \begin{math}
    \begin{aligned}
        & \frac{4654389 e^{10}}{163840} - \frac{132635 e^{8}}{2048} + \frac{21975 e^{6}}{256} \\ 
        & - \frac{861 e^{4}}{16} + \frac{49 e^{2}}{4} 
    \end{aligned}
    \end{math}
    \vspace{2mm}
    \end{minipage} \\ \hline

    $-2\dot{\theta}_{j} + 4n$ & 2, 2, 0, \;2 & $\frac{2}{9}$ & $\frac{1}{9}$ & $\frac{1}{9}$ & $\frac{1}{18}$ & 
    Same as above & 
    \begin{minipage}{80mm}
    \vspace{2mm}    \begin{math}
    \begin{aligned}
        & - \frac{43773 e^{10}}{80} + \frac{83551 e^{8}}{144} - \frac{1955 e^{6}}{6} + \frac{289 e^{4}}{4} 
    \end{aligned}
    \end{math}
    \vspace{2mm}
    \end{minipage} \\ \hline

    $-2\dot{\theta}_{j} + 5n$ & 2, 2, 0, \;3 & $\frac{5}{18}$ & $\frac{1}{9}$ & $\frac{1}{9}$ & $\frac{1}{18}$ & 
    Same as above & 
    \begin{minipage}{80mm}
    \vspace{2mm}    \begin{math}
    \begin{aligned}
        & \frac{587225375 e^{10}}{196608} - \frac{27483625 e^{8}}{18432} + \frac{714025 e^{6}}{2304} 
    \end{aligned}
    \end{math}
    \vspace{2mm}
    \end{minipage} \\ \hline

    $-2\dot{\theta}_{j} + 6n$ & 2, 2, 0, \;4 & $\frac{1}{3}$ & $\frac{1}{9}$ & $\frac{1}{9}$ & $\frac{1}{18}$ & 
    Same as above & 
    \begin{minipage}{80mm}
    \vspace{2mm}    \begin{math}
    \begin{aligned}
        & - \frac{7369791 e^{10}}{1280} + \frac{284089 e^{8}}{256} 
    \end{aligned}
    \end{math}
    \vspace{2mm}
    \end{minipage} \\ \hline

    $-2\dot{\theta}_{j} + 7n$ & 2, 2, 0, \;5 & $\frac{7}{18}$ & $\frac{1}{9}$ & $\frac{1}{9}$ & $\frac{1}{18}$ & 
    Same as above & 
    \begin{minipage}{80mm}
    \vspace{2mm}    \begin{math}
    \begin{aligned}
        & \frac{52142352409 e^{10}}{14745600} 
    \end{aligned}
    \end{math}
    \vspace{2mm}
    \end{minipage} \\ \hline

    $-2\dot{\theta}_{j} - 5n$ & 2, 2, 1, -5 & $- \frac{5}{18}$ & $0$ & $\frac{1}{9}$ & $\frac{1}{18}$ & 
    \begin{math}
    \begin{aligned}
        & 36 \sin^{4}{\left(I_{j}/2 \right)} \cos^{4}{\left(I_{j}/2 \right)} 
    \end{aligned}
    \end{math} & 
    \begin{minipage}{80mm}
    \vspace{2mm}    \begin{math}
    \begin{aligned}
        & \frac{3143529 e^{10}}{65536} 
    \end{aligned}
    \end{math}
    \vspace{2mm}
    \end{minipage} \\ \hline

    $-2\dot{\theta}_{j} - 4n$ & 2, 2, 1, -4 & $- \frac{2}{9}$ & $0$ & $\frac{1}{9}$ & $\frac{1}{18}$ & 
    Same as above & 
    \begin{minipage}{80mm}
    \vspace{2mm}    \begin{math}
    \begin{aligned}
        & \frac{9933 e^{10}}{1280} + \frac{5929 e^{8}}{256} 
    \end{aligned}
    \end{math}
    \vspace{2mm}
    \end{minipage} \\ \hline

    $-2\dot{\theta}_{j} - 3n$ & 2, 2, 1, -3 & $- \frac{1}{6}$ & $0$ & $\frac{1}{9}$ & $\frac{1}{18}$ & 
    Same as above & 
    \begin{minipage}{80mm}
    \vspace{2mm}    \begin{math}
    \begin{aligned}
        & \frac{6019881 e^{10}}{327680} + \frac{20829 e^{8}}{2048} + \frac{2809 e^{6}}{256} 
    \end{aligned}
    \end{math}
    \vspace{2mm}
    \end{minipage} \\ \hline

    $-2\dot{\theta}_{j} - 2n$ & 2, 2, 1, -2 & $- \frac{1}{9}$ & $0$ & $\frac{1}{9}$ & $\frac{1}{18}$ & 
    Same as above & 
    \begin{minipage}{80mm}
    \vspace{2mm}    \begin{math}
    \begin{aligned}
        & \frac{12027 e^{10}}{640} + \frac{1661 e^{8}}{128} + \frac{63 e^{6}}{8} + \frac{81 e^{4}}{16} 
    \end{aligned}
    \end{math}
    \vspace{2mm}
    \end{minipage} \\ \hline

    $-2\dot{\theta}_{j} - n$ & 2, 2, 1, -1 & $- \frac{1}{18}$ & $0$ & $\frac{1}{9}$ & $\frac{1}{18}$ & 
    Same as above & 
    \begin{minipage}{80mm}
    \vspace{2mm}    \begin{math}
    \begin{aligned}
        & \frac{3240741 e^{10}}{163840} + \frac{28403 e^{8}}{2048} + \frac{2295 e^{6}}{256} \\ 
        & + \frac{81 e^{4}}{16} + \frac{9 e^{2}}{4} 
    \end{aligned}
    \end{math}
    \vspace{2mm}
    \end{minipage} \\ \hline

    $-2\dot{\theta}_{j}$ & 2, 2, 1, \;0 & $0$ & $0$ & $\frac{1}{9}$ & $\frac{1}{18}$ & 
    Same as above & 
    \begin{minipage}{80mm}
    \vspace{2mm}    \begin{math}
    \begin{aligned}
        & - \frac{1}{\left(e^{2} - 1\right)^{3}} 
    \end{aligned}
    \end{math}
    \vspace{2mm}
    \end{minipage} \\ \hline

    $-2\dot{\theta}_{j} + n$ & 2, 2, 1, \;1 & $\frac{1}{18}$ & $0$ & $\frac{1}{9}$ & $\frac{1}{18}$ & 
    Same as above & 
    \begin{minipage}{80mm}
    \vspace{2mm}    \begin{math}
    \begin{aligned}
        & \frac{3240741 e^{10}}{163840} + \frac{28403 e^{8}}{2048} + \frac{2295 e^{6}}{256} \\ 
        & + \frac{81 e^{4}}{16} + \frac{9 e^{2}}{4} 
    \end{aligned}
    \end{math}
    \vspace{2mm}
    \end{minipage} \\ \hline

    $-2\dot{\theta}_{j} + 2n$ & 2, 2, 1, \;2 & $\frac{1}{9}$ & $0$ & $\frac{1}{9}$ & $\frac{1}{18}$ & 
    Same as above & 
    \begin{minipage}{80mm}
    \vspace{2mm}    \begin{math}
    \begin{aligned}
        & \frac{12027 e^{10}}{640} + \frac{1661 e^{8}}{128} + \frac{63 e^{6}}{8} + \frac{81 e^{4}}{16} 
    \end{aligned}
    \end{math}
    \vspace{2mm}
    \end{minipage} \\ \hline

    $-2\dot{\theta}_{j} + 3n$ & 2, 2, 1, \;3 & $\frac{1}{6}$ & $0$ & $\frac{1}{9}$ & $\frac{1}{18}$ & 
    Same as above & 
    \begin{minipage}{80mm}
    \vspace{2mm}    \begin{math}
    \begin{aligned}
        & \frac{6019881 e^{10}}{327680} + \frac{20829 e^{8}}{2048} + \frac{2809 e^{6}}{256} 
    \end{aligned}
    \end{math}
    \vspace{2mm}
    \end{minipage} \\ \hline

    $-2\dot{\theta}_{j} + 4n$ & 2, 2, 1, \;4 & $\frac{2}{9}$ & $0$ & $\frac{1}{9}$ & $\frac{1}{18}$ & 
    Same as above & 
    \begin{minipage}{80mm}
    \vspace{2mm}    \begin{math}
    \begin{aligned}
        & \frac{9933 e^{10}}{1280} + \frac{5929 e^{8}}{256} 
    \end{aligned}
    \end{math}
    \vspace{2mm}
    \end{minipage} \\ \hline

    $-2\dot{\theta}_{j} + 5n$ & 2, 2, 1, \;5 & $\frac{5}{18}$ & $0$ & $\frac{1}{9}$ & $\frac{1}{18}$ & 
    Same as above & 
    \begin{minipage}{80mm}
    \vspace{2mm}    \begin{math}
    \begin{aligned}
        & \frac{3143529 e^{10}}{65536} 
    \end{aligned}
    \end{math}
    \vspace{2mm}
    \end{minipage} \\ \hline

    $-2\dot{\theta}_{j} - 7n$ & 2, 2, 2, -5 & $- \frac{7}{18}$ & $- \frac{1}{9}$ & $\frac{1}{9}$ & $\frac{1}{18}$ & 
    \begin{math}
    \begin{aligned}
        & 9 \sin^{8}{\left(I_{j}/2 \right)} 
    \end{aligned}
    \end{math} & 
    \begin{minipage}{80mm}
    \vspace{2mm}    \begin{math}
    \begin{aligned}
        & \frac{52142352409 e^{10}}{14745600} 
    \end{aligned}
    \end{math}
    \vspace{2mm}
    \end{minipage} \\ \hline

    $-2\dot{\theta}_{j} - 6n$ & 2, 2, 2, -4 & $- \frac{1}{3}$ & $- \frac{1}{9}$ & $\frac{1}{9}$ & $\frac{1}{18}$ & 
    Same as above & 
    \begin{minipage}{80mm}
    \vspace{2mm}    \begin{math}
    \begin{aligned}
        & - \frac{7369791 e^{10}}{1280} + \frac{284089 e^{8}}{256} 
    \end{aligned}
    \end{math}
    \vspace{2mm}
    \end{minipage} \\ \hline

    $-2\dot{\theta}_{j} - 5n$ & 2, 2, 2, -3 & $- \frac{5}{18}$ & $- \frac{1}{9}$ & $\frac{1}{9}$ & $\frac{1}{18}$ & 
    Same as above & 
    \begin{minipage}{80mm}
    \vspace{2mm}    \begin{math}
    \begin{aligned}
        & \frac{587225375 e^{10}}{196608} - \frac{27483625 e^{8}}{18432} + \frac{714025 e^{6}}{2304} 
    \end{aligned}
    \end{math}
    \vspace{2mm}
    \end{minipage} \\ \hline

    $-2\dot{\theta}_{j} - 4n$ & 2, 2, 2, -2 & $- \frac{2}{9}$ & $- \frac{1}{9}$ & $\frac{1}{9}$ & $\frac{1}{18}$ & 
    Same as above & 
    \begin{minipage}{80mm}
    \vspace{2mm}    \begin{math}
    \begin{aligned}
        & - \frac{43773 e^{10}}{80} + \frac{83551 e^{8}}{144} - \frac{1955 e^{6}}{6} + \frac{289 e^{4}}{4} 
    \end{aligned}
    \end{math}
    \vspace{2mm}
    \end{minipage} \\ \hline

    $-2\dot{\theta}_{j} - 3n$ & 2, 2, 2, -1 & $- \frac{1}{6}$ & $- \frac{1}{9}$ & $\frac{1}{9}$ & $\frac{1}{18}$ & 
    Same as above & 
    \begin{minipage}{80mm}
    \vspace{2mm}    \begin{math}
    \begin{aligned}
        & \frac{4654389 e^{10}}{163840} - \frac{132635 e^{8}}{2048} + \frac{21975 e^{6}}{256} \\ 
        & - \frac{861 e^{4}}{16} + \frac{49 e^{2}}{4} 
    \end{aligned}
    \end{math}
    \vspace{2mm}
    \end{minipage} \\ \hline

    $-2\dot{\theta}_{j} - 2n$ & 2, 2, 2, \;0 & $- \frac{1}{9}$ & $- \frac{1}{9}$ & $\frac{1}{9}$ & $\frac{1}{18}$ & 
    Same as above & 
    \begin{minipage}{80mm}
    \vspace{2mm}    \begin{math}
    \begin{aligned}
        & - \frac{3481 e^{10}}{19200} + \frac{2881 e^{8}}{2304} - \frac{155 e^{6}}{36} + \frac{63 e^{4}}{8} \\ 
        & - 5 e^{2} + 1 
    \end{aligned}
    \end{math}
    \vspace{2mm}
    \end{minipage} \\ \hline

    $-2\dot{\theta}_{j} - n$ & 2, 2, 2, \;1 & $- \frac{1}{18}$ & $- \frac{1}{9}$ & $\frac{1}{9}$ & $\frac{1}{18}$ & 
    Same as above & 
    \begin{minipage}{80mm}
    \vspace{2mm}    \begin{math}
    \begin{aligned}
        & \frac{2639 e^{10}}{491520} + \frac{113 e^{8}}{18432} + \frac{13 e^{6}}{768} - \frac{e^{4}}{16} \\ 
        & + \frac{e^{2}}{4} 
    \end{aligned}
    \end{math}
    \vspace{2mm}
    \end{minipage} \\ \hline

    $-2\dot{\theta}_{j} + n$ & 2, 2, 2, \;3 & $\frac{1}{18}$ & $- \frac{1}{9}$ & $\frac{1}{9}$ & $\frac{1}{18}$ & 
    Same as above & 
    \begin{minipage}{80mm}
    \vspace{2mm}    \begin{math}
    \begin{aligned}
        & \frac{619 e^{10}}{983040} + \frac{11 e^{8}}{18432} + \frac{e^{6}}{2304} 
    \end{aligned}
    \end{math}
    \vspace{2mm}
    \end{minipage} \\ \hline

    $-2\dot{\theta}_{j} + 2n$ & 2, 2, 2, \;4 & $\frac{1}{9}$ & $- \frac{1}{9}$ & $\frac{1}{9}$ & $\frac{1}{18}$ & 
    Same as above & 
    \begin{minipage}{80mm}
    \vspace{2mm}    \begin{math}
    \begin{aligned}
        & \frac{7 e^{10}}{2880} + \frac{e^{8}}{576} 
    \end{aligned}
    \end{math}
    \vspace{2mm}
    \end{minipage} \\ \hline

    $-2\dot{\theta}_{j} + 3n$ & 2, 2, 2, \;5 & $\frac{1}{6}$ & $- \frac{1}{9}$ & $\frac{1}{9}$ & $\frac{1}{18}$ & 
    Same as above & 
    \begin{minipage}{80mm}
    \vspace{2mm}    \begin{math}
    \begin{aligned}
        & \frac{6561 e^{10}}{1638400} 
    \end{aligned}
    \end{math}
    \vspace{2mm}
    \end{minipage} \\ \hline
\end{longtable}
\end{longrotatetable}
\normalsize

\section{Spin Synchronous Dissipation at Zero Inclination}\label{sec:app:sync}

Table \ref{tab:dissipation_order10} makes no assumption about the object of interest's rotation rate ($\dot{\theta}$ may or may not be equal to $n$ or a rational multiple of $n$). If, however, the object has reached its spin-synchronous state then the number of active modes will dramatically decrease (many will be equal to one another or zero). A further simplification can be made if the object is assumed to have zero obliquity ($I_{j} = 0$). This allows for much simpler (and computationally cheaper) calculations of tidal heating and the potential derivatives when compared to using Eqs. \ref{eq:app:potential_derivative_eq} and \ref{eq:app:heating_eq}. We provide these reduced formulae here for completeness\footnote{The same assumptions discussed in Section \ref{sec:method:orbital} apply to these equations as well.}. \par

For the following equations, we first define the \textit{tidal susceptibility} as,
\begin{equation}\label{eq:app:sync:tidal_suscept}
    T_{j,\;k} = \frac{3}{2}\frac{G M_{k}^{2} R_{j}^{5}}{a^{6}}.
\end{equation}

Tidal heating is then,

\begin{equation}\label{eq:app:sync:heating}
\begin{aligned}
    \avgd{\dot{E}_{j}} = T_{j,\;k} n \Bigg[& \left(\frac{2555911 e^{10}}{122880} - \frac{63949 e^{8}}{2304} + \frac{551 e^{6}}{12} - \frac{101 e^{4}}{4} + 7 e^{2}\right) \text{K}_{j}{\left(\left|{n}\right| \right)} \\
    & + \left(- \frac{171083 e^{10}}{320} + \frac{339187 e^{8}}{576} - \frac{3847 e^{6}}{12} + \frac{605 e^{4}}{8}\right) \text{K}_{j}{\left(2 \left|{n}\right| \right)} \\
    & + \left(\frac{368520907 e^{10}}{81920} - \frac{1709915 e^{8}}{768} + \frac{2855 e^{6}}{6}\right) \text{K}_{j}{\left(3 \left|{n}\right| \right)} \\
    &  + \left(- \frac{66268493 e^{10}}{5760} + \frac{2592379 e^{8}}{1152}\right) \text{K}_{j}{\left(4 \left|{n}\right| \right)} \\
    & + \frac{6576742601 e^{10}}{737280}\text{K}_{j}{\left(5 \left|{n}\right| \right)} \Bigg].
\end{aligned}
\end{equation}

The derivative of the tidal potential with respect to the mean anomaly is, 
\begin{equation}\label{eq:app:sync:dUdM}
\begin{aligned}
    \partialDerv{\avgd{U_{j}}}{\meanAnom} = \frac{T_{j,\;k}}{M_{k}} \text{Sgn}(n) \Bigg[&  \left(\frac{6046043 e^{10}}{122880} - \frac{426355 e^{8}}{4608} + \frac{12647 e^{6}}{96} - 79 e^{4} + 19 e^{2}\right) \text{K}_{j}{\left(\left|{n}\right| \right)} \\
    & + \left(- \frac{69235 e^{10}}{64} + \frac{673391 e^{8}}{576} - \frac{7757 e^{6}}{12} + \frac{1183 e^{4}}{8}\right) \text{K}_{j}{\left(2 \left|{n}\right| \right)} \\
    & + \left(\frac{122639619 e^{10}}{16384} - \frac{5710133 e^{8}}{1536} + \frac{75431 e^{6}}{96}\right) \text{K}_{j}{\left(3 \left|{n}\right| \right)} \\
    & + \left(- \frac{198865133 e^{10}}{11520} + \frac{7741555 e^{8}}{2304}\right) \text{K}_{j}{\left(4 \left|{n}\right| \right)} \\
    & + \frac{9183857269 e^{10}}{737280}\text{K}_{j}{\left(5 \left|{n}\right| \right)} \Bigg].
\end{aligned}
\end{equation}

For an object with no obliquity the derivative of the tidal potential with respect to its orbital node is equal to the derivative with respect to the pericenter:
\begin{equation}\label{eq:app:sync:dUdw}
\begin{aligned}
    \partialDerv{\avgd{U_{j}}}{\Omega_{j}} = \partialDerv{\avgd{U_{j}}}{\pericenter} = \frac{T_{j,\;k}}{M_{k}} \text{Sgn}(n) \Bigg[& \left(\frac{872533 e^{10}}{30720} - \frac{298457 e^{8}}{4608} + \frac{8239 e^{6}}{96} - \frac{215 e^{4}}{4} + 12 e^{2}\right) \text{K}_{j}{\left(\left|{n}\right| \right)} \\
    & + \left(- \frac{43773 e^{10}}{80} + \frac{83551 e^{8}}{144} - \frac{1955 e^{6}}{6} + \frac{289 e^{4}}{4}\right) \text{K}_{j}{\left(2 \left|{n}\right| \right)} \\
    & + \left(\frac{61169297 e^{10}}{20480} - \frac{2290303 e^{8}}{1536} + \frac{9917 e^{6}}{32}\right) \text{K}_{j}{\left(3 \left|{n}\right| \right)} \\
    & + \left(- \frac{66328147 e^{10}}{11520} + \frac{2556797 e^{8}}{2304}\right) \text{K}_{j}{\left(4 \left|{n}\right| \right)} \\
    & + \frac{651778667 e^{10}}{184320}\text{K}_{j}{\left(5 \left|{n}\right| \right)} \Bigg].
\end{aligned}
\end{equation}

An important feature of the dissipation equations is that, by including higher orders of eccentricity, even though here we are only considering the spin-synchronous case, we are still left with multiple tidal modes which act as inputs to the rheological model (all are integer multiples of the absolute value of the orbital motion). It is also important to note that the sign of the orbital motion\footnote{The sign of the orbital motion is positive for prograde orbits and negative for retrograde ones. Here ``prograde'' requires some reference direction which we always choose to be the host's spin vector.} (designated by $\text{Sgn}(n)$) is present for the tidal potential derivatives. Therefore the orbital evolution formulae depend upon the orbital \textit{direction} while the interior heating does not.

\newpage\clearpage
\section{Eccentricity and Inclination Functions}\label{sec:app:eccen_inclin}

The eccentricity and inclination functions used in Eqs. \ref{eq:dissipation} are presented below for reference. Please refer to Chapter 6 of \citet{MD} for an introductory discussion of these functions and how they arise in the tidal potential equation. However, the definitions of the functions in that text lack some of the nuances that are discussed below. \par

Throughout this section, $l$, $m$, $p$, $q$ refer to the Fourier summation indices used in the Darwin-Kaula derivation of the tidal potential. Eccentricity and obliquity\footnote{A slight misnomer exists with respect to the naming of the ``inclination'' functions. These functions have a long tradition, which we continue in this manuscript, of being named for inclination when, in the context of tides, they should actually be used in conjunction with the relative \textit{obliquity} of the object in question. An inclined orbit \textit{will} change the relative obliquity of the \textit{host} object. For example, Triton's large inclination will affect \textit{Neptune's} obliquity tides, not Triton's. Inclination can, however, influence the satellite if the host has a significant oblateness, which we do not consider in this work.} are denoted by, respectively, $e$ and $I$. For a dual dissipating system, both objects share the same eccentricity, but may have different obliquities ($I_{h}$, $I_{s}$) relative to their mutual orbital plane.

\subsection{Eccentricity Functions}
The eccentricity functions are related to the \textit{Hansen coefficients} \citep{Kaula1964},
\begin{equation}\label{eq:app:eccen_hansen}
    G_{lpq}(e) = H_{l-2p+q}^{-l-1, \; l-2p} (e).
\end{equation}

The Hansen coefficients $H_{k}^{n,\;m}$ can be calculated for two different regimes (note the integer $n$ should not be confused with the orbital mean motion discussed elsewhere in this article). Exact solutions exist for $k=0$, otherwise we must rely on truncating powers of eccentricity \citep{Hughes1981}. \par

For $k=0$, the Hansen coefficients have the properties $H_{0}^{n,\;m} = H_{0}^{n,\;-m}$ and that $H_{0}^{n,\;m} = 0$ for $m > |n|$ \citep{Hughes1981}. Thus, unique solutions can be found by restricting $m$ such that $0 \leq m \leq |n|$. Two additional sub-regimes emerge based on the value of $n$ \citep[see Appendix A in][]{LaskarBoue2010}. For $n = -1$ and $m \in \{0, 1\}$,

\begin{subequations}\label{eq:app:hansen:k0_nneg_props}
    \begin{align}
        H_{0}^{-1, 0} &= 1, \label{eq:app:hansen:k0_nneg_props:propA} \\ 
        H_{0}^{-1, 1} &= \frac{\sqrt{1 - e^{2}} - 1}{e}. \label{eq:app:hansen:k0_nneg_props:propB}
    \end{align}
\end{subequations}

\noindent For all other $n < -1$ and associated $m$, the coefficients can be calculated using the terminating summation,

\begin{equation}\label{eq:app:hansen:k0_nneg_eq}
    H_{0}^{-n', m} = \frac{1}{\left(1 - e^{2}\right)^{n' - 3/2}} \sum_{j=0}^{\lfloor (n'-2-m)/2 \rfloor}\frac{(n'-2)!}{j!(m + j)!(n' - 2 - m - 2j)!}\left(\frac{e}{2}\right)^{m + 2j},
\end{equation}

\noindent where we replaced $n$ (which is still assumed to be \replaced{$< -1$}{less than -1} at this stage) with the strictly positive $n'$ defined as, $n' = -n$. \par

For the tidal dissipation calculations used in this work, $n = -l -1$. Therefore, because $l \geq 2$, $n$ will always be less than or equal to -3. However, for completeness, we present the work of \citet{LaskarBoue2010} who showed that for $n \geq 0$ (and $0 \leq m \leq n$),

\begin{equation}\label{eq:app:hansen:k0_npos}
    H_{0}^{n,\;m} = \left(-1\right)^{m} \frac{(1 + n + m)!}{(1 + n)!} \sum_{j=0}^{\lfloor (1 + n - m)/2 \rfloor}\frac{(1 + n - m)!}{j!(m + j)!(1 + n - m - 2j)!}\left(\frac{e}{2}\right)^{m + 2j}.
\end{equation}

For non-zero values of $k$ and all $n \in \mathbb{Z}$ the Hansen coefficients can be found by \citep{Veras2019},

\begin{equation}\label{eq:app:hansen:knon0}
    H_{k}^{n,\;m} = \left(1 + z^{2}\right)^{-(n+1)}\sum_{j=0}^{\infty}(-z)^{j}\sum_{h=0}^{j}\binom{1 + n + m}{j-h}\binom{1 + n - m}{h}J_{k-m+j-2h}(ke),
\end{equation}

\noindent where $z = (1 - \sqrt{1-e^{2}}) / e$ and $J_{z}(x)$ represents the Bessel function of the first kind defined as \citep[e.g.,][]{Giacaglia1987},

\begin{equation}\label{eq:app:hansen:bessel}
    J_{z}(x) = \left(\sqrt{-1}\right)^{|z| - z}\sum_{j\geq0}^{\infty}\frac{(-1)^{j}}{j!(|z|+j)!}\left(\frac{x}{2}\right)^{2j + |z|}.
\end{equation}

In this regime ($k \neq 0$), and only this regime, the Hansen coefficients may also be approximated via the Newcomb operators as presented in \citet[][see Eqs. 6.39--6.42, on page 232.]{MD}. \citet{Cherniack1972} found the Newcomb operators to be a computationally efficient way to estimate the Hansen coefficients. However, in this work, we pre-calculate the eccentricity function coefficients; therefore, mathematical accuracy outweighs computational efficiency and we choose to forego the Newcomb operators in favor of Eq. \ref{eq:app:hansen:knon0}. \par

In order to perform tidal calculations, the infinite summations in Eqs. \ref{eq:app:hansen:knon0} and \ref{eq:app:hansen:bessel} require us to truncate the equations to a predetermined power of $e$. The implications of different truncation levels are discussed in Section \ref{sec:result:trunc}.

\subsection{Inclination Functions}
The inclination functions (used in Eq. \ref{eq:dissipation}) as written down by \citet{Kaula1961} required the inefficient calculation of a triple summation. \citet{Allan1965} re-derived the functions in a simpler form which we use here with the exception that the author included an erroneous factor of $\left(\sqrt{-1}\right)^{l-m}$ (this was carried into the 2000 edition of \citet{MD}) that was corrected in later revisions \citep{GoodingWagner2008, Veras2019}. The final, corrected, definition is given by, 

\begin{align}\label{eq:app:inclin_func}
    \begin{aligned}
    F_{lmp}(I) = \;& \frac{(l+m)!}{2^{l}p!(l-p)!}\sum_{\lambda=\sigma_{lmp}}^{\upsilon_{lmp}}(-1)^{\lambda} \\
    &\times \binom{2l-2p}{\lambda}\binom{2p}{l-m-\lambda}\left(\cos{\frac{I}{2}}\right)^{3l-m-2p-2\lambda}\left(\sin{\frac{I}{2}}\right)^{m-l+2p+2\lambda},
    \end{aligned}
\end{align}

\noindent where $\sigma_{lmp} = \text{max}(0,\;l-m-2p)$ and $\upsilon_{lmp} = \text{min}(l-m,\;2l-2p)$. \par

Unlike the eccentricity functions, the inclination functions do not have infinite summations and can always be written as an exact solution containing sines and cosines. A discussion of the history of these functions, as well as methods to improve computation times, can be found in \citet{GoodingWagner2008}\footnote{The inclination functions found in \citet{GoodingWagner2008}\, $F_{lm}^{k}$, differ slightly in definition to $F_{lmp}$ used in Eq. \ref{eq:app:inclin_func}. This difference is discussed in \ibid.}.

\newpage\clearpage
\section{Moving Beyond the Quadrupole}\label{sec:app:high_l}

Throughout this study we have only considered the quadrupole terms in the tidal dissipation equation ($l=2$). Tidal dissipation drops off quickly at higher orders of $l$ due to the radius over orbital separation scaling factor in Eq. \ref{eq:dissipation}: $(R / a)^{2l + 1}$. For this reason, many tidal studies choose to exclude $l > 2$. However, care should be taken because this choice should not be made purely based on the orbital separation and radius. For high values of eccentricity (or obliquity), terms beyond the quadrupole may become important especially for planets experiencing NSR. This reasoning comes from the fact that eccentricity functions (and to a lesser extent, inclination functions) calculated for $l>3$ can become quite large due to the presence of large coefficients. These functions may then act as large multipliers potentially negating the diminishing effect of a small radius to orbital separation ratio. \par

\begin{figure}[hbt!]
\centering
\includegraphics[width=.9\textwidth]{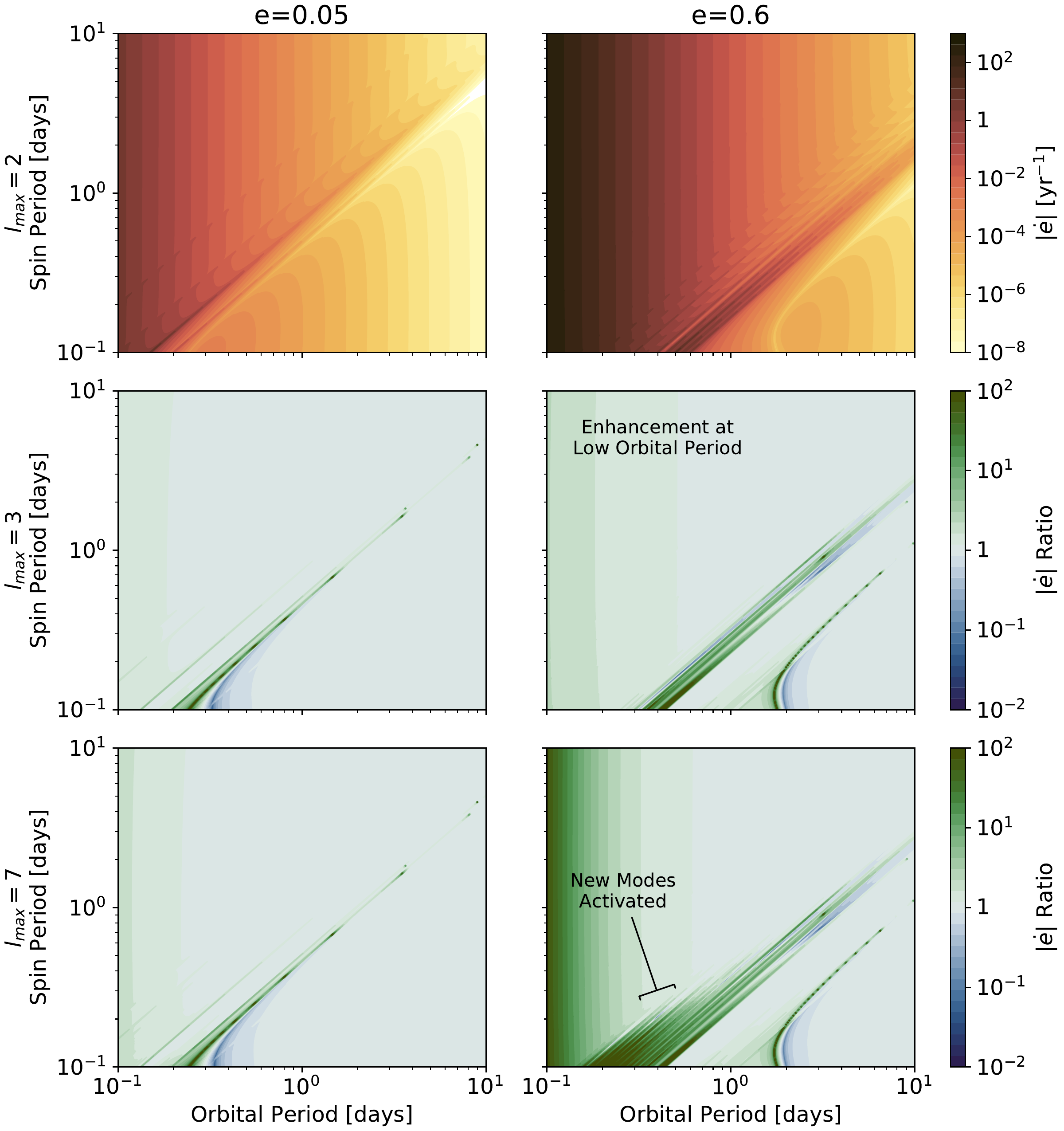}
\caption{Top row: To demonstrate the role of higher non-quadrupole values of $l$, the absolute value of $\dot{e}$ is calculated for $l_{\text{max}}=2$ (as is used elsewhere in this study) while varying Pluto's rotation period (y-axis) and orbital period (x-axis). Charon's spin rate is assumed to be synchronous with the orbital motion and therefore also varies across the x-axis. Two different eccentricity values are used: $e=0.05$ for column one and $e=0.6$ for column two. Eccentricity changes are quickest (darker regions) at low orbital period and near spin-orbit resonances. Second row: ratio between $\dot{e}$ calculated for $l_{\text{max}}=3$ and $l_{\text{max}}=2$ (top row). Last row: ratio between $l_{\text{max}}=7$ and $l_{\text{max}}=2$. Differences between $l_{\text{max}} = 2$ and $l_{\text{max}} > 2$ appear at low orbital period and near higher-order spin-orbit resonances. Higher eccentricity enhances these differences, particularly by the activation of new tidal modes. Utilizing higher orders of $l$ can generate up to 2 orders of magnitude differences in $\dot{e}$. Fine-scale structure seen in these plots is real and of high complexity.}
\label{fig:appen:dedt}
\end{figure}

To show this, we calculate $\dot{e}$ for $l_{\text{max}}=3$ and $l_{\text{max}}=7$ (using the $e^{20}$ truncation level)\footnote{In order to make these calculations we must calculate all tidal dissipation terms starting from $l=2$ to $l=l_{\text{max}}$. Since the number of tidal modes also increases, non-linearly, with $l$ then the calculations can quickly become quite computationally expensive. Therefore, it is important to determine the minimum $l$ a problem warrants.}. We then divide these results by their respective values calculated for $l_{\text{max}}=2$ to emphasize when the higher-order $l$ results diverge from quadrupole (Figure \ref{fig:appen:dedt}). For these calculations we use the planetary properties of Pluto and Charon (see Table \ref{tab:methods:planet_params}) since they are close enough to one another that their $R/a$ is already quite large (Pluto's $R/a \approx 0.06$ compared to Io's $R/a \approx 0.004$). \par

We find that using higher orders of $l$, particularly for high eccentricity, can generate differences of up to two orders of magnitude from the traditional $l_{\text{max}} = 2$. These differences are highly sensitive to the spin state of the planet. At $l>2$, the dissipation equations (Eqs. \ref{eq:dissipation}) depend on a greater number of tidal modes. More of these modes become active at high eccentricity. This can be seen by the increase in number of diagonal line features in the lower-right subplot of Figure \ref{fig:appen:dedt}. Depending on the specifics of the problem, some of these modes may also lead to additional spin-orbit resonance trappings. \par

In conclusion, terms beyond the quadrupole should be considered for objects which are close to their tidal host (relative to their radius) \textit{and} for objects with a large eccentricity or in NSR. \par




\newpage

\bibliography{references}{}
\bibliographystyle{aasjournal}

\listofchanges

\end{document}